\documentclass[preprint]{elsarticle}

\makeatletter
\def\ps@pprintTitle{
 \let\@oddhead\@empty
 \let\@evenhead\@empty
 \def\@oddfoot{\centerline{\thepage}}
 \let\@evenfoot\@oddfoot}
\makeatother

\usepackage{hyperref}
\usepackage{amssymb,amsmath,amsthm,pifont,scalerel,mathtools,setspace,subcaption,tikz,tikz-cd,tabto}

\theoremstyle{definition}
\newtheorem{theorem}{Theorem}[section]
\newtheorem{lemma}[theorem]{Lemma}
\newtheorem{corollary}[theorem]{Corollary}
\newtheorem{definition}[theorem]{Definition}
\newtheorem{conjecture}[theorem]{Conjecture}

\newcommand{\abs}[1]{\ensuremath{\lvert#1\rvert}}
\newcommand{\overbar}[1]{\mkern 1.5mu\overline{\mkern-1.5mu#1\mkern-1.5mu}\mkern 1.5mu}
\newcommand{\restr}[2]{{#1\!\!\mid_{\scaleto{#2}{6pt}}}}

\makeatletter
\newcommand{\xRrightarrow}[2][]{\ext@arrow 0359\Rrightarrowfill@{#1}{#2}}
\newcommand{\Rrightarrowfill@}{\arrowfill@\equiv\equiv\Rrightarrow}
\makeatother

\usepackage{tabularx,environ}
\makeatletter
\newcommand{\probinstance}[1]{\gdef\@probinstance{#1}}%
\newcommand{\probquestion}[1]{\gdef\@probquestion{#1}}
\NewEnviron{prob}{
  \BODY
  \par\addvspace{.25\baselineskip}
  \noindent
  \begin{tabularx}{\textwidth}{@{\hspace{\parindent}} l X c}
    \textbf{Instance:} & \@probinstance \\
    \textbf{Question:} & \@probquestion
  \end{tabularx}
  \par\addvspace{.25\baselineskip}
}
\makeatother

\begin{document}

\begin{frontmatter}

\title{Parallel Hyperedge Replacement Grammars}

\author[1]{Graham Campbell\fnref{fn1}}
\ead{g.j.campbell2@newcastle.ac.uk}

\address[1]{School of Mathematics, Statistics and Physics, Newcastle University, Newcastle upon Tyne, United Kingdom}

\fntext[fn1]{Supported by a Doctoral Training Grant No. (2281162) from the Engineering and Physical Sciences Research Council (EPSRC) in the UK.}

\begin{abstract}
In 2018, it was shown that all finitely generated virtually Abelian groups have multiple context-free word problems, and it is still an open problem as to where to precisely place the word problems of hyperbolic groups in the formal language hierarchy. Motivated by this, we introduce a new language class, the parallel hyperedge replacement string languages, containing all multiple context-free and ET0L languages. We show that parallel hyperedge replacement grammars can be ``synchronised'', which allows us to establish many useful formal language closure results relating to both the hypergraph and string languages generated by various families of parallel hyperedge replacement grammars, laying the foundations for future work in this area.
\end{abstract}

\begin{keyword}
Hyperedge Replacement\sep Lindenmayer Systems\sep E0L Languages\sep ET0L Languages\sep Multiple Context-Free Languages\sep Word Problems
\end{keyword}

\end{frontmatter}

\section{Introduction}

Lindenmayer systems (L systems) originated in the late 1960s to model the development of multicellular organisms \cite{Lindenmayer68a,Lindenmayer68b,Lindenmayer71a}, and since then, the string generational power of various families of L system has been extensively studied \cite{Rozenberg-Salomaa80a,Kari-Rozenberg-Salomaa97a}. The family of extended table zero-interaction Lindenmayer (ET0L) string languages is particularly important due to its closure properties and equivalence to many other families of L systems. On the other side of the picture, hyperedge replacement appeared in the early 1970s \cite{Feder71a,Pavlidis72a} as a generalisation of context-free string rewriting for hypergraphs, and it was shown that the string generational power of hyperedge replacement grammars is the same as that of multiple context-free grammars \cite{Seki-Matsumura-Fujii-Kasami91a,Engelfriet-Heyker91a,Weir92a}.

Looking purely at string language generation, one can view ET0L grammars as a generalisation of context-free grammars with parallel rewriting, and similarly, one can view the class of string generating hyperedge replacement grammars as a generalisation of context-free grammars where intermediate states can be hypergraphs, rather than just strings. In this paper, we combine both of these ideas, to explore the string generational power of grammars that can exploit the additional power of both hypergraphs and parallel rewriting.

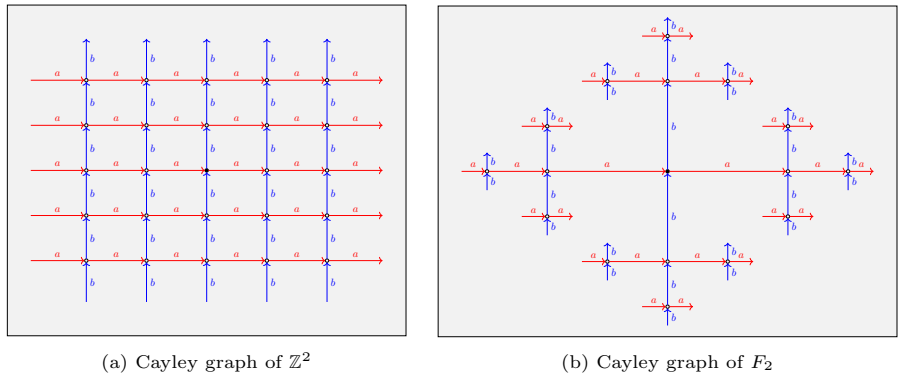
\begin{figure}[!ht]
\begin{subfigure}{.464\textwidth}
\centering
\scalebox{0.4}{
\begin{tikzpicture}[every node/.style={align=center}]

    \draw (-6.625,-5.5) -- (-6.625,5.5) -- (6.625,5.5) -- (6.625,-5.5) -- cycle [fill=black!5];

    \node (UUULLL)    at (-6,4.5)                                                    {\,};
    \node (UUULL)     at (-4,4.5)                                                    {\,};
    \node (UUUL)      at (-2,4.5)                                                    {\,};
    \node (UUUC)      at (0,4.5)                                                     {\,};
    \node (UUUR)      at (2,4.5)                                                     {\,};
    \node (UUURR)     at (4,4.5)                                                     {\,};
    \node (UUURRR)    at (6,4.5)                                                     {\,};

    \node (UULLL)     at (-6,3.0)                                                    {\,};
    \node (UULL)      at (-4,3.0)     [draw, circle, thick, fill=black!5, scale=0.3] {\,};
    \node (UUL)       at (-2,3.0)     [draw, circle, thick, fill=black!5, scale=0.3] {\,};
    \node (UUC)       at (0,3.0)      [draw, circle, thick, fill=black!5, scale=0.3] {\,};
    \node (UUR)       at (2,3.0)      [draw, circle, thick, fill=black!5, scale=0.3] {\,};
    \node (UURR)      at (4,3.0)      [draw, circle, thick, fill=black!5, scale=0.3] {\,};
    \node (UURRR)     at (6,3.0)                                                     {\,};

    \node (ULLL)      at (-6,1.5)                                                    {\,};
    \node (ULL)       at (-4,1.5)     [draw, circle, thick, fill=black!5, scale=0.3] {\,};
    \node (UL)        at (-2,1.5)     [draw, circle, thick, fill=black!5, scale=0.3] {\,};
    \node (UC)        at (0,1.5)      [draw, circle, thick, fill=black!5, scale=0.3] {\,};
    \node (UR)        at (2,1.5)      [draw, circle, thick, fill=black!5, scale=0.3] {\,};
    \node (URR)       at (4,1.5)      [draw, circle, thick, fill=black!5, scale=0.3] {\,};
    \node (URRR)      at (6,1.5)                                                     {\,};

    \node (LLL)       at (-6,0.0)                                                    {\,};
    \node (LL)        at (-4,0.0)     [draw, circle, thick, fill=black!5, scale=0.3] {\,};
    \node (L)         at (-2,0.0)     [draw, circle, thick, fill=black!5, scale=0.3] {\,};
    \node (C)         at (0,0.0)      [draw, circle, thick, fill=black, scale=0.3]   {\,};
    \node (R)         at (2,0.0)      [draw, circle, thick, fill=black!5, scale=0.3] {\,};
    \node (RR)        at (4,0.0)      [draw, circle, thick, fill=black!5, scale=0.3] {\,};
    \node (RRR)       at (6,0.0)                                                     {\,};

    \node (DLLL)      at (-6,-1.5)                                                   {\,};
    \node (DLL)       at (-4,-1.5)    [draw, circle, thick, fill=black!5, scale=0.3] {\,};
    \node (DL)        at (-2,-1.5)    [draw, circle, thick, fill=black!5, scale=0.3] {\,};
    \node (DC)        at (0,-1.5)     [draw, circle, thick, fill=black!5, scale=0.3] {\,};
    \node (DR)        at (2,-1.5)     [draw, circle, thick, fill=black!5, scale=0.3] {\,};
    \node (DRR)       at (4,-1.5)     [draw, circle, thick, fill=black!5, scale=0.3] {\,};
    \node (DRRR)      at (6,-1.5)                                                    {\,};

    \node (DDLLL)     at (-6,-3.0)                                                   {\,};
    \node (DDLL)      at (-4,-3.0)    [draw, circle, thick, fill=black!5, scale=0.3] {\,};
    \node (DDL)       at (-2,-3.0)    [draw, circle, thick, fill=black!5, scale=0.3] {\,};
    \node (DDC)       at (0,-3.0)     [draw, circle, thick, fill=black!5, scale=0.3] {\,};
    \node (DDR)       at (2,-3.0)     [draw, circle, thick, fill=black!5, scale=0.3] {\,};
    \node (DDRR)      at (4,-3.0)     [draw, circle, thick, fill=black!5, scale=0.3] {\,};
    \node (DDRRR)     at (6,-3.0)                                                    {\,};

    \node (DDDLLL)    at (-6,-4.5)                                                   {\,};
    \node (DDDLL)     at (-4,-4.5)                                                   {\,};
    \node (DDDL)      at (-2,-4.5)                                                   {\,};
    \node (DDDC)      at (0,-4.5)                                                    {\,};
    \node (DDDR)      at (2,-4.5)                                                    {\,};
    \node (DDDRR)     at (4,-4.5)                                                    {\,};
    \node (DDDRRR)    at (6,-4.5)                                                    {\,};

    \draw (UULLL)     edge[->,thick,color=red]  node[above] {$a$}  (UULL)
          (UULL)      edge[->,thick,color=red]  node[above] {$a$}  (UUL)
          (UUL)       edge[->,thick,color=red]  node[above] {$a$}  (UUC)
          (UUC)       edge[->,thick,color=red]  node[above] {$a$}  (UUR)
          (UUR)       edge[->,thick,color=red]  node[above] {$a$}  (UURR)
          (UURR)      edge[->,thick,color=red]  node[above] {$a$}  (UURRR);

    \draw (ULLL)      edge[->,thick,color=red]  node[above] {$a$}  (ULL)
          (ULL)       edge[->,thick,color=red]  node[above] {$a$}  (UL)
          (UL)        edge[->,thick,color=red]  node[above] {$a$}  (UC)
          (UC)        edge[->,thick,color=red]  node[above] {$a$}  (UR)
          (UR)        edge[->,thick,color=red]  node[above] {$a$}  (URR)
          (URR)       edge[->,thick,color=red]  node[above] {$a$}  (URRR);

    \draw (LLL)       edge[->,thick,color=red]  node[above] {$a$}  (LL)
          (LL)        edge[->,thick,color=red]  node[above] {$a$}  (L)
          (L)         edge[->,thick,color=red]  node[above] {$a$}  (C)
          (C)         edge[->,thick,color=red]  node[above] {$a$}  (R)
          (R)         edge[->,thick,color=red]  node[above] {$a$}  (RR)
          (RR)        edge[->,thick,color=red]  node[above] {$a$}  (RRR);

    \draw (DLLL)      edge[->,thick,color=red]  node[above] {$a$}  (DLL)
          (DLL)       edge[->,thick,color=red]  node[above] {$a$}  (DL)
          (DL)        edge[->,thick,color=red]  node[above] {$a$}  (DC)
          (DC)        edge[->,thick,color=red]  node[above] {$a$}  (DR)
          (DR)        edge[->,thick,color=red]  node[above] {$a$}  (DRR)
          (DRR)       edge[->,thick,color=red]  node[above] {$a$}  (DRRR);

    \draw (DDLLL)     edge[->,thick,color=red]  node[above] {$a$}  (DDLL)
          (DDLL)      edge[->,thick,color=red]  node[above] {$a$}  (DDL)
          (DDL)       edge[->,thick,color=red]  node[above] {$a$}  (DDC)
          (DDC)       edge[->,thick,color=red]  node[above] {$a$}  (DDR)
          (DDR)       edge[->,thick,color=red]  node[above] {$a$}  (DDRR)
          (DDRR)      edge[->,thick,color=red]  node[above] {$a$}  (DDRRR);

    \draw (DDDLL)     edge[->,thick,color=blue] node[right] {$b$}  (DDLL)
          (DDLL)      edge[->,thick,color=blue] node[right] {$b$}  (DLL)
          (DLL)       edge[->,thick,color=blue] node[right] {$b$}  (LL)
          (LL)        edge[->,thick,color=blue] node[right] {$b$}  (ULL)
          (ULL)       edge[->,thick,color=blue] node[right] {$b$}  (UULL)
          (UULL)      edge[->,thick,color=blue] node[right] {$b$}  (UUULL);

    \draw (DDDL)      edge[->,thick,color=blue] node[right] {$b$}  (DDL)
          (DDL)       edge[->,thick,color=blue] node[right] {$b$}  (DL)
          (DL)        edge[->,thick,color=blue] node[right] {$b$}  (L)
          (L)         edge[->,thick,color=blue] node[right] {$b$}  (UL)
          (UL)        edge[->,thick,color=blue] node[right] {$b$}  (UUL)
          (UUL)       edge[->,thick,color=blue] node[right] {$b$}  (UUUL);

    \draw (DDDC)      edge[->,thick,color=blue] node[right] {$b$}  (DDC)
          (DDC)       edge[->,thick,color=blue] node[right] {$b$}  (DC)
          (DC)        edge[->,thick,color=blue] node[right] {$b$}  (C)
          (C)         edge[->,thick,color=blue] node[right] {$b$}  (UC)
          (UC)        edge[->,thick,color=blue] node[right] {$b$}  (UUC)
          (UUC)       edge[->,thick,color=blue] node[right] {$b$}  (UUUC);

    \draw (DDDR)      edge[->,thick,color=blue] node[right] {$b$}  (DDR)
          (DDR)       edge[->,thick,color=blue] node[right] {$b$}  (DR)
          (DR)        edge[->,thick,color=blue] node[right] {$b$}  (R)
          (R)         edge[->,thick,color=blue] node[right] {$b$}  (UR)
          (UR)        edge[->,thick,color=blue] node[right] {$b$}  (UUR)
          (UUR)       edge[->,thick,color=blue] node[right] {$b$}  (UUUR);

    \draw (DDDRR)     edge[->,thick,color=blue] node[right] {$b$}  (DDRR)
          (DDRR)      edge[->,thick,color=blue] node[right] {$b$}  (DRR)
          (DRR)       edge[->,thick,color=blue] node[right] {$b$}  (RR)
          (RR)        edge[->,thick,color=blue] node[right] {$b$}  (URR)
          (URR)       edge[->,thick,color=blue] node[right] {$b$}  (UURR)
          (UURR)      edge[->,thick,color=blue] node[right] {$b$}  (UUURR);

\end{tikzpicture}
}
\caption{Cayley graph of $\mathbb{Z}^2$}
\end{subfigure}
\begin{subfigure}{.534\textwidth}
\centering
\scalebox{0.4}{
\begin{tikzpicture}[every node/.style={align=center}]

    \draw (-7.625,-5.5) -- (-7.625,5.5) -- (7.625,5.5) -- (7.625,-5.5) -- cycle [fill=black!5];

    \node (C)         at (0,0)        [draw, circle, thick, fill=black, scale=0.3]   {\,};
    \node (R)         at (4,0)        [draw, circle, thick, fill=black!5, scale=0.3] {\,};
    \node (U)         at (0,3)        [draw, circle, thick, fill=black!5, scale=0.3] {\,};
    \node (L)         at (-4,0)       [draw, circle, thick, fill=black!5, scale=0.3] {\,};
    \node (D)         at (0,-3)       [draw, circle, thick, fill=black!5, scale=0.3] {\,};
    \draw (C)         edge[->,thick,color=red]  node[above] {$a$} (R)
          (C)         edge[->,thick,color=blue] node[right] {$b$} (U)
          (L)         edge[->,thick,color=red]  node[above] {$a$} (C)
          (D)         edge[->,thick,color=blue] node[right] {$b$} (C);

    \node (RR)        at (6,0)        [draw, circle, thick, fill=black!5, scale=0.3] {\,};
    \node (RU)        at (4,1.5)      [draw, circle, thick, fill=black!5, scale=0.3] {\,};
    \node (RD)        at (4,-1.5)     [draw, circle, thick, fill=black!5, scale=0.3] {\,};
    \draw (R)         edge[->,thick,color=red]  node[above] {$a$} (RR)
          (R)         edge[->,thick,color=blue] node[right] {$b$} (RU)
          (RD)        edge[->,thick,color=blue] node[right] {$b$} (R);

    \node (RRR)       at (7,0)        {\,};
    \node (RRU)       at (6,0.75)     {\,};
    \node (RRD)       at (6,-0.75)    {\,};
    \draw (RR)        edge[->,thick,color=red]  node[above] {$a$} (RRR)
          (RR)        edge[->,thick,color=blue] node[right] {$b$} (RRU)
          (RRD)       edge[->,thick,color=blue] node[right] {$b$} (RR);

    \node (RUR)       at (5,1.5)      {\,};
    \node (RUU)       at (4,2.25)     {\,};
    \node (RUL)       at (3,1.5)      {\,};
    \draw (RU)        edge[->,thick,color=red]  node[above] {$a$} (RUR)
          (RU)        edge[->,thick,color=blue] node[right] {$b$} (RUU)
          (RUL)       edge[->,thick,color=red]  node[above] {$a$} (RU);

    \node (RDR)       at (5,-1.5)     {\,};
    \node (RDL)       at (3,-1.5)     {\,};
    \node (RDD)       at (4,-2.25)    {\,};
    \draw (RD)        edge[->,thick,color=red]  node[above] {$a$} (RDR)
          (RDD)       edge[->,thick,color=blue] node[right] {$b$} (RD)
          (RDL)       edge[->,thick,color=red]  node[above] {$a$} (RD);

    \node (RR)        at (-6,0)       [draw, circle, thick, fill=black!5, scale=0.3] {\,};
    \node (RU)        at (-4,1.5)     [draw, circle, thick, fill=black!5, scale=0.3] {\,};
    \node (RD)        at (-4,-1.5)    [draw, circle, thick, fill=black!5, scale=0.3] {\,};
    \draw (L)         edge[<-,thick,color=red]  node[above] {$a$} (RR)
          (L)         edge[->,thick,color=blue] node[right] {$b$} (RU)
          (RD)        edge[->,thick,color=blue] node[right] {$b$} (L);

    \node (RRR)       at (-7,0)       {\,};
    \node (RRU)       at (-6,0.75)    {\,};
    \node (RRD)       at (-6,-0.75)   {\,};
    \draw (RR)        edge[<-,thick,color=red]  node[above] {$a$} (RRR)
          (RR)        edge[->,thick,color=blue] node[right] {$b$} (RRU)
          (RRD)       edge[->,thick,color=blue] node[right] {$b$} (RR);

    \node (RUR)       at (-5,1.5)     {\,};
    \node (RUU)       at (-4,2.25)    {\,};
    \node (RUL)       at (-3,1.5)     {\,};
    \draw (RU)        edge[<-,thick,color=red]  node[above] {$a$} (RUR)
          (RU)        edge[->,thick,color=blue] node[right] {$b$} (RUU)
          (RUL)       edge[<-,thick,color=red]  node[above] {$a$} (RU);

    \node (RDR)       at (-5,-1.5)    {\,};
    \node (RDL)       at (-3,-1.5)    {\,};
    \node (RDD)       at (-4,-2.25)   {\,};
    \draw (RD)        edge[<-,thick,color=red]  node[above] {$a$} (RDR)
          (RDD)       edge[->,thick,color=blue] node[right] {$b$} (RD)
          (RDL)       edge[<-,thick,color=red]  node[above] {$a$} (RD);

    \node (UR)        at (2,3)        [draw, circle, thick, fill=black!5, scale=0.3] {\,};
    \node (UU)        at (0,4.5)      [draw, circle, thick, fill=black!5, scale=0.3] {\,};
    \node (UL)        at (-2,3)       [draw, circle, thick, fill=black!5, scale=0.3] {\,};
    \draw (U)         edge[->,thick,color=red]  node[above] {$a$} (UR)
          (U)         edge[->,thick,color=blue] node[right] {$b$} (UU)
          (UL)        edge[->,thick,color=red]  node[above] {$a$} (U);

    \node (URR)        at (3,3)       {\,};
    \node (URU)        at (2,3.75)    {\,};
    \node (URD)        at (2,2.25)    {\,};
    \draw (UR)         edge[->,thick,color=red]  node[above] {$a$} (URR)
          (UR)         edge[->,thick,color=blue] node[right] {$b$} (URU)
          (URD)        edge[->,thick,color=blue] node[right] {$b$} (UR);

    \node (UUR)        at (1,4.5)     {\,};
    \node (UUU)        at (0,5.25)    {\,};
    \node (UUL)        at (-1,4.5)    {\,};
    \draw (UU)         edge[->,thick,color=red]  node[above] {$a$} (UUR)
          (UU)         edge[->,thick,color=blue] node[right] {$b$} (UUU)
          (UUL)        edge[->,thick,color=red]  node[above] {$a$} (UU);

    \node (ULU)        at (-2,3.75)   {\,};
    \node (ULL)        at (-3,3)      {\,};
    \node (ULD)        at (-2,2.25)   {\,};
    \draw (UL)         edge[->,thick,color=blue] node[right] {$b$} (ULU)
          (ULL)        edge[->,thick,color=red]  node[above] {$a$} (UL)
          (ULD)        edge[->,thick,color=blue] node[right] {$b$} (UL);

    \node (UR)        at (2,-3)       [draw, circle, thick, fill=black!5, scale=0.3] {\,};
    \node (UU)        at (0,-4.5)     [draw, circle, thick, fill=black!5, scale=0.3] {\,};
    \node (UL)        at (-2,-3)      [draw, circle, thick, fill=black!5, scale=0.3] {\,};
    \draw (D)         edge[->,thick,color=red]  node[above] {$a$} (UR)
          (D)         edge[<-,thick,color=blue] node[right] {$b$} (UU)
          (UL)        edge[->,thick,color=red]  node[above] {$a$} (D);

    \node (URR)        at (3,-3)      {\,};
    \node (URU)        at (2,-3.75)   {\,};
    \node (URD)        at (2,-2.25)   {\,};
    \draw (UR)         edge[->,thick,color=red]  node[above] {$a$} (URR)
          (UR)         edge[<-,thick,color=blue] node[right] {$b$} (URU)
          (URD)        edge[<-,thick,color=blue] node[right] {$b$} (UR);

    \node (UUR)        at (1,-4.5)    {\,};
    \node (UUU)        at (0,-5.25)   {\,};
    \node (UUL)        at (-1,-4.5)   {\,};
    \draw (UU)         edge[->,thick,color=red]  node[above] {$a$} (UUR)
          (UU)         edge[<-,thick,color=blue] node[right] {$b$} (UUU)
          (UUL)        edge[->,thick,color=red]  node[above] {$a$} (UU);

    \node (ULU)        at (-2,-3.75)  {\,};
    \node (ULL)        at (-3,-3)     {\,};
    \node (ULD)        at (-2,-2.25)  {\,};
    \draw (UL)         edge[<-,thick,color=blue] node[right] {$b$} (ULU)
          (ULL)        edge[->,thick,color=red]  node[above] {$a$} (UL)
          (ULD)        edge[<-,thick,color=blue] node[right] {$b$} (UL);

\end{tikzpicture}
}
\caption{Cayley graph of $F_2$}
\end{subfigure}
\caption{Example Cayley graphs}
\label{fig:cayley}
\end{figure}

A major motivation of this work is the study of word problems of finitely generated groups. In general, the word problem is the question that asks if two strings (words) represent the same element in some structure. In the case of groups, this is the equivalent to asking if a given string represents the identity element, since if \(u\), \(v\) are strings, then they are equal in a group if and only if \(u v^{-1}\) represents the identity in the group. Thus, given a presentation \(\langle X \mid R \rangle\) for a group \(G\), the word problem is equivalent to the membership problem for the string language \(\mathrm{WP}_X(G) = \{ w \in (X \cup X^{-1})^* \mid w =_G 1_G\}\). Recall that the Cayley graph of a group with respect to a fixed presentation has the group elements as vertices and an edge between two vertices \(g, h \in G\) whenever there is a generator \(s \in X\) such that \(g s = h\). Viewing things geometrically, the word problem of a group can be identified with the set of loops based at the identity in the Cayley graph. A sketch of a finite portion of the Cayley graphs of $\mathbb{Z}^2$ and $F_2$ with their usual generating sets is provided in Figure \ref{fig:cayley}.

\vspace{0.25em}

A natural question to ask is how hard the word problem is, in general, and for specific families of groups. Unsurprisingly, both the universal word problem and the word problem are undecidable in general, even for finite presentations \cite{Novikov55a}. It is well known that a presentation defines a finite group if and only if it admits a regular word problem \cite{Anisimov71a}, and defines a finitely generated virtually free group if and only if it admits a deterministic context-free word problem if and only if it admits a context-free word problem \cite{Muller-Schupp83a}. The multiple context-free (MCF) languages sit strictly in between the context-free and context-sensitive languages \cite{Seki-Matsumura-Fujii-Kasami91a}. In 2015, a major breakthrough of Salvati was published, showing that the word problem of \(\mathbb{Z}^2\) is an MCF language \cite{Salvati15a}, and in 2018, Ho extended this result to all finitely generated virtually Abelian groups \cite{Ho18a}. This is interesting since the MCF languages are exactly the string languages generated by hyperedge replacement grammars \cite{Engelfriet-Heyker91a,Weir92a}. It remains an open problem as to which other families of groups admit MCF word problems, however, we do at least know that the fundamental group of a hyperbolic three-manifold does not admit an MCF word problem \cite{Gilman-Kropholler-Schleimer18a}.

\begin{figure}[!ht]
\begin{subfigure}{.499\textwidth}
\centering
\scalebox{0.9}{
\begin{tikzpicture}
  \node (b) at (0,1.8) {$\mathcal{C}\mathcal{S}$};
  \node (d) at (2,0.9) {$\mathcal{I}\mathcal{N}\mathcal{D}\mathcal{E}\mathcal{X}$};
  \node (e) at (-2,0.45) {$\mathcal{M}\mathcal{C}\mathcal{F}$};
  \node (f) at (2,0) {$\mathcal{E}\mathcal{T}\mathcal{O}\mathcal{L}$};
  \node (g) at (0,-0.9) {$\mathcal{C}\mathcal{F}$};
  \node (h) at (0,-1.8) {$\mathcal{D}\mathcal{C}\mathcal{F}$};
  \node (i) at (0,-2.7) {$\mathcal{R}\mathcal{E}\mathcal{G}$};
  \draw (b) -- (e) -- (g);
  \draw (b) -- (d) -- (f) -- (g);
  \draw (g) -- (h) -- (i);
\end{tikzpicture}
}
\caption{String language hierarchy}
\end{subfigure}
\begin{subfigure}{.499\textwidth}
\centering
\scalebox{0.9}{
\begin{tikzpicture}
  \node (b) at (0,1.8) {$\mathcal{C}\mathcal{S} \cap \mathcal{G}\mathcal{P}$};
  \node (d) at (2,0.9) {$\mathcal{I}\mathcal{N}\mathcal{D}\mathcal{E}\mathcal{X} \cap \mathcal{G}\mathcal{P}$};
  \node (e) at (-2,0.45) {$\mathcal{M}\mathcal{C}\mathcal{F} \cap \mathcal{G}\mathcal{P}$};
  \node (f) at (2,0) {$\mathcal{E}\mathcal{T}\mathcal{O}\mathcal{L} \cap \mathcal{G}\mathcal{P}$};
  \node (g) at (0,-1.3) {$\mathcal{D}\mathcal{C}\mathcal{F} \cap \mathcal{G}\mathcal{P} = \mathcal{C}\mathcal{F} \cap \mathcal{G}\mathcal{P}$};
  \node (h) at (0,-2.7) {$\mathcal{R}\mathcal{E}\mathcal{G} \cap \mathcal{G}\mathcal{P}$};
  \draw (e) -- (b);
  \draw (e) -- (g);
  \draw (b) -- (d);
  \draw[dashed] (d) -- (f) -- (g);
  \draw (g) -- (h);
\end{tikzpicture}
}
\caption{Group language hierarchy}
\end{subfigure}
\caption{Previously known formal language hierarchies}
\label{fig:knownlh}
\end{figure}
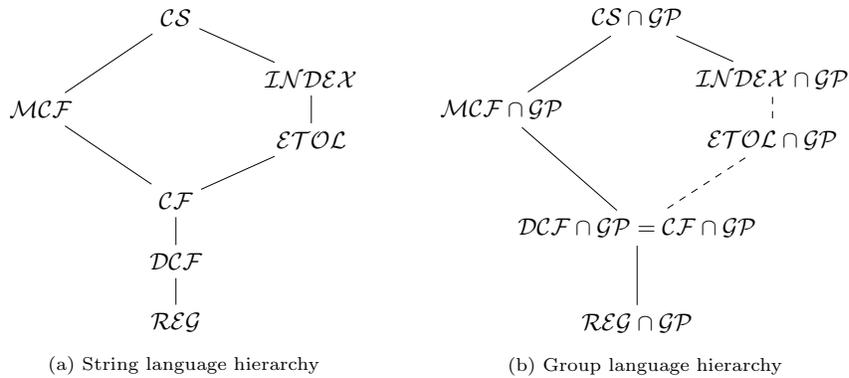

There are of course, lots of other well-behaved language classes sitting in between the context-free and context-sensitive classes, such as the indexed languages \cite{Aho68a} or the subclass of ET0L languages \cite{Rozenberg-Salomaa80a}. It is not known if there are any groups with indexed word problems, other than the virtually free groups, but it is known that a particular subclass of the indexed languages, not contained in ET0L, only contains word problems of virtually free groups \cite{Gilman-Shapiro98a}. We also do not know if any hyperbolic groups have ET0L word problems \cite{Ciobanu-Elder-Ferov18a} (other than the virtually free groups), such as the fundamental group of the double torus. It is conjectured that every ET0L group language is admitted by a virtually free group \cite{Ciobanu-Elder-Ferov18a}. Figure \ref{fig:knownlh} shows the (group) language hierarchy, where necessarily strict inclusion uses a solid line, and \(\mathcal{G}\mathcal{P}\) denotes the class of all group languages (the class of word problems of all finitely generated groups). Descriptions of the other language classes can be found in Section \ref{sec:prelim}.

\vspace{0.1em}

\begin{figure}[!ht]
\begin{subfigure}{.499\textwidth}
\centering
\scalebox{0.9}{
\begin{tikzpicture}
  \node[align=center] (a) at (-2,3.6) {$\mathcal{R}\mathcal{E}\mathcal{C}$};
  \node[align=center] (b) at (0,2.7) {$\mathcal{C}\mathcal{S}$};
  \node[align=center] (y) at (0,1.56) {$\mathcal{P}\mathcal{H}\mathcal{R}\mathcal{S}$};
  \node[align=center] (x) at (0,0.66) {$\mathcal{P}\mathcal{H}\mathcal{R}\mathcal{S}^{\mathrm{rf}}$};
  \node[align=center] (d) at (2,0.9) {$\mathcal{I}\mathcal{N}\mathcal{D}\mathcal{E}\mathcal{X}$};
  \node[align=center] (e) at (-2,0) {$\mathcal{M}\mathcal{C}\mathcal{F}$};
  \node[align=center] (f) at (2,0) {$\mathcal{E}\mathcal{T}\mathcal{O}\mathcal{L}$};
  \node[align=center] (g) at (0,-0.9) {$\mathcal{C}\mathcal{F}$};
  \node[align=center] (h) at (0,-1.8) {$\mathcal{D}\mathcal{C}\mathcal{F}$};
  \node[align=center] (i) at (0,-2.7) {$\mathcal{R}\mathcal{E}\mathcal{G}$};
  \draw (a) -- (b);
  \draw (y) -- (a);
  \draw[dashed] (x) -- (y);
  \draw (e) -- (x);
  \draw (f) -- (x);
  \draw (b) -- (e) -- (g);
  \draw (b) -- (d) -- (f) -- (g);
  \draw (g) -- (h) -- (i);
\end{tikzpicture}
}
\caption{Proved string language hierarchy}
\end{subfigure}
\begin{subfigure}{.499\textwidth}
\centering
\scalebox{0.9}{
\begin{tikzpicture}
  \node[align=center] (a) at (0,3.6) {$\mathcal{R}\mathcal{E}\mathcal{C} \cap \mathcal{G}\mathcal{P}$};
  \node[align=center] (b) at (0,2.7) {$\mathcal{C}\mathcal{S} \cap \mathcal{G}\mathcal{P}$};
  \node[align=center] (c) at (0,1.35) {$\mathcal{P}\mathcal{H}\mathcal{R}\mathcal{S}^{\mathrm{rf}} \cap \mathcal{G}\mathcal{P} \overset{\text{?}}{=} \mathcal{P}\mathcal{H}\mathcal{R}\mathcal{S} \cap \mathcal{G}\mathcal{P}$};
  \node[align=center] (d) at (0,0) {$\mathcal{M}\mathcal{C}\mathcal{F} \cap \mathcal{G}\mathcal{P}$};
  \node[align=center] (e) at (0,-1.3) {$\mathcal{D}\mathcal{C}\mathcal{F} \cap \mathcal{G}\mathcal{P} \overset{\text{\ding{51}}}{=} \mathcal{C}\mathcal{F} \cap \mathcal{G}\mathcal{P}$\\$\overset{\text{?}}{=} \mathcal{E}\mathcal{T}\mathcal{O}\mathcal{L} \cap \mathcal{G}\mathcal{P} \overset{\text{?}}{=} \mathcal{I}\mathcal{N}\mathcal{D}\mathcal{E}\mathcal{X} \cap \mathcal{G}\mathcal{P}$};
  \node[align=center] (f) at (0,-2.7) {$\mathcal{R}\mathcal{E}\mathcal{G} \cap \mathcal{G}\mathcal{P}$};
  \draw (a) -- (b) -- (c) -- (d) -- (e) -- (f);
\end{tikzpicture}
}
\caption{Conjectured group language hierarchy}
\end{subfigure}
\caption{New formal language hierarchies}
\label{fig:newlh}
\end{figure}
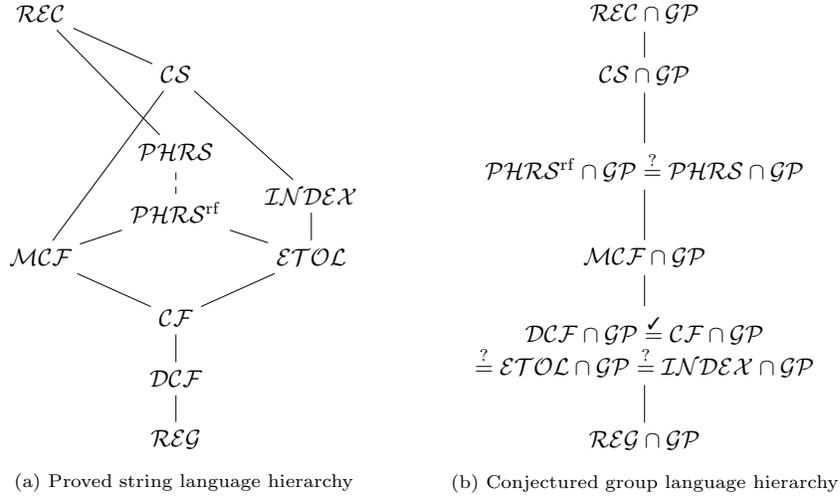

In this paper, we define and study a new string language class, combining ideas from both ET0L and hyperedge replacement grammars. We call our new class the parallel hyperedge replacement string (PHRS) languages, and show that the class strictly contains both the classes of MCF and ET0L languages, that it is a hyper-algebraically closed super abstract family of languages, and that PHRS group languages are closed under free product. While parallel hyperedge replacement has been considered before, most notably by Habel and Kreowski (separately) \cite{Habel92b,Kreowski92a,Kreowski93a}, the work is not extensive and does not consider repetition-freeness, rational control, number of tables, or string generational power. Figure \ref{fig:newlh} summarises how the PHRS and repetition-free PHRS languages fit into the string language hierarchy and also how we conjecture the hierarchy collapses when we restrict to group languages. At the end of the paper, we provide much more detailed diagrams, showing both hypergraph and string language hierarchies. Our long term goal is to place the word problem for as many hyperbolic groups as possible in the PHRS class. Knowledge of (geometric) group theory and word problems is not required to read and understand this paper - it is purely motivational!

This paper is based on, and replaces, the author's TERMGRAPH 2020 workshop proceedings paper \cite{Campbell21a}, with significant revisions, improvements, and new results. Most notably, in the workshop paper, the number of tables was not considered at all. Consideration of the number of tables was a major undertaking, leading to the strengthening of many of the existing results, many of which requiring entirely new proofs, also leading to the creation of Subsection \ref{subsec:subc} looking at synchronisation of grammars and substitution closure. Moreover, we were able to additionally show that the PHRS languages are hyper-algebraically closed. The second most notable addition is showing that the PHRS membership problem is decidable (Subsection \ref{subsec:emptiness}) which provides us with an algorithm for computing membership of PHRS languages (Subsection \ref{subsec:phrsmp}). Finally, without the space limitations of a workshop paper, we were able to provide a more self-contained preliminaries section (Section \ref{sec:prelim}), some additional examples in the main part of the paper, and a more detailed handling of WPHRS languages.

\vspace{0.1em}

\section{Preliminaries} \label{sec:prelim}

Before we get started, we must give some preliminary notations and known results. By \(\mathbb{N}\) we mean the natural numbers including zero, by \(\underline{n}\) we mean \(\{1, \dots, n\}\), and \(\oplus\) denotes relational override. Given a set \(S\), denote by \(\mathcal{P}(S)\) the set of all subsets of \(S\) and \(\mathcal{P}_0(S) = \mathcal{P}(S) \setminus \{\emptyset\}\). A (finite) sequence on a set \(S\) is a function \(\sigma: \underline{n} \to S\), we view strings as sequences on an alphabet. We denote by \(S^*\) all sequences on a set \(S\), and denote subsequence (scattered subsequence) by \(\sqsubseteq\) (\(\sqsubseteq_{\mathrm{sc}}\)).

\subsection{String Languages} \label{subsec:strings}

In this paper, all alphabets will be finite. We can equivalently view \(A^*\) as the free monoid on \(A\). A string language \(L \subseteq A^*\) is simply a set of strings over some alphabet \(A\). It is elementary that monoid homomorphisms \(\varphi: A^* \to B^*\) are totally determined by their values on \(A\). We call \(\varphi\) non-erasing if \(\epsilon \not\in \varphi(A)\), a coding (letter-to-letter homomorphism) if \(\varphi(A) \subseteq B\), and a weak coding if \(\varphi(A) \subseteq B \cup \{\epsilon\}\).

By a family of string languages, we mean an isomorphism-closed (closed under renaming of symbols), non-trivial (contains languages other than \(\emptyset, \{\epsilon\}\)) class of string languages. Given a family of string languages \(\mathcal{F}\) and alphabets \(A, B\), an \(\mathcal{F}\)-substitution of strings on \(A\) is a function \(h: A \to \mathcal{P}_0(B^*)\) such that for each \(a \in A\), \(h(a) \in \mathcal{F}\). When \(\mathcal{F}\) is the family of all finite string languages, we call \(h\) a finite substitution of strings. If \(h\) is such that for each \(a \in A\), \(a \in h(a)\), then we call \(h\) nested, and if \(\epsilon \not\in h(A)\), then we call \(h\) non-erasing. Given a string \(w = x_1 \cdots x_n\) over \(A\), write \(h(w)\) for the language \(\{w' \in B^* \mid \exists w_1 \in h(x_1), \dots, \exists w_n \in h(x_n), w' = w_1 \cdots w_n\}\), and given a string language \(L\) over \(A\), write \(h(L)\) for the language \(\bigcup_{w \in L} h(w)\).

A family of string languages \(\mathcal{F}\) is called a cone if it is closed under homomorphisms, inverse homomorphisms, and intersection with regular languages. That is, if \(L \subseteq A^*\) is an \(\mathcal{F}\) language, \(L' \subseteq B^*\) is an \(\mathcal{F}\) language, \(K \subseteq A^*\) is a regular language, and \(\varphi: A^* \to B^*\) is a homomorphism, then \(\varphi(L) \in \mathcal{F}\), \(\varphi^{-1}(L') \in \mathcal{F}\), and \(L \cap K \in \mathcal{F}\). The rational operations are union (\(L_1 \cup L_2\)), concatenation (\(L_1 L_2 = \{w_1 w_2 \mid w_1 \in L_1, w_2 \in L_2\}\)), and Kleene plus (\(L^+ = \bigcup_{i \geq 1} \{w^i \mid w \in L\}\)). \(\mathcal{F}\) is called a full abstract family of languages (full AFL) if it is both a cone and is closed under the rational operations. That is, if \(L_1, L_2 \subseteq A^*\) are \(\mathcal{F}\) languages, then \(L_1 \cup L_2 \in \mathcal{F}\), \(L_1 L_2 \in \mathcal{F}\), and \(L_1^+ \in \mathcal{F}\). Call \(\mathcal{F}\) substitution closed if for any \(\mathcal{F}\) language \(L \subseteq A^*\) and \(\mathcal{F}\)-substitution \(h\) of strings on \(A\), \(h(L) \in \mathcal{F}\). Call \(\mathcal{F}\) (nested) iterated substitution closed if for any \(\mathcal{F}\) language \(L \subseteq A^*\) and a (nested) \(\mathcal{F}\)-substitution \(h\) of strings on \(A\), \(\bigcup_{n \in \mathbb{N}} h^n(L) \in \mathcal{F}\), where \(h^n(L) = h( \cdots h(L) \cdots )\). A full AFL is called a super AFL if it is additionally substitution and nested iterated substitution closed.

We denote by \(\mathcal{R}\mathcal{E}\mathcal{G}\), \(\mathcal{D}\mathcal{C}\mathcal{F}\), \(\mathcal{C}\mathcal{F}\), \(\mathcal{C}\mathcal{S}\), \(\mathcal{R}\mathcal{E}\mathcal{C}\), \(\mathcal{R}\mathcal{E}\) the families of regular, deterministic context-free (DCF), context-free (CF), context sensitive, recursive, and recursively enumerable string languages, respectively. One can consult any undergraduate textbook on formal languages and automata such as \cite{Hopcroft-Motwani-Ullman06a} for an introduction to these language families. These language families enjoy the following hierarchy: \(\mathcal{R}\mathcal{E}\mathcal{G} \subsetneq \mathcal{D}\mathcal{C}\mathcal{F} \subsetneq \mathcal{C}\mathcal{F} \subsetneq \mathcal{C}\mathcal{S} \subsetneq \mathcal{R}\mathcal{E}\mathcal{C} \subsetneq \mathcal{R}\mathcal{E}\).

\subsection{MCF Languages} \label{subsec:mcf}

The multiple context-free languages were defined by Seki, Matsumura, Fujii, and Kasami in the early 90s \cite{Seki-Matsumura-Fujii-Kasami91a} as a conservative extension of the context-free language, showing that the family was a substitution closed full AFL, and closure under nested iterated substitution is also obvious by trivial modification of Kr\'al's proof for context-free languages \cite{Kral70a} (Theorem \ref{thm:mcfafl}). Various alternative descriptions of the MCF languages have now been established (Theorem \ref{thm:mcfequiv}). The actual definition of MCF grammars will not matter to us, since we will only be interested in the hyperedge replacement characterisation. We denote by \(\mathcal{M}\mathcal{C}\mathcal{F}_k\) and \(\mathcal{M}\mathcal{C}\mathcal{F}\) the order at most \(k\) multiple context-free languages and the multiple context-free languages, respectively. These families enjoy the following hierarchy: \(\mathcal{C}\mathcal{F} = \mathcal{M}\mathcal{C}\mathcal{F}_1 \subsetneq \cdots \subsetneq \mathcal{M}\mathcal{C}\mathcal{F}_k \subsetneq \mathcal{M}\mathcal{C}\mathcal{F}_{k+1} \subsetneq \cdots \subsetneq \mathcal{M}\mathcal{C}\mathcal{F} \subsetneq \mathcal{C}\mathcal{S}\).

\begin{theorem}[MCF Closure Properties \cite{Kral70a, Seki-Matsumura-Fujii-Kasami91a}] \label{thm:mcfafl}
For \(k \geq 1\), \(\mathcal{M}\mathcal{C}\mathcal{F}_{k}\) and \(\mathcal{M}\mathcal{C}\mathcal{F}\) are super AFLs.
\end{theorem}

\begin{theorem}[MCF Equivalence] \label{thm:mcfequiv}
The following families of languages are equivalent, for any \(k \geq 1\):
\begin{enumerate}
\item \(\mathcal{H}\mathcal{R}\mathcal{S}_{2k} = \mathcal{H}\mathcal{R}\mathcal{S}_{2k+1} = \mathcal{H}\mathcal{R}\mathcal{S}_{2k}^{\mathrm{rf}} = \mathcal{H}\mathcal{R}\mathcal{S}_{2k+1}^{\mathrm{rf}}\): the string languages of (repetition-free) hyperedge replacement grammars of order \(2k\) or \(2k + 1\) (see Subsection \ref{subsec:hr});
\item \(\mathrm{OUT}(\mathcal{D}\mathcal{T}\mathcal{W}\mathcal{T}_{k})\): the output languages of deterministic tree walking transducers of crossing number at most \(k\) (see \cite{Aho-Ullman72a});
\item \(\mathcal{L}\mathcal{C}\mathcal{F}\mathcal{R}_{k}\): the string languages of linear context-free rewriting systems of rank at most \(k\) (see \cite{Shanker-Weir-Joshi87a});
\item \(\mathcal{M}\mathcal{C}\mathcal{F}_{k}\): the languages of \(k\)-multiple context-free grammars (see \cite{Seki-Matsumura-Fujii-Kasami91a});
\item \(\mathcal{R}\mathcal{T}\mathcal{S}\mathcal{A}_{k}\): the languages of \(k\)-restricted tree stack automata (see \cite{Denkinger16a}).
\end{enumerate}
\end{theorem}

\begin{proof}
\(\mathcal{H}\mathcal{R}\mathcal{S}_{k} = \mathcal{H}\mathcal{R}\mathcal{S}_{k}^{\mathrm{rf}}\) for all \(k \geq 2\) is due to Theorem \ref{thm:hrrf} and \(\mathcal{H}\mathcal{R}\mathcal{S}_{2k+1}^{\mathrm{rf}} \subseteq \mathrm{OUT}(\mathcal{D}\mathcal{T}\mathcal{W}\mathcal{T}_{k}) \subseteq \mathcal{H}\mathcal{R}\mathcal{S}_{2k}\) for all \(k \geq 1\) is due to Engelfriet and Heyker (1991) \cite{Engelfriet-Heyker91a}, which gives us the all the equalities in \((1)\) and \((1) = (2)\). \((2) = (3)\) is due to Weir (1992) \cite{Weir92a}, \((3) = (4)\) is due to Seki, Matsumura, Fujii, and Kasami (1991) \cite{Seki-Matsumura-Fujii-Kasami91a}, and \((4) = (5)\) is due to Denkinger (2016) \cite{Denkinger16a}.
\end{proof}

Call a set \(S \subseteq \mathbb{N}^d\) linear if it is of the form \(\{p + a_1 p_1 + \cdots a_n p_n \mid a_1, \dots a_n \in \mathbb{N}\}\) for some fixed \(p, p_1, \dots p_n \in \mathbb{N}^d\). Call \(S\) semilinear if it is a finite union of linear sets. The Parikh vector of a string \(w\) over an ordered alphabet \(A = (a_1, \dots, a_d)\) is \(\psi_A(w) = (\abs{w}_{a_1}, \dots, \abs{w}_{a_d})\) where \(\abs{w}_{a_i}\) counts the number of occurrences of \(a_i\) in \(w\). A string language \(L\) is called linear (semilinear) if the set of Parikh vectors \(\psi_A(L)\) is linear (semilinear). Two string languages \(L_1, L_2 \subseteq A^*\) are called letter-equivalent if \(\psi_A(L_1) = \psi_A(L_2)\). In 1961, Parikh circulated the following famous results, re-published in 1966, showing that all context-free languages are semilinear, and that being semilinear amounts to being letter-equivalent to a regular language:

\begin{theorem}[Parikh's Theorem I \cite{Parikh66a}]
If \(L \in \mathcal{C}\mathcal{F}\), then \(L\) is semilinear.
\end{theorem}

\begin{theorem}[Parikh's Theorem II \cite{Parikh66a}]
A string language \(L\) is semilinear if and only if it is letter-equivalent to a regular language. 
\end{theorem}

\begin{corollary} \label{cor:semilinreg}
If \(L \subseteq \{a\}^*\) is semilinear, then \(L \in \mathcal{R}\mathcal{E}\mathcal{G}\).
\end{corollary}

In 1991, Seki et al. showed that all MCF languages are semilinear:

\begin{theorem}[MCF Semilinear \cite{Seki-Matsumura-Fujii-Kasami91a}] \label{thm:mcfsemilin}
If \(L \in \mathcal{M}\mathcal{C}\mathcal{F}\), then \(L\) is semilinear.
\end{theorem}

\subsection{ET0L Languages} \label{subsec:et0l}

Lindenmayer systems (L systems) were introduced in 1968 by Aristid Lindenmayer \cite{Lindenmayer68a,Lindenmayer68b,Lindenmayer71a}. We direct the reader to \cite{Rozenberg-Salomaa80a} for a comprehensive introduction to the topic. We are interested in the family of string languages called the ET0L languages. These languages are described by a specific type of L system.

Given a set \(H\) of substitutions of strings on \(\Sigma\) and a language \(L \subseteq \Sigma^*\), we define \(\mathrm{ITER}_H(L) = \bigcup_{n \in \mathbb{N}} \{h_n( \cdots h_1(L) \cdots ) \mid h_1, \dots, h_n \in H\}\). A table over \(\Sigma\) is a left-total finite binary relation \(T \subseteq \Sigma \times \Sigma^{*}\), and can be associated to a finite substitution \(h_T\) where \(h_T(a) = \{b \in \Sigma^* \mid (a, b) \in T\}\) for each \(a \in \Sigma\). An ET0L grammar is a tuple \(\mathcal{G} = (\Sigma, A, S, \mathcal{T})\) where \(\Sigma\) is an alphabet, \(A \subseteq \Sigma\) is the terminal alphabet, \(S \in \Sigma\) is the start symbol and \(\mathcal{T}\) is a finite set of tables over \(\Sigma\). The language generated by \(\mathcal{G}\) is \(\mathrm{L}(\mathcal{G}) = \mathrm{ITER}_{\{h_{T} \mid T \in \mathcal{T}\}}(\{S\}) \cap A^*\). A language \(L \subseteq A^*\) is called an ET0L language if there exists an ET0L grammar \(\mathcal{G}\) such that \(\mathrm{L}(\mathcal{G}) = L\), and is called E0L if the grammar contains only one table. It will be convenient to think of table entries as rules and substitutions as parallel  rewriting, and we will take this view in Section \ref{sec:phr}. We denote the E0L, ET0L, and indexed language families by \(\mathcal{E}\mathcal{O}\mathcal{L}\), \(\mathcal{E}\mathcal{T}\mathcal{O}\mathcal{L}\), and \(\mathcal{I}\mathcal{N}\mathcal{D}\mathcal{E}\mathcal{X}\), respectively, and we have the following strict inclusions: \(\mathcal{C}\mathcal{F} \subsetneq \mathcal{E}\mathcal{O}\mathcal{L} \subsetneq \mathcal{E}\mathcal{T}\mathcal{O}\mathcal{L} \subsetneq \mathcal{I}\mathcal{N}\mathcal{D}\mathcal{E}\mathcal{X} \subsetneq \mathcal{C}\mathcal{S}\).

Call an ET0L grammar propagating if each table is contained in \(\Sigma \times \Sigma^{+}\) (rather than just \(\Sigma \times \Sigma^{*}\)). That is, rules have non-empty right-hand sides. ET0L grammars can be assumed to be propagating:

\begin{theorem}[Propagating ET0L Generative Power \cite{Rozenberg-Salomaa80a}] \label{thm:emptyrhs}
Given an ET0L grammar \(\mathcal{G}\), one can effectively construct a propagating ET0L grammar \(\mathcal{G}'\) with the same number of tables such that \(\mathrm{L}(\mathcal{G}) \setminus \{\epsilon\} = \mathrm{L}(\mathcal{G}')\).
\end{theorem}

An extremely useful notion in L systems is that of a synchronised system. Call an ET0L grammar \((\Sigma, A, S, \mathcal{T})\) synchronised if any string \(w, \in A^*\) and any table \(T \in \mathcal{T}\), \(h_T(w) \not\in A^*\). ET0L grammars can always be synchronised:

\begin{theorem}[Synchronisation of ET0L Grammars \cite{Rozenberg-Salomaa80a}] \label{thm:et0lsync}
Given a (propagating) ET0L grammar \(\mathcal{G}\), one can effectively construct a synchronised (propagating) ET0L grammar \(\mathcal{G}'\) with the same number of tables such that \(\mathrm{L}(\mathcal{G}) = \mathrm{L}(\mathcal{G}')\).
\end{theorem}

A family of string languages \(\mathcal{F}\) is called hyper-algebraically closed if for any \(\mathcal{F}\) language \(L \subseteq A^*\) and finite set \(H\) of \(\mathcal{F}\)-substitutions of strings on \(A\), \(\mathrm{ITER}_H(L) \in \mathcal{F}\). Clearly hyper-algebraic closure implies closure under (nested) iterated substitution, viewed as a special case with \(\abs{H} = 1\).

\begin{theorem}[ET0L Closure Properties \cite{Rozenberg-Salomaa74a}] \label{thm:etolafl}
    \(\mathcal{E}\mathcal{T}\mathcal{O}\mathcal{L}\) is a hyper-algebraically closed super AFL. Moreover, it is the smallest hyper-algebraically closed super AFL. \(\mathcal{E}\mathcal{O}\mathcal{L}\) is closed under finite substitution, intersection with regular languages, and rational operations.
\end{theorem}

For any \(n \geq 1\), the order \(n\) Dyck language \(D_n\) is generated by the context-free grammar with start symbol \(S\), terminal set \(\{a_1, a'_1, \dots, a_n, a'_n\}\), and rules \(S \to \epsilon, S \to SS, S \to a_1 S a'_1, \dots, S \to a_n S a'_n\). In 1977, Ehrenfeucht and Rozenberg showed that for all \(n \geq 8\), \(D_n\) is not an EDT0L language \cite{Ehrenfeucht-Rozenberg77a}. With not too much work, this gives us the following result:

\begin{theorem}[MCF and ET0L Incomparable] \label{thm:incompar}
\,
\begin{enumerate}
    \item \(K = \{w h(w) \mid w \in D_8\}\) is a \(2\)-MCF language which is not ET0L, where \(\overbar{\Sigma}\) is a disjoint copy of \(\Sigma\) and \(h: \Sigma^* \to \overbar{\Sigma}^*\) is defined by sending each \(a \in \Sigma\) to its copy \(\overbar{a} \in \overbar{\Sigma}\).
    \item \(L = \{a^{2^n} \mid n \geq 0\}\) is an E0L language but not MCF. Moreover, \(L\) is not semilinear.
\end{enumerate}
\end{theorem}

\begin{proof}
The first part follows from Theorem 8 of \cite{Nishida-Seki00a}, which holds for any language in place of \(D_8\) which is context-free but not EDT0L.

For the second part, it is easy to see that \(\mathcal{G} = (\{a\}, \{a\}, a, \{\{(a, aa)\}\})\) is an ET0L grammar using one table with \(\mathrm{L}(\mathcal{G}) = L\). Corollary \ref{cor:semilinreg} tells us \(L\) is semilinear if and only if it is regular, and clearly that language is not regular, so it cannot be semilinear. Theorem \ref{thm:mcfsemilin} tells us all MCF languages are semilinear, so it must be the case that \(L\) is not MCF.
\end{proof}

\subsection{Hyperedge Replacement} \label{subsec:hr}

This subsection is mostly based on \cite{Habel92b,Drewes-Kreowski-Habel97a}. By a signature we mean a pair \(\mathcal{C} = (\Sigma, \mathrm{type})\) where \(\Sigma\) is some alphabet, called the label set, and \(\mathrm{type}: \Sigma \to \mathbb{N}\) is a typing function which assigns to each label an arity called its type. We usually will assume some arbitrary but fixed signature \(\mathcal{C} = (\Sigma, \mathrm{type})\).

A hypergraph is a tuple \(H = (V_H, E_H, \textrm{att}_H, \textrm{lab}_H, \textrm{ext}_H)\) where \(V_H\) is a finite set of nodes, \(E_H\) is a finite set of hyperedges, \(\textrm{att}_H: E_H \to V_H^*\) is the attachment function, \(\textrm{lab}_H: E_H \to \Sigma\) is the labelling function, and \(\textrm{ext}_H \in V_H^*\) are the external nodes, such that labelling is compatible with typing (\(\textrm{type} \circ \textrm{lab}_H = \abs{\cdot} \circ \textrm{att}_H\)). In an abuse of notation, we write \(\mathrm{type}(H) = \abs{\textrm{ext}_H}\) for the type of \(H\), and define \(\textrm{type}_H: E_H \to \mathbb{N}\) by \(\textrm{type}_H = \textrm{type} \circ \textrm{lab}_H\) for the type of a hyperedge. For any hyperedge \(e \in E_H\), whenever \(m = \mathrm{type}_H(e)\) we call \(e\) a type \(m\) hyperedge, and call \(e\) proper whenever \(\textrm{att}_H(e)\) is injective (contains no repeated nodes). Call \(H\) proper if every \(e \in E_H\) is proper, repetition-free if \(\textrm{ext}_H\) is injective, and well-formed if it is both proper and repetition-free. The class of all hypergraphs (repetition-free hypergraphs) over \(\mathcal{C}\) is denoted \(\mathcal{H}_\mathcal{C}\) (\(\mathcal{H}_\mathcal{C}^{\mathrm{rf}}\)). A hypergraph morphism \(g: G \to H\) between \(G, H \in \mathcal{H}_\mathcal{C}\) is a pair of functions \((g_V: V_G \to V_H, g_E: E_G \to E_H)\) such that \(\textrm{att}_H \circ g_E = g_V^* \circ \textrm{att}_G\), \(\textrm{lab}_H \circ g_E = \textrm{lab}_G\), and \(g_V \circ \textrm{ext}_G \sqsubseteq_{\mathrm{sc}} \textrm{ext}_H\). \(g\) is called external node reflecting if \(g_V \circ \textrm{ext}_G = \textrm{ext}_H\), is called injective (surjective, bijective) whenever both \(g_V\) and \(g_E\) are injective (surjective, bijective), and is called hyperedge-injective if \(g_E\) is injective. We say two hypergraphs \(G, H \in \mathcal{H}_\mathcal{C}\) are isomorphic (\(G \cong H\)) there is a bijective external node reflecting hypergraph morphism \(G \to H\).

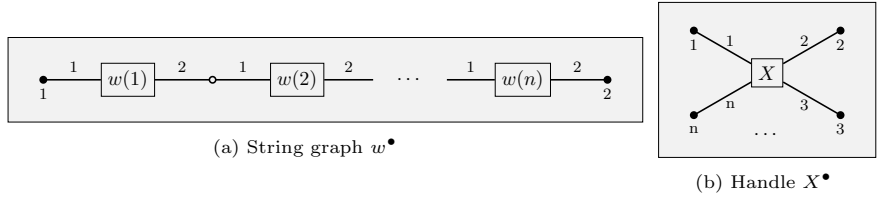
\begin{figure}[!ht]
\begin{subfigure}{.666666\textwidth}
\centering
\scalebox{0.75}{
\begin{tikzpicture}[every node/.style={align=center}]
    \draw (-0.625,-0.75) -- (-0.625,0.75) -- (10.625,0.75) -- (10.625,-0.75) -- cycle [fill=black!5];
    \node (a) at (0.0,0)       [draw, circle, thick, fill=black, scale=0.3]      {\,};
    \node (n) at (0,-0.25)                                                       {\footnotesize{1}};
    \node (b) at (1.5,0)       [draw, rectangle, minimum size=5mm]               {$w(1)$};
    \node (c) at (3.0,0)       [draw, circle, thick, fill=black!5, scale=0.3]    {\,};
    \node (d) at (4.5,0)       [draw, rectangle, minimum size=5mm]               {$w(2)$};
    \node (e) at (6.0,0)                                                         {\,};
    \node (f) at (6.5,0)                                                         {$\cdots$};
    \node (g) at (7.0,0)                                                         {\,};
    \node (h) at (8.5,0)       [draw, rectangle, minimum size=5mm]               {$w(n)$};
    \node (i) at (10,0)        [draw, circle, thick, fill=black, scale=0.3]      {\,};
    \node (m) at (10,-0.25)                                                      {\footnotesize{2}};

    \draw (a) edge[-,thick] node [above] {\footnotesize{1}} (b)
          (b) edge[-,thick] node [above] {\footnotesize{2}} (c)
          (c) edge[-,thick] node [above] {\footnotesize{1}} (d)
          (d) edge[-,thick] node [above] {\footnotesize{2}} (e)
          (g) edge[-,thick] node [above] {\footnotesize{1}} (h)
          (h) edge[-,thick] node [above] {\footnotesize{2}} (i);
\end{tikzpicture}}
\caption{String graph $w^{\bullet}$}
\end{subfigure}
\begin{subfigure}{.32\textwidth}
\centering
\scalebox{0.75}{
\begin{tikzpicture}[every node/.style={align=center}]
    \draw (-1.925,-1.5) -- (-1.925,1.25) -- (1.925,1.25) -- (1.925,-1.5) -- cycle [fill=black!5];
    \node (X) at (0.0,0.0)     [draw, rectangle, minimum size=5mm]               {$X$};
    \node (A) at (-1.3,0.75)   [draw, circle, thick, fill=black, scale=0.3]      {\,};
    \node (a) at (-1.3,0.5)                                                      {\footnotesize{1}};
    \node (B) at (1.3,0.75)    [draw, circle, thick, fill=black, scale=0.3]      {\,};
    \node (b) at (1.3,0.5)                                                       {\footnotesize{2}};
    \node (C) at (1.3,-0.75)   [draw, circle, thick, fill=black, scale=0.3]      {\,};
    \node (c) at (1.3,-1.0)                                                      {\footnotesize{3}};
    \node (D) at (0.0,-1.1)                                                      {$\cdots$};
    \node (E) at (-1.3,-0.75)  [draw, circle, thick, fill=black, scale=0.3]      {\,};
    \node (e) at (-1.3,-1.0)                                                     {\footnotesize{n}};

    \draw (X) edge[-,thick] node [right,xshift=-1mm,yshift=1.25mm] {\footnotesize{1}} (A)
          (X) edge[-,thick] node [left,xshift=1mm,yshift=1.25mm] {\footnotesize{2}} (B)
          (X) edge[-,thick] node [left,xshift=1mm,yshift=-1.25mm] {\footnotesize{3}} (C)
          (X) edge[-,thick] node [right,xshift=-1mm,yshift=-1.25mm] {\footnotesize{n}} (E);
\end{tikzpicture}}
\caption{Handle $X^{\bullet}$}
\end{subfigure}
\caption{Example hypergraphs}
\label{fig:eggraphs}
\end{figure}

Given a string \(w \in \Sigma^*\) of length \(n\), its string graph is \(w^{\bullet} =\) \((\{v_0, \dots, v_n\},\) \(\{e_1, \dots, e_n\}), \mathrm{att}, \mathrm{lab}, v_0 v_n)\) where \(\mathrm{att}(e_i) = v_{i-1} v_i\) and \(\mathrm{lab}(e_i) = w(i)\) for all \(i \in \underline{n}\) (Figure \ref{fig:eggraphs}(a)). If \(H \cong w^{\bullet}\) for some \(w \in \Sigma^*\), we call \(H\) a string graph representing \(w\). We also use the superscript bullet to denote the handle of a label. If \(X \in \Sigma\) is of type \(n\), then the handle of \(X\) is the hypergraph \(X^{\bullet} = (\{v_1, \dots, v_n\}, \{e\},\) \(\mathrm{att}, \mathrm{lab}, v_1 \cdots v_n)\) where \(\mathrm{att}(e) = v_1 \cdots v_n\) and \(\mathrm{lab}(e) = X\) (Figure \ref{fig:eggraphs}(b)). These two definitions coincide for a type \(2\) label, considered either as a string of length \(1\) or as a label, so there can be no confusion.

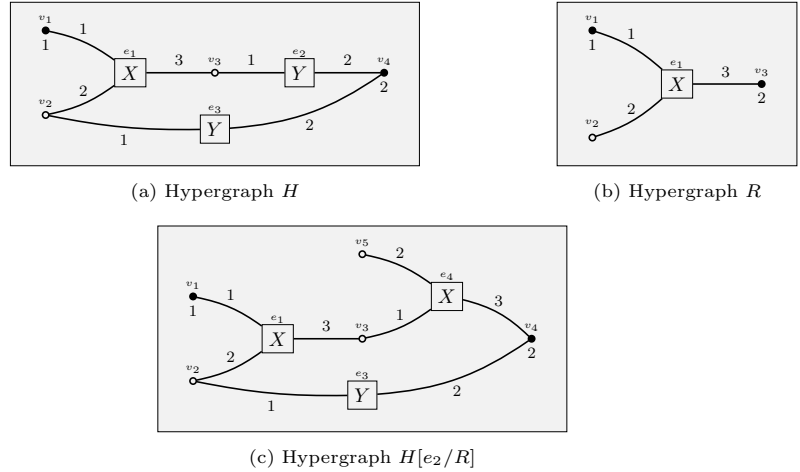
\begin{figure}[!ht]
\begin{subfigure}{.666666\textwidth}
\centering
\scalebox{0.75}{
\begin{tikzpicture}[every node/.style={align=center}]
    \draw (-0.625,-0.9) -- (-0.625,2.0) -- (6.625,2.0) -- (6.625,-0.9) -- cycle [fill=black!5];
    \node (A) at (0,1.7)                                                         {\tiny{$v_1$}};
    \node (a) at (0,1.5)       [draw, circle, thick, fill=black, scale=0.3]      {\,};
    \node (n) at (0,1.25)                                                        {\footnotesize{1}};
    \node (B) at (0,0.2)                                                         {\tiny{$v_2$}};
    \node (b) at (0,0)         [draw, circle, thick, fill=black!5, scale=0.3]    {\,};
    \node (F) at (1.5,1.1)                                                       {\tiny{$e_1$}};
    \node (x) at (1.5,0.75)    [draw, rectangle, minimum size=5mm]               {$X$};
    \node (C) at (3,0.95)                                                        {\tiny{$v_3$}};
    \node (c) at (3,0.75)      [draw, circle, thick, fill=black!5, scale=0.3]    {\,};
    \node (F) at (4.5,1.1)                                                       {\tiny{$e_2$}};
    \node (y) at (4.5,0.75)    [draw, rectangle, minimum size=5mm]               {$Y$};
    \node (D) at (6,0.95)                                                        {\tiny{$v_4$}};
    \node (d) at (6,0.75)      [draw, circle, thick, fill=black, scale=0.3]      {\,};
    \node (G) at (6,0.5)                                                         {\footnotesize{2}};
    \node (F) at (3,0.1)                                                         {\tiny{$e_3$}};
    \node (z) at (3,-0.25)     [draw, rectangle, minimum size=5mm]               {$Y$};

    \draw (a) edge[-,thick,bend left=12] node [above] {\footnotesize{1}} (x)
          (b) edge[-,thick,bend right=12] node [above] {\footnotesize{2}} (x)
          (x) edge[-,thick] node [above] {\footnotesize{3}} (c)
          (c) edge[-,thick] node [above] {\footnotesize{1}} (y)
          (y) edge[-,thick] node [above] {\footnotesize{2}} (d)
          (b) edge[-,thick,bend right=6] node [below] {\footnotesize{1}} (z)
          (z) edge[-,thick,bend right=15] node [below] {\footnotesize{2}} (d);
\end{tikzpicture}
}
\caption{Hypergraph $H$}
\end{subfigure}
\begin{subfigure}{.32\textwidth}
\centering
\scalebox{0.75}{
\begin{tikzpicture}[every node/.style={align=center}]
    \draw (-0.625,-0.9) -- (-0.625,2.0) -- (3.625,2.0) -- (3.625,-0.9) -- cycle [fill=black!5];
    \node (A) at (0.0,1.7)                                                      {\tiny{$v_1$}};
    \node (a) at (0.0,1.5)    [draw, circle, thick, fill=black, scale=0.3]      {\,};
    \node (n) at (0.0,1.25)                                                     {\footnotesize{1}};
    \node (B) at (0.0,-0.15)                                                    {\tiny{$v_2$}};
    \node (b) at (0.0,-0.4)   [draw, circle, thick, fill=black!5, scale=0.3]    {\,};
    \node (F) at (1.5,0.9)                                                      {\tiny{$e_1$}};
    \node (x) at (1.5,0.55)   [draw, rectangle, minimum size=5mm]               {$X$};
    \node (C) at (3.0,0.75)                                                     {\tiny{$v_3$}};
    \node (c) at (3.0,0.55)  [draw, circle, thick, fill=black, scale=0.3]       {\,};
    \node (G) at (3.0,0.3)                                                      {\footnotesize{2}};

    \draw (a) edge[-,thick,bend left=12] node [above] {\footnotesize{1}} (x)
          (b) edge[-,thick,bend right=12] node [above] {\footnotesize{2}} (x)
          (x) edge[-,thick] node [above] {\footnotesize{3}} (c);
\end{tikzpicture}
}
\caption{Hypergraph $R$}
\end{subfigure}
\begin{subfigure}{.99\textwidth}
\centering
\vspace{0.75em}
\scalebox{0.75}{
\begin{tikzpicture}[every node/.style={align=center}]
    \draw (-0.625,-0.9) -- (-0.625,2.75) -- (6.625,2.75) -- (6.625,-0.9) -- cycle [fill=black!5];
    \node (A) at (0,1.7)                                                         {\tiny{$v_1$}};
    \node (a) at (0,1.5)       [draw, circle, thick, fill=black, scale=0.3]      {\,};
    \node (n) at (0,1.25)                                                        {\footnotesize{1}};
    \node (B) at (0,0.2)                                                         {\tiny{$v_2$}};
    \node (b) at (0,0)         [draw, circle, thick, fill=black!5, scale=0.3]    {\,};
    \node (F) at (1.5,1.1)                                                       {\tiny{$e_1$}};
    \node (x) at (1.5,0.75)    [draw, rectangle, minimum size=5mm]               {$X$};
    \node (C) at (3,0.95)                                                        {\tiny{$v_3$}};
    \node (c) at (3,0.75)      [draw, circle, thick, fill=black!5, scale=0.3]    {\,};
    \node (F) at (4.5,1.85)                                                      {\tiny{$e_4$}};
    \node (y) at (4.5,1.50)    [draw, rectangle, minimum size=5mm]               {$X$};
    \node (D) at (6,0.95)                                                        {\tiny{$v_4$}};
    \node (d) at (6,0.75)      [draw, circle, thick, fill=black, scale=0.3]      {\,};
    \node (F) at (3,2.45)                                                        {\tiny{$v_5$}};
    \node (e) at (3,2.25)      [draw, circle, thick, fill=black!5, scale=0.3]    {\,};
    \node (G) at (6,0.5)                                                         {\footnotesize{2}};
    \node (F) at (3,0.1)                                                         {\tiny{$e_3$}};
    \node (z) at (3,-0.25)     [draw, rectangle, minimum size=5mm]               {$Y$};

    \draw (a) edge[-,thick,bend left=12] node [above] {\footnotesize{1}} (x)
          (b) edge[-,thick,bend right=12] node [above] {\footnotesize{2}} (x)
          (x) edge[-,thick] node [above] {\footnotesize{3}} (c)
          (c) edge[-,thick,bend right=12] node [above] {\footnotesize{1}} (y)
          (e) edge[-,thick,bend left=12] node [above] {\footnotesize{2}} (y)
          (y) edge[-,thick,bend left=15] node [above] {\footnotesize{3}} (d)
          (b) edge[-,thick,bend right=6] node [below] {\footnotesize{1}} (z)
          (z) edge[-,thick,bend right=15] node [below] {\footnotesize{2}} (d);
\end{tikzpicture}
}
\caption{Hypergraph $H[e_2/R]$}
\end{subfigure}
\caption{Example hyperedge replacement}
\label{fig:egrepl}
\end{figure}

Let \(H \in \mathcal{H}_{\mathcal{C}}\) be a hypergraph and \(B \subseteq E_H\) be a selection of hyperedges. Then \(\sigma: B \to \mathcal{H}_{\mathcal{C}}\) is called a replacement function if \(\mathrm{type} \circ \sigma = \restr{\mathrm{type}_H}{B}\). The replacement of \(B\) in \(H\) using \(\sigma\) is denoted by \(H[\sigma]\), and is the hypergraph obtained from \(H\) by removing \(B\) from \(E_H\), disjointly adding the nodes and hyperedges of \(\sigma(e)\), for each \(e \in B\), and identifying the \(i\)-th external node of \(\sigma(e)\) with the \(i\)-th attachment node of \(e\), for each \(e \in B\) and \(i \in \underline{\mathrm{type}_H(e)}\). The external nodes of \(H[\sigma]\) remain exactly those of \(H\) and all hyperedges keep their original attachments and labels. \(H[\sigma]\) exists exactly when \(\sigma: B \to \mathcal{H}_{\mathcal{C}}\) is a replacement function, and is unique up to isomorphism. If \(B = \{e_1, \dots, e_n\}\) and \(R_i = \sigma(e_i)\) for all \(i \in \underline{n}\), then we write \(H[e_1/R_1, \dots, e_n/R_n]\) in place of \(H[\sigma]\). Figure \ref{fig:egrepl} shows an example replacement.

Let \(N \subseteq \Sigma\) be a set of non-terminals. A type \(n\) rule over \(N\) is a pair \((L, R)\) with \(L \in N\), \(R \in \mathcal{H}_{\mathcal{C}}\), and \(\mathrm{type}(L) = \mathrm{type}(R) = n\). Call a rule \((L, R)\) repetition-free (proper, well-formed) if \(R\) is repetition-free (proper, well-formed). Given a hypergraph \(H \in \mathcal{H}_{\mathcal{C}}\) and a set of rules \(\mathcal{R}\), if \(e \in E_H\) and \((\mathrm{lab}_H(e), R) \in \mathcal{R}\), then we say that \(H\) directly derives \(H' \cong H[e/R]\) and write \(H \Rightarrow_{\mathcal{R}} H'\). For a given hyperedge \(e\) and choice of rule, \(H'\) is unique up to isomorphism. Clearly \(\Rightarrow_{\mathcal{R}}\) is a binary relation on \(\mathcal{H}_{\mathcal{C}}\). We say \(H \in \mathcal{H}_{\mathcal{C}}\) derives \(H'\) if there is a sequence \(H \Rightarrow_{\mathcal{R}} H_1 \Rightarrow_{\mathcal{R}} \cdots \Rightarrow_{\mathcal{R}} H_k = H'\) for some \(k \geq 1\) or \(H \cong H'\). We write \(H \Rightarrow_{\mathcal{R}}^k H'\) or \(H \Rightarrow_{\mathcal{R}}^* H'\). Clearly, (direct) derivations cannot delete nodes, and (direct) derivations made using repetition-free rules cannot merge nodes. We have the following result for repetition-free rules:

\begin{theorem}[HR Context-Freeness \cite{Habel92b}] \label{thm:cflem}
Let \(\mathcal{R}\) be a set of repetition-free rules over \(N\), \(H \in \mathcal{H}_{\mathcal{C}}\), \(X \in N\), and \(k \in \mathbb{N}\). Then there is a derivation \(X^{\bullet} \Rightarrow^{k+1} H\) if and only if there is a rule \((X, R) \in \mathcal{R}\) and a mapping \(\sigma: \mathrm{lab}_R^{-1}(N) \to \mathcal{H}_{\mathcal{C}}\) such that \(H = R[\sigma]\), \(\forall e \in \mathrm{lab}_R^{-1}(N), \mathrm{lab}_R(e)^{\bullet} \Rightarrow^{k(e)} \sigma(e)\), and \(\sum_{e \in \mathrm{lab}_R^{-1}(N)} k(e) = k\).
\end{theorem}

A hyperedge replacement grammar of order \(k\) (\(k\)-HR grammar) is a tuple \(\mathcal{G} = (\mathcal{C}, N, S, \mathcal{R})\) where \(\mathcal{C} = (\Sigma, \mathrm{type})\) is a signature, \(N \subseteq \Sigma\) is the set of non-terminal labels, \(S \in N\) is the start symbol, and \(\mathcal{R}\) is a finite set of rules over \(N\), with \(\mathrm{max}(\{\mathrm{type}(r) \mid r \in \mathcal{R}\}) \leq k\). We call \(\Sigma \setminus N\) the terminal labels and call \(\mathcal{G}\) repetition-free (proper, well-formed) if all its rules are repetition-free (proper, well-formed). The language generated by \(\mathcal{G}\) is \(\mathrm{L}(\mathcal{G}) = \{H \in \mathcal{H}_{\mathcal{C}} \mid S^{\bullet} \Rightarrow_{\mathcal{R}}^* H \textrm{ with } \mathrm{lab}_H^{-1}(N) = \emptyset\} \subseteq \mathcal{H}_{\mathcal{C}}\). \(L \subseteq \mathcal{H}_{\mathcal{C}}\) is called a (repetition-free) hyperedge replacement language of order \(k\) ((repetition-free) \(k\)-HR language) if there is a (repetition-free) \(k\)-HR grammar such that \(\mathrm{L}(\mathcal{G}) = L\). The class of (repetition-free) HR languages is the union of all (repetition-free) \(k\)-HR languages for \(k \geq 0\). Denote these \(\mathcal{H}\mathcal{R}_k\) and \(\mathcal{H}\mathcal{R}\) (\(\mathcal{H}\mathcal{R}_k^{\mathrm{rf}}\) and \(\mathcal{H}\mathcal{R}^{\mathrm{rf}}\)). All such languages are closed under hypergraph isomorphism and are homogeneous (all hypergraphs have the same type). We define the type of a homogeneous language to be the type of its members.

\begin{theorem}[Repetition-Free HR Generational Power \cite{Engelfriet-Heyker91a}] \label{thm:hrrf}
Given an HR grammar \(\mathcal{G}\) over \(\mathcal{C}\), one can effectively construct a repetition-free HR grammar \(\mathcal{G}'\) with \(\mathrm{L}(\mathcal{G}') = \mathrm{L}(\mathcal{G}) \cap \mathcal{H}_{\mathcal{C}}^{\mathrm{rf}}\).
\end{theorem}

\begin{theorem}[HR Linear-Growth \cite{Habel92b}] \label{thm:lingrowth}
Given an infinite HR language \(L\), there exists an infinite sequence of hypergraphs in \(L\), say \(H_0, H_1, H_2, \dots\) and constants \(c,d \in \mathbb{N}\) with \(c+d \geq 1\), such that for all \(i \in \mathbb{N}\), \(\abs{V_{H_{i+1}}} = \abs{V_{H_{i}}} + c\) and \(\abs{E_{H_{i+1}}} = \abs{E_{H_{i}}} + d\).
\end{theorem}

Given a class of hypergraph languages \(\mathcal{F}\), signatures \(\mathcal{C} = (\Sigma, \mathrm{type})\) and \(\mathcal{D}\), and \(A \subseteq \Sigma\), an \(\mathcal{F}\)-substitution of hypergraphs on \(A\) is a function \(s: A \to \mathcal{P}_0(\mathcal{H}_{\mathcal{D}})\) such that for each label \(X \in A\), \(s(X) \in \mathcal{F}\) and \(s(X)\) is of type \(\mathrm{type}(X)\). When \(\mathcal{F}\) is the class of all finite languages of hypergraphs, we call \(s\) a finite substitution of hypergraphs. When \(s\) is such that for each label \(X \in A\), \(A^{\bullet} \in s(A)\), then we call \(s\) nested. Given a hypergraph \(H\) over \(A\) with \(E_H = \{e_1, \dots, e_n\}\), write \(s(H)\) for the language \(\{H' \in \mathcal{H}_{\mathcal{D}} \mid \exists R_1 \in s(\mathrm{lab}_H(e_1)), \dots, \exists R_n \in s(\mathrm{lab}_H(e_n)), H' \cong H[e_1/R_1, \dots, e_n/R_n]\}\), and given a hypergraph language \(L\) over \(A\), write \(s(L)\) for the language \(\bigcup_{H \in L} s(H)\). Call \(\mathcal{F}\) substitution closed if for any \(\mathcal{F}\) language over \(A\) and \(\mathcal{F}\)-substitution \(s\) of hypergraphs on \(A\), \(s(L) \in \mathcal{F}\). Call \(\mathcal{F}\) (nested) iterated substitution closed if for any \(\mathcal{F}\) language over \(A\) and (nested) \(\mathcal{F}\)-substitution \(s\) of hypergraphs on \(A\), \(\bigcup_{n \in \mathbb{N}} s^n(L) \in \mathcal{F}\), where \(s^n(L) = s( \cdots s(L) \cdots )\).

\begin{theorem}[HR Languages Substitution Closed \cite{Habel92b}] \label{thm:hrsub}
For \(k \geq 0\), the \(k\)-HR languages are substitution closed and nested iterated substitution closed.
\end{theorem}

The partial function \(\mathrm{STR}: \mathcal{H}_{\mathcal{C}} \rightharpoonup \Sigma^*\) sends string graphs to the strings they represent, and is undefined elsewhere. A language \(L \subseteq \mathcal{H}_{\mathcal{C}}\) is said to be a string graph language if it only contains string graphs. Given an HR grammar \(\mathcal{G}\) that generates a string graph language, we write \(\mathrm{STR}(\mathrm{L}(\mathcal{G}))\) for the actual string language it generates. A string language \(L \subseteq A^*\) is called a (repetition-free) hyperedge replacement string language of order \(k\) ((repetition-free) \(k\)-HRS language) if there is a (repetition-free) \(k\)-HR grammar \(\mathcal{G}\) such that \(\mathcal{G}\) generates a string graph language and \(\mathrm{STR}(\mathrm{L}(\mathcal{G})) = L \setminus \{\epsilon\}\). The class of (repetition-free) HRS languages is the union of all (repetition-free) \(k\)-HRS languages for \(k \geq 2\). Denote these \(\mathcal{H}\mathcal{R}\mathcal{S}_k\) and \(\mathcal{H}\mathcal{R}\mathcal{S}\) (\(\mathcal{H}\mathcal{R}\mathcal{S}_k^{\mathrm{rf}}\) and \(\mathcal{H}\mathcal{R}\mathcal{S}^{\mathrm{rf}}\)).

\section{PHR Languages} \label{sec:phr}

In this section, we will formally define parallel hyperedge replacement and parallel hyperedge replacement grammars, showing that they generalise hyperedge replacement. Our definitions are equivalent to those defined by Habel in Chapter VIII.3 of \cite{Habel92b}. After that, we investigate the effects of restricting the maximum number of tables, show that rational control of traces does not change generative power when the number of tables is not restricted, establish some substitution closure results, and finally, show that the emptiness problem for parallel hyperedge replacement grammars is decidable.

\subsection{Definitions and Foundations}

The most fundamental notion to us is that of a parallel direct derivation, where every hyperedge is necessarily replaced. In order to ensure progress can always be made, we are only interested in sets of rules that are tables:

\begin{definition}[Table]
Given \(\mathcal{C} = (\Sigma, \mathrm{type})\), a table \(T\) over \(\Sigma\) is a finite set of rules over \(\Sigma\) such that for each \(L \in \Sigma\) there is at least one \(R \in \mathcal{H}_{\mathcal{C}}\) with \((L, R) \in T\). Call \(T\) repetition-free (proper, well-formed) if all its rules are repetition-free (proper, well-formed).
\end{definition}

\begin{definition}[Parallel Direct Derivation]
Given \(\mathcal{C} = (\Sigma, \mathrm{type})\), \(H \in \mathcal{H}_{\mathcal{C}}\) with \(E_H = \{e_1, \dots e_n\}\), and \(T\) a table over \(\Sigma\), if for each \(e_i \in E_H\), there is a \(R_i \in \mathcal{H}_{\mathcal{C}}\) such that \((\mathrm{lab}_H(e_i), R_i) \in T\), then we say that \(H\) parallelly directly derives \(H' \cong H[e_1/R_1, \dots, e_n/R_n]\), and write \(H \Rrightarrow_{T} H'\).
\end{definition}

Direct derivations enjoy a restriction property, and it is easy to see that parallel direct derivations also enjoy this same property. Given hypergraphs \(G\) and \(H\), write \(G \subseteq H\) if there is an injective hypergraph morphism \(G \to H\) and \(G \sqsubseteq H\) is there is a hyperedge-injective hypergraph morphism \(G \to H\).

\begin{lemma}[PHR Derivation Restriction]
Given \(\mathcal{C} = (\Sigma, \mathrm{type})\), a table \(T\) over \(\Sigma\), and hypergraphs \(H, H' \in \mathcal{H}_{\mathcal{C}}\) such that \(H \Rrightarrow_{T} H'\) and \(G \sqsubseteq H\), then there is a hypergraph \(G' \in \mathcal{H}_{\mathcal{C}}\) such that \(G \Rrightarrow_{T} G'\) and \(G' \sqsubseteq H'\). Moreover, if \(T\) is repetition-free, we can replace both occurrences of \(\sqsubseteq\) with \(\subseteq\).
\end{lemma}

An equivalent way to view a parallel direct derivation is the application of a finite substitution of hypergraphs. To any table \(T\) over \(\Sigma\), we can associate a finite substitution of hypergraphs \(s_{T}\) where \(s_{T}(L) = \{R \mid (L, R) \in T\}\) for each \(L \in \Sigma\). Next, we define parallel derivations to be a sequence of parallel direct derivations. Equivalently, we could view these as the application of a composition of substitutions, like in Section \ref{subsec:et0l}.

\begin{definition}[Parallel Derivation]
Given \(\mathcal{C} = (\Sigma, \mathrm{type})\), \(H, H' \in \mathcal{H}_{\mathcal{C}}\), and a finite set of tables \(\mathcal{T} = \{T_i \mid i \in I\}\) over \(\Sigma\) indexed by \(I\), we say \(H\) parallelly derives \(H'\) if there is a sequence \(H \Rrightarrow_{T_{{i_1}}} H_1 \Rrightarrow_{T_{i_2}} \cdots \Rrightarrow_{T_{i_k}} H_k = H'\) or \(H \cong H'\). We write \(H \Rrightarrow_{\mathcal{T}}^{i_1 i_2 \cdots i_k} H'\), \(H \Rrightarrow_{\mathcal{T}}^k H'\), or \(H \Rrightarrow_{\mathcal{T}}^{*} H'\). Call \(i_1 i_2 \cdots i_k \in I^*\) the trace of the derivation, defined to be \(\epsilon\) when \(H \cong H'\).
\end{definition}

Rather than replacing only non-terminals, as is usual in hyperedge replacement grammars, we allow all hyperedges to be replaced, and have a special set of terminal symbols to allow us to say when it is that a hypergraph is terminally labelled, just like ET0L grammars. This decision is more than just a matter of taste. Disallowing rewriting of terminal symbols can change generational power of systems with parallel rewriting. For example, the E0L language \(\{a^{n^2} \mid n \geq 1\}\) would not be E0L if we disallowed rewriting of terminal symbols, however, it would at least be ET0L without needing to re-write terminals to something other than themselves. Combining this observation with Theorem \ref{thm:phrset0l} tells us that \(\mathcal{P}\mathcal{H}\mathcal{R}_{2,1}\) would be different if we were to disallow replacement of terminals.

\begin{definition}[PHR Grammar]
A parallel hyperedge replacement grammar of order \(k\) using at most \(l\) tables is a tuple \(\mathcal{G} = (\mathcal{C}, A, S, \mathcal{T})\) where \(\mathcal{C} = (\Sigma, \mathrm{type})\) is a signature, \(A \subseteq \Sigma\) is the set of terminal labels, \(S \in \Sigma\) is the start symbol, and \(\mathcal{T} = \{T_i \mid i \in I\}\) is a non-empty, finite set of tables over \(\Sigma\) indexed by \(I\) with \(\mathrm{max}(\{\mathrm{type}(r) \mid r \in \bigcup_{T_i\in\mathcal{T}} T_i\}) \leq k\) and \(\abs{\mathcal{T}} \leq l\). Call \(\Sigma \setminus A\) the non-terminal labels and call \(\mathcal{G}\) repetition-free (proper, well-formed) if all its tables are repetition-free (proper, well-formed). Also call \(\mathcal{G}\) a \((k,l)\)-PHR grammar, \(k\)-PHR grammar, or PHR grammar. The language generated by \(\mathcal{G}\) is \(\mathrm{L}(\mathcal{G}) = \{H \in \mathcal{H}_{\mathcal{C}} \mid S^{\bullet} \Rrightarrow_{\mathcal{T}}^{*} H \textrm{ with } \mathrm{lab}_H^{-1}(A) = E_H\} \subseteq \mathcal{H}_{\mathcal{C}}\).
\end{definition}

\begin{definition}[PHR Language]
\(L \subseteq \mathcal{H}_{\mathcal{C}}\) is called a (repetition-free) parallel hyperedge replacement language of order \(k\) using at most \(l\) tables if there is a (repetition-free) \((k,l)\)-PHR grammar \(\mathcal{G}\) such that \(\mathrm{L}(\mathcal{G}) = L\). The class of (repetition-free) \(k\)-PHR languages is the union of all (repetition-free) \((k,l)\)-PHR languages for \(l \geq 1\), and the class of (repetition-free) PHR languages is the union of all (repetition-free) \(k\)-PHR languages for \(k \geq 0\). Denote these \(\mathcal{P}\mathcal{H}\mathcal{R}_{k,l}\), \(\mathcal{P}\mathcal{H}\mathcal{R}_{k}\), and \(\mathcal{P}\mathcal{H}\mathcal{R}\) (\(\mathcal{P}\mathcal{H}\mathcal{R}_{k,l}^{\mathrm{rf}}\), \(\mathcal{P}\mathcal{H}\mathcal{R}_{k}^{\mathrm{rf}}\), and \(\mathcal{P}\mathcal{H}\mathcal{R}^{\mathrm{rf}}\)).
\end{definition}

Just like languages generated by hyperedge replacement, parallel hyperedge replacement languages are closed under hypergraph isomorphism and are homogeneous in the sense that all hypergraphs in a language have the same type.

Our next theorem confirms that PHR languages strictly contain the HR languages, as expected. We note that our statement is actually stronger than Theorem VIII.3.3 of Habel's book \cite{Habel92b}, which sketches a proof for only \(k \geq 2\), and did not explicitly consider the number of tables required.

\begin{theorem}[PHR Generalises HR] \label{thm:phrgen}
For all \(k \geq 0\), \(\mathcal{H}\mathcal{R}_{k} \subsetneq \mathcal{P}\mathcal{H}\mathcal{R}_{k,1}\) and \(\mathcal{H}\mathcal{R}_{k}^{\mathrm{rf}} \subsetneq \mathcal{P}\mathcal{H}\mathcal{R}_{k,1}^{\mathrm{rf}}\).
\end{theorem}

\begin{proof}
Suppose \(\mathcal{C} = (\Sigma, \mathrm{type})\) and \(\mathcal{G} = (\mathcal{C}, N, S, \mathcal{R})\) is a (repetition-free) \(k\)-HR grammar with \(\mathcal{R} = \{r_1, \dots, r_n\}\). Then we construct a (repetition-free) \((k,1)\)-PHR grammar \(\mathcal{G}' = (\mathcal{C}, A, S, \mathcal{T})\) with \(A = \Sigma \setminus N\) and \(\mathcal{T} = \{T_1\}\) where \(T_1 = \mathcal{R} \cup \{(X, X^{\bullet}) \mid X \in \Sigma\}\). Clearly every parallel direct derivation with start hypergraph \(G\) can be decomposed into at most \(\abs{E_G}\) direct derivations where if an edge is replaced by itself, we omit it, and if a genuine replacement from \(\mathcal{R}\) occurs, we use that. So, by induction on derivation length, we see that every parallel derivation in \(\mathcal{G}'\) can actually be written as a derivation in \(\mathcal{G}\). Similarly, every direct derivation in \(\mathcal{G}\) can be lifted to a parallel derivation in \(\mathcal{G}'\) by replacing all but one edge by itself. So, by induction on derivation length, we see that every derivation in \(\mathcal{G}\) can be written as a parallel derivation in \(\mathcal{G}'\). Thus, together with the fact that the terminal symbols and start symbol coincide, we have that \(\mathrm{L}(\mathcal{G}) = \mathrm{L}(\mathcal{G}')\).

To see strictness, we are inspired by the fact that the string language \(\{a^{2^n} \mid n \in \mathbb{N}\}\) is E0L but not MCF (Theorem \ref{thm:incompar}). We will show there is a repetition-free \((0,1)\)-PHR language that is not \(k\)-HR for any \(k \geq 0\). Let \(\mathcal{G} = (\mathcal{C}, A, S, \mathcal{T})\) be the repetition-free \((0,1)\)-PHR grammar with \(\mathcal{C} = (\{\square\},\{(\square, 0)\})\), \(A = \{\square\}\), \(S = \square\), and \(\mathcal{T} = \{T_1\}\) where \(T_1 = \{(\square, \square^{\bullet} \sqcup \square^{\bullet})\}\) (\(\sqcup\) denotes disjoint union of hypergraphs). Clearly \(\mathrm{L}(\mathcal{G})\) is the language of hypergraphs over \(\mathcal{C}\) with \(2^n\) hyperedges. By Theorem \ref{thm:lingrowth}, \(\mathrm{L}(\mathcal{G})\) is not a HR language, as required.
\end{proof}

\begin{corollary} \label{cor:nonlingrowth}
PHR languages need not have only linear growth, in the sense of Theorem \ref{thm:lingrowth}.
\end{corollary}

Finishing this subsection, we remark that it is easy to see that proper PHR languages can be generated by proper PHR grammars:

\begin{lemma}[Constructing Proper PHR Grammars] \label{lem:properphr}
For any \(k \geq 0, l \geq 1\), given a (repetition-free) \((k, l)\)-PHR grammar \(\mathcal{G}\) generating a proper language, one can effectively construct a (repetition-free) proper \((k, l)\)-PHR grammar \(\mathcal{G}'\) such that \(\mathrm{L}(\mathcal{G}) = \mathrm{L}(\mathcal{G}')\).
\end{lemma}

Lemma \ref{lem:properphr} will be useful to us when we look at the string generational power of PHR grammars in Section \ref{sec:phrs}, since string graphs are proper hypergraphs.

\subsection{Example PHR Languages} \label{subsec:examples}

\begin{figure}[!ht]
\centering
\scalebox{0.75}{
\begin{tikzpicture}[every node/.style={align=center}]
    \draw (-0.625,-0.5) -- (-0.625,0.5) -- (0.625,0.5) -- (0.625,-0.5) -- cycle [fill=black!5];
    \node (X) at (0.0,0.0) [draw, rectangle, minimum size=5mm] {$\square$};

    \node (X) at (1.0,0.0) [minimum size=5mm] {$\Rrightarrow$};

    \draw (1.375,-1.0) -- (1.375,1.0) -- (2.625,1.0) -- (2.625,-1.0) -- cycle [fill=black!5];
    \node (X) at (2.0,0.5) [draw, rectangle, minimum size=5mm] {$\square$};
    \node (X) at (2.0,-0.5) [draw, rectangle, minimum size=5mm] {$\square$};

    \node (X) at (3.0,0.0) [minimum size=5mm] {$\Rrightarrow$};

    \draw (3.375,-1.0) -- (3.375,1.0) -- (5.625,1.0) -- (5.625,-1.0) -- cycle [fill=black!5];
    \node (X) at (4.0,0.5) [draw, rectangle, minimum size=5mm] {$\square$};
    \node (X) at (4.0,-0.5) [draw, rectangle, minimum size=5mm] {$\square$};
    \node (X) at (5.0,0.5) [draw, rectangle, minimum size=5mm] {$\square$};
    \node (X) at (5.0,-0.5) [draw, rectangle, minimum size=5mm] {$\square$};

    \node (X) at (6.0,0.0) [minimum size=5mm] {$\Rrightarrow$};

    \draw (6.375,-1.0) -- (6.375,1.0) -- (10.625,1.0) -- (10.625,-1.0) -- cycle [fill=black!5];
    \node (X) at (7.0,0.5) [draw, rectangle, minimum size=5mm] {$\square$};
    \node (X) at (7.0,-0.5) [draw, rectangle, minimum size=5mm] {$\square$};
    \node (X) at (8.0,0.5) [draw, rectangle, minimum size=5mm] {$\square$};
    \node (X) at (8.0,-0.5) [draw, rectangle, minimum size=5mm] {$\square$};
    \node (X) at (9.0,0.5) [draw, rectangle, minimum size=5mm] {$\square$};
    \node (X) at (9.0,-0.5) [draw, rectangle, minimum size=5mm] {$\square$};
    \node (X) at (10.0,0.5) [draw, rectangle, minimum size=5mm] {$\square$};
    \node (X) at (10.0,-0.5) [draw, rectangle, minimum size=5mm] {$\square$};
\end{tikzpicture}}
\caption{Example parallel derivation}
\label{fig:egder}
\end{figure}
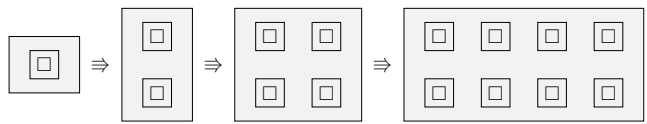

The proof of Theorem \ref{thm:phrgen} tells us all HR languages are PHR languages, and also provides us with an example repetition-free \((0, 1)\)-PHR language which is not HR. An example derivation in this grammar is shown in Figure \ref{fig:egder}. At each stage, the type \(0\) hyperedges are simultaneously replaced by two such hyperedges.

\begin{figure}[!ht]
\begin{subfigure}{.245\textwidth}
\centering
\scalebox{0.75}{
\begin{tikzpicture}[every node/.style={align=center}]
    \draw (-0.5,-0.5) -- (-0.5,0.5) -- (2.875,0.5) -- (2.875,-0.5) -- cycle [fill=black!5];
    \node (a) at (0.0,0)                                                         {$S$};
    \node (b) at (0.5,0)                                                         {$\to$};
    \node (c) at (1.75,0)       [draw, circle, thick, fill=black, scale=0.3]      {\,};
    \node (d) at (2.0,0)                                                        {\footnotesize{1}};
\end{tikzpicture}}
\caption{Rule $r_0$}
\end{subfigure}
\begin{subfigure}{.245\textwidth}
\centering
\scalebox{0.75}{
\begin{tikzpicture}[every node/.style={align=center}]
    \draw (-0.5,-1.5) -- (-0.5,1.5) -- (2.875,1.5) -- (2.875,-1.5) -- cycle [fill=black!5];
    \node (a) at (0.0,0)                                                         {$S$};
    \node (b) at (0.5,0)                                                         {$\to$};
    \node (c) at (1.75,1)      [draw, circle, thick, fill=black, scale=0.3]      {\,};
    \node (d) at (2.0,1)                                                         {\footnotesize{1}};
    \node (e) at (1.25,0)      [draw, rectangle, minimum size=5mm]               {$X$};
    \node (f) at (1.25,-1)     [draw, circle, thick, fill=black!5, scale=0.3]  {\,};
    \node (g) at (2.25,0)      [draw, rectangle, minimum size=5mm]               {$X$};
    \node (h) at (2.25,-1)     [draw, circle, thick, fill=black!5, scale=0.3]  {\,};

    \draw (c) edge[-,thick, bend right=30, pos=0.6] node [right] {\footnotesize{1}} (e)
          (c) edge[-,thick, bend left=30, pos=0.6] node [right] {\footnotesize{1}} (g)
          (e) edge[-,thick] node [right] {\footnotesize{2}} (f)
          (g) edge[-,thick] node [right] {\footnotesize{2}} (h);
\end{tikzpicture}}
\caption{Rule $r_1$}
\end{subfigure}
\begin{subfigure}{.245\textwidth}
\centering
\scalebox{0.75}{
\begin{tikzpicture}[every node/.style={align=center}]
    \draw (-0.5,-2.5) -- (-0.5,2.5) -- (2.875,2.5) -- (2.875,-2.5) -- cycle [fill=black!5];
    \node (a) at (0.0,0)                                                         {$X$};
    \node (b) at (0.5,0)                                                         {$\to$};
    \node (c) at (1.75,2)      [draw, circle, thick, fill=black, scale=0.3]      {\,};
    \node (d) at (2.0,2)                                                         {\footnotesize{1}};
    \node (e) at (1.75,1)      [draw, rectangle, minimum size=5mm]               {$Y$};
    \node (f) at (1.75,0)      [draw, circle, thick, fill=black, scale=0.3]      {\,};
    \node (g) at (2.0,0)                                                         {\footnotesize{2}};
    \node (h) at (1.25,-1)     [draw, rectangle, minimum size=5mm]               {$X$};
    \node (i) at (1.25,-2)     [draw, circle, thick, fill=black!5, scale=0.3]  {\,};
    \node (j) at (2.25,-1)     [draw, rectangle, minimum size=5mm]               {$X$};
    \node (k) at (2.25,-2)     [draw, circle, thick, fill=black!5, scale=0.3]  {\,};

    \draw (c) edge[-,thick] node [right] {\footnotesize{1}} (e)
          (e) edge[-,thick] node [right] {\footnotesize{2}} (f)
          (f) edge[-,thick, bend right=30, pos=0.6] node [right] {\footnotesize{1}} (h)
          (f) edge[-,thick, bend left=30, pos=0.6] node [right] {\footnotesize{1}} (j)
          (h) edge[-,thick] node [right] {\footnotesize{2}} (i)
          (j) edge[-,thick] node [right] {\footnotesize{2}} (k);
\end{tikzpicture}}
\caption{Rule $r_2$}
\vspace{0.75em}
\end{subfigure}
\begin{subfigure}{.245\textwidth}
\centering
\scalebox{0.75}{
\begin{tikzpicture}[every node/.style={align=center}]
    \draw (-0.5,-1.5) -- (-0.5,1.5) -- (2.875,1.5) -- (2.875,-1.5) -- cycle [fill=black!5];
    \node (a) at (0.0,0)                                                         {$X$};
    \node (b) at (0.5,0)                                                         {$\to$};
    \node (c) at (1.75,1)      [draw, circle, thick, fill=black, scale=0.3]      {\,};
    \node (d) at (2.0,1)                                                         {\footnotesize{1}};
    \node (e) at (1.75,0)      [draw, rectangle, minimum size=5mm]               {$\square$};
    \node (f) at (1.75,-1)     [draw, circle, thick, fill=black, scale=0.3]      {\,};
    \node (g) at (2.0,-1)                                                        {\footnotesize{2}};

    \draw (c) edge[-,thick] node [right] {\footnotesize{1}} (e)
          (e) edge[-,thick] node [right] {\footnotesize{2}} (f);
\end{tikzpicture}}
\caption{Rule $r_3$}
\end{subfigure}
\begin{subfigure}{.245\textwidth}
\centering
\scalebox{0.75}{
\begin{tikzpicture}[every node/.style={align=center}]
    \draw (-0.5,-1.5) -- (-0.5,1.5) -- (2.875,1.5) -- (2.875,-1.5) -- cycle [fill=black!5];
    \node (a) at (0.0,0)                                                         {$Y$};
    \node (b) at (0.5,0)                                                         {$\to$};
    \node (c) at (1.75,1)      [draw, circle, thick, fill=black, scale=0.3]      {\,};
    \node (d) at (2.0,1)                                                         {\footnotesize{1}};
    \node (e) at (1.75,0)      [draw, rectangle, minimum size=5mm]               {$Y$};
    \node (f) at (1.75,-1)     [draw, circle, thick, fill=black, scale=0.3]      {\,};
    \node (g) at (2.0,-1)                                                        {\footnotesize{2}};

    \draw (c) edge[-,thick] node [right] {\footnotesize{1}} (e)
          (e) edge[-,thick] node [right] {\footnotesize{2}} (f);
\end{tikzpicture}}
\caption{Rule $r_4$}
\end{subfigure}
\begin{subfigure}{.245\textwidth}
\centering
\scalebox{0.75}{
\begin{tikzpicture}[every node/.style={align=center}]
    \draw (-0.5,-1.5) -- (-0.5,1.5) -- (2.875,1.5) -- (2.875,-1.5) -- cycle [fill=black!5];
    \node (a) at (0.0,0)                                                         {$Y$};
    \node (b) at (0.5,0)                                                         {$\to$};
    \node (c) at (1.75,1)      [draw, circle, thick, fill=black, scale=0.3]      {\,};
    \node (d) at (2.0,1)                                                         {\footnotesize{1}};
    \node (e) at (1.75,0)      [draw, rectangle, minimum size=5mm]               {$\square$};
    \node (f) at (1.75,-1)     [draw, circle, thick, fill=black, scale=0.3]      {\,};
    \node (g) at (2.0,-1)                                                        {\footnotesize{2}};

    \draw (c) edge[-,thick] node [right] {\footnotesize{1}} (e)
          (e) edge[-,thick] node [right] {\footnotesize{2}} (f);
\end{tikzpicture}}
\caption{Rule $r_5$}
\end{subfigure}
\begin{subfigure}{.245\textwidth}
\centering
\scalebox{0.75}{
\begin{tikzpicture}[every node/.style={align=center}]
    \draw (-0.5,-1.5) -- (-0.5,1.5) -- (2.875,1.5) -- (2.875,-1.5) -- cycle [fill=black!5];
    \node (a) at (0.0,0)                                                         {$\square$};
    \node (b) at (0.5,0)                                                         {$\to$};
    \node (c) at (1.75,1)      [draw, circle, thick, fill=black, scale=0.3]      {\,};
    \node (d) at (2.0,1)                                                         {\footnotesize{1}};
    \node (e) at (1.75,0)      [draw, rectangle, minimum size=5mm]               {$F$};
    \node (f) at (1.75,-1)     [draw, circle, thick, fill=black, scale=0.3]      {\,};
    \node (g) at (2.0,-1)                                                        {\footnotesize{2}};

    \draw (c) edge[-,thick] node [right] {\footnotesize{1}} (e)
          (e) edge[-,thick] node [right] {\footnotesize{2}} (f);
\end{tikzpicture}}
\caption{Rule $r_6$}
\end{subfigure}
\begin{subfigure}{.245\textwidth}
\centering
\scalebox{0.75}{
\begin{tikzpicture}[every node/.style={align=center}]
    \draw (-0.5,-1.5) -- (-0.5,1.5) -- (2.875,1.5) -- (2.875,-1.5) -- cycle [fill=black!5];
    \node (a) at (0.0,0)                                                         {$F$};
    \node (b) at (0.5,0)                                                         {$\to$};
    \node (c) at (1.75,1)      [draw, circle, thick, fill=black, scale=0.3]      {\,};
    \node (d) at (2.0,1)                                                         {\footnotesize{1}};
    \node (e) at (1.75,0)      [draw, rectangle, minimum size=5mm]               {$F$};
    \node (f) at (1.75,-1)     [draw, circle, thick, fill=black, scale=0.3]      {\,};
    \node (g) at (2.0,-1)                                                        {\footnotesize{2}};

    \draw (c) edge[-,thick] node [right] {\footnotesize{1}} (e)
          (e) edge[-,thick] node [right] {\footnotesize{2}} (f);
\end{tikzpicture}}
\caption{Rule $r_7$}
\end{subfigure}
\caption{Full binary tree rules}
\label{fig:fbtreerules}
\end{figure}
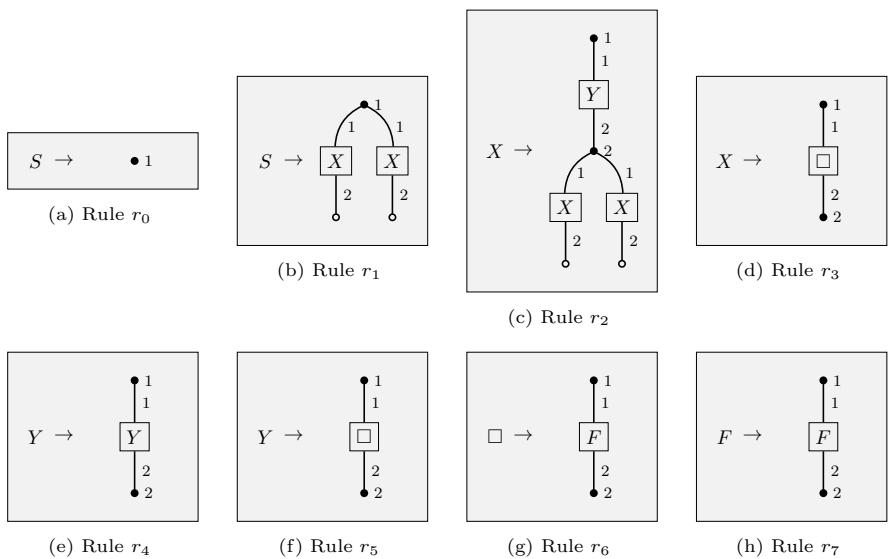

For our next example, recall that one can view rooted unlabelled trees as directed graphs, with the root an external node. Such a tree is called a full binary tree if each node has exactly zero or two children, and if a node has no children, then all other nodes on the same level must also have no children. We denote the language of all such hypergraphs \(\mathrm{FBT}\), using a type \(2\) symbol \(\square\) to label all the edges. It is easy to check that the repetition-free \((2,1)\)-PHR grammar \(\mathcal{G} = ((\Sigma, \mathrm{type}), \{\square\}, S, \{T_1\})\) where \(\Sigma = \{S, X, Y, F, \square\}\), \(\mathrm{type}(S) = 1\), \(\mathrm{type}(X) = \mathrm{type}(Y) = \mathrm{type}(F) = \mathrm{type}(\square) = 2\), \(T_1 = \{r_0, \dots, r_7\}\), and the eight rules are defined in Figure \ref{fig:fbtreerules}, generates \(\mathrm{FBT}\).

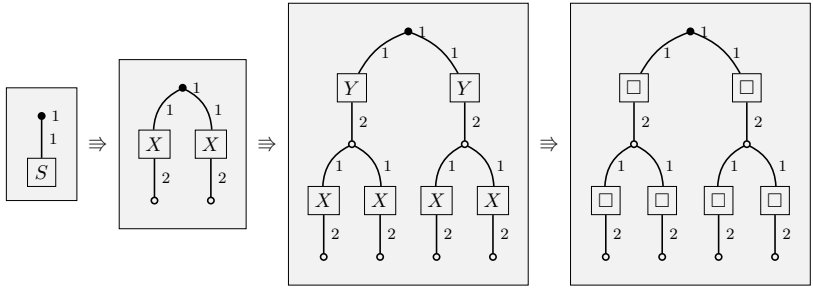
\begin{figure}[!ht]
\centering
\scalebox{0.75}{
\begin{tikzpicture}[every node/.style={align=center}]

    \draw (-0.625,-1.0) -- (-0.625,1.0) -- (0.625,1.0) -- (0.625,-1.0) -- cycle [fill=black!5];

    \node (a) at (0.0,0.5)      [draw, circle, thick, fill=black, scale=0.3]      {\,};
    \node (b) at (0.25,0.5)                                                         {\footnotesize{1}};
    \node (c) at (0.0,-0.5)      [draw, rectangle, minimum size=5mm]               {$S$};

    \draw (a) edge[-,thick] node [right] {\footnotesize{1}} (c);

    \node (X) at (1.0,0.0) [minimum size=5mm] {$\Rrightarrow$};

    \draw (1.375,-1.5) -- (1.375,1.5) -- (3.625,1.5) -- (3.625,-1.5) -- cycle [fill=black!5];

    \node (a) at (2.5,1.0)      [draw, circle, thick, fill=black, scale=0.3]      {\,};
    \node (b) at (2.75,1.0)                                                         {\footnotesize{1}};
    \node (c) at (2.0,0.0)      [draw, rectangle, minimum size=5mm]               {$X$};
    \node (d) at (2.0,-1.0)      [draw, circle, thick, fill=black!5, scale=0.3]      {\,};
    \node (e) at (3.0,0.0)      [draw, rectangle, minimum size=5mm]               {$X$};
    \node (f) at (3.0,-1.0)      [draw, circle, thick, fill=black!5, scale=0.3]      {\,};

    \draw (a) edge[-,thick, bend right=30, pos=0.6] node [right] {\footnotesize{1}} (c)
          (a) edge[-,thick, bend left=30, pos=0.6] node [right] {\footnotesize{1}} (e)
          (c) edge[-,thick] node [right] {\footnotesize{2}} (d)
          (e) edge[-,thick] node [right] {\footnotesize{2}} (f);

    \node (X) at (4.0,0.0) [minimum size=5mm] {$\Rrightarrow$};

    \draw (4.375,-2.5) -- (4.375,2.5) -- (8.625,2.5) -- (8.625,-2.5) -- cycle [fill=black!5];

    \node (a) at (6.5,2.0)      [draw, circle, thick, fill=black, scale=0.3]      {\,};
    \node (b) at (6.75,2.0)                                                         {\footnotesize{1}};
    \node (c) at (5.5,1.0)      [draw, rectangle, minimum size=5mm]               {$Y$};
    \node (d) at (5.5,0.0)      [draw, circle, thick, fill=black!5, scale=0.3]      {\,};
    \node (e) at (7.5,1.0)      [draw, rectangle, minimum size=5mm]               {$Y$};
    \node (f) at (7.5,0.0)      [draw, circle, thick, fill=black!5, scale=0.3]      {\,};

    \node (g) at (5.0,-1.0)      [draw, rectangle, minimum size=5mm]               {$X$};
    \node (h) at (5.0,-2.0)      [draw, circle, thick, fill=black!5, scale=0.3]      {\,};
    \node (i) at (6.0,-1.0)      [draw, rectangle, minimum size=5mm]               {$X$};
    \node (j) at (6.0,-2.0)      [draw, circle, thick, fill=black!5, scale=0.3]      {\,};

    \node (k) at (7.0,-1.0)      [draw, rectangle, minimum size=5mm]               {$X$};
    \node (l) at (7.0,-2.0)      [draw, circle, thick, fill=black!5, scale=0.3]      {\,};
    \node (m) at (8.0,-1.0)      [draw, rectangle, minimum size=5mm]               {$X$};
    \node (n) at (8.0,-2.0)      [draw, circle, thick, fill=black!5, scale=0.3]      {\,};

    \draw (a) edge[-,thick, bend right=20, pos=0.6] node [right] {\footnotesize{1}} (c)
          (a) edge[-,thick, bend left=20, pos=0.6] node [right] {\footnotesize{1}} (e)
          (c) edge[-,thick] node [right] {\footnotesize{2}} (d)
          (e) edge[-,thick] node [right] {\footnotesize{2}} (f);

    \draw (d) edge[-,thick, bend right=30, pos=0.6] node [right] {\footnotesize{1}} (g)
          (d) edge[-,thick, bend left=30, pos=0.6] node [right] {\footnotesize{1}} (i)
          (g) edge[-,thick] node [right] {\footnotesize{2}} (h)
          (i) edge[-,thick] node [right] {\footnotesize{2}} (j);

    \draw (f) edge[-,thick, bend right=30, pos=0.6] node [right] {\footnotesize{1}} (k)
          (f) edge[-,thick, bend left=30, pos=0.6] node [right] {\footnotesize{1}} (m)
          (k) edge[-,thick] node [right] {\footnotesize{2}} (l)
          (m) edge[-,thick] node [right] {\footnotesize{2}} (n);

    \node (X) at (9.0,0.0) [minimum size=5mm] {$\Rrightarrow$};

    \draw (9.375,-2.5) -- (9.375,2.5) -- (13.625,2.5) -- (13.625,-2.5) -- cycle [fill=black!5];

    \node (a) at (11.5,2.0)      [draw, circle, thick, fill=black, scale=0.3]      {\,};
    \node (b) at (11.75,2.0)                                                         {\footnotesize{1}};
    \node (c) at (10.5,1.0)      [draw, rectangle, minimum size=5mm]               {$\square$};
    \node (d) at (10.5,0.0)      [draw, circle, thick, fill=black!5, scale=0.3]      {\,};
    \node (e) at (12.5,1.0)      [draw, rectangle, minimum size=5mm]               {$\square$};
    \node (f) at (12.5,0.0)      [draw, circle, thick, fill=black!5, scale=0.3]      {\,};

    \node (g) at (10.0,-1.0)      [draw, rectangle, minimum size=5mm]               {$\square$};
    \node (h) at (10.0,-2.0)      [draw, circle, thick, fill=black!5, scale=0.3]      {\,};
    \node (i) at (11.0,-1.0)      [draw, rectangle, minimum size=5mm]               {$\square$};
    \node (j) at (11.0,-2.0)      [draw, circle, thick, fill=black!5, scale=0.3]      {\,};

    \node (k) at (12.0,-1.0)      [draw, rectangle, minimum size=5mm]               {$\square$};
    \node (l) at (12.0,-2.0)      [draw, circle, thick, fill=black!5, scale=0.3]      {\,};
    \node (m) at (13.0,-1.0)      [draw, rectangle, minimum size=5mm]               {$\square$};
    \node (n) at (13.0,-2.0)      [draw, circle, thick, fill=black!5, scale=0.3]      {\,};

    \draw (a) edge[-,thick, bend right=20, pos=0.6] node [right] {\footnotesize{1}} (c)
          (a) edge[-,thick, bend left=20, pos=0.6] node [right] {\footnotesize{1}} (e)
          (c) edge[-,thick] node [right] {\footnotesize{2}} (d)
          (e) edge[-,thick] node [right] {\footnotesize{2}} (f);

    \draw (d) edge[-,thick, bend right=30, pos=0.6] node [right] {\footnotesize{1}} (g)
          (d) edge[-,thick, bend left=30, pos=0.6] node [right] {\footnotesize{1}} (i)
          (g) edge[-,thick] node [right] {\footnotesize{2}} (h)
          (i) edge[-,thick] node [right] {\footnotesize{2}} (j);

    \draw (f) edge[-,thick, bend right=30, pos=0.6] node [right] {\footnotesize{1}} (k)
          (f) edge[-,thick, bend left=30, pos=0.6] node [right] {\footnotesize{1}} (m)
          (k) edge[-,thick] node [right] {\footnotesize{2}} (l)
          (m) edge[-,thick] node [right] {\footnotesize{2}} (n);

\end{tikzpicture}}
\caption{Example positive derivation}
\label{fig:fbtreeder1}
\end{figure}

\begin{figure}[!ht]
\centering
\scalebox{0.75}{
\begin{tikzpicture}[every node/.style={align=center}]

    \draw (-0.625,-1.0) -- (-0.625,1.0) -- (0.625,1.0) -- (0.625,-1.0) -- cycle [fill=black!5];

    \node (a) at (0.0,0.5)      [draw, circle, thick, fill=black, scale=0.3]      {\,};
    \node (b) at (0.25,0.5)                                                         {\footnotesize{1}};
    \node (c) at (0.0,-0.5)      [draw, rectangle, minimum size=5mm]               {$S$};

    \draw (a) edge[-,thick] node [right] {\footnotesize{1}} (c);

    \node (X) at (1.0,0.0) [minimum size=5mm] {$\Rrightarrow$};

    \draw (1.375,-1.5) -- (1.375,1.5) -- (3.625,1.5) -- (3.625,-1.5) -- cycle [fill=black!5];

    \node (a) at (2.5,1.0)      [draw, circle, thick, fill=black, scale=0.3]      {\,};
    \node (b) at (2.75,1.0)                                                         {\footnotesize{1}};
    \node (c) at (2.0,0.0)      [draw, rectangle, minimum size=5mm]               {$X$};
    \node (d) at (2.0,-1.0)      [draw, circle, thick, fill=black!5, scale=0.3]      {\,};
    \node (e) at (3.0,0.0)      [draw, rectangle, minimum size=5mm]               {$X$};
    \node (f) at (3.0,-1.0)      [draw, circle, thick, fill=black!5, scale=0.3]      {\,};

    \draw (a) edge[-,thick, bend right=30, pos=0.6] node [right] {\footnotesize{1}} (c)
          (a) edge[-,thick, bend left=30, pos=0.6] node [right] {\footnotesize{1}} (e)
          (c) edge[-,thick] node [right] {\footnotesize{2}} (d)
          (e) edge[-,thick] node [right] {\footnotesize{2}} (f);

    \node (X) at (4.0,0.0) [minimum size=5mm] {$\Rrightarrow$};

    \draw (4.375,-2.5) -- (4.375,2.5) -- (8.625,2.5) -- (8.625,-2.5) -- cycle [fill=black!5];

    \node (a) at (6.5,2.0)      [draw, circle, thick, fill=black, scale=0.3]      {\,};
    \node (b) at (6.75,2.0)                                                         {\footnotesize{1}};
    \node (c) at (5.5,1.0)      [draw, rectangle, minimum size=5mm]               {$Y$};
    \node (d) at (5.5,0.0)      [draw, circle, thick, fill=black!5, scale=0.3]      {\,};
    \node (e) at (7.5,1.0)      [draw, rectangle, minimum size=5mm]               {$\square$};
    \node (f) at (7.5,0.0)      [draw, circle, thick, fill=black!5, scale=0.3]      {\,};

    \node (g) at (5.0,-1.0)      [draw, rectangle, minimum size=5mm]               {$X$};
    \node (h) at (5.0,-2.0)      [draw, circle, thick, fill=black!5, scale=0.3]      {\,};
    \node (i) at (6.0,-1.0)      [draw, rectangle, minimum size=5mm]               {$X$};
    \node (j) at (6.0,-2.0)      [draw, circle, thick, fill=black!5, scale=0.3]      {\,};

    \draw (a) edge[-,thick, bend right=20, pos=0.6] node [right] {\footnotesize{1}} (c)
          (a) edge[-,thick, bend left=20, pos=0.6] node [right] {\footnotesize{1}} (e)
          (c) edge[-,thick] node [right] {\footnotesize{2}} (d)
          (e) edge[-,thick] node [right] {\footnotesize{2}} (f);

    \draw (d) edge[-,thick, bend right=30, pos=0.6] node [right] {\footnotesize{1}} (g)
          (d) edge[-,thick, bend left=30, pos=0.6] node [right] {\footnotesize{1}} (i)
          (g) edge[-,thick] node [right] {\footnotesize{2}} (h)
          (i) edge[-,thick] node [right] {\footnotesize{2}} (j);

    \node (X) at (9.0,0.0) [minimum size=5mm] {$\Rrightarrow$};

    \draw (9.375,-2.5) -- (9.375,2.5) -- (13.625,2.5) -- (13.625,-2.5) -- cycle [fill=black!5];

    \node (a) at (11.5,2.0)      [draw, circle, thick, fill=black, scale=0.3]      {\,};
    \node (b) at (11.75,2.0)                                                         {\footnotesize{1}};
    \node (c) at (10.5,1.0)      [draw, rectangle, minimum size=5mm]               {$\square$};
    \node (d) at (10.5,0.0)      [draw, circle, thick, fill=black!5, scale=0.3]      {\,};
    \node (e) at (12.5,1.0)      [draw, rectangle, minimum size=5mm]               {$F$};
    \node (f) at (12.5,0.0)      [draw, circle, thick, fill=black!5, scale=0.3]      {\,};

    \node (g) at (10.0,-1.0)      [draw, rectangle, minimum size=5mm]               {$\square$};
    \node (h) at (10.0,-2.0)      [draw, circle, thick, fill=black!5, scale=0.3]      {\,};
    \node (i) at (11.0,-1.0)      [draw, rectangle, minimum size=5mm]               {$\square$};
    \node (j) at (11.0,-2.0)      [draw, circle, thick, fill=black!5, scale=0.3]      {\,};

    \draw (a) edge[-,thick, bend right=20, pos=0.6] node [right] {\footnotesize{1}} (c)
          (a) edge[-,thick, bend left=20, pos=0.6] node [right] {\footnotesize{1}} (e)
          (c) edge[-,thick] node [right] {\footnotesize{2}} (d)
          (e) edge[-,thick] node [right] {\footnotesize{2}} (f);

    \draw (d) edge[-,thick, bend right=30, pos=0.6] node [right] {\footnotesize{1}} (g)
          (d) edge[-,thick, bend left=30, pos=0.6] node [right] {\footnotesize{1}} (i)
          (g) edge[-,thick] node [right] {\footnotesize{2}} (h)
          (i) edge[-,thick] node [right] {\footnotesize{2}} (j);

\end{tikzpicture}}
\caption{Example negative derivation}
\label{fig:fbtreeder2}
\end{figure}
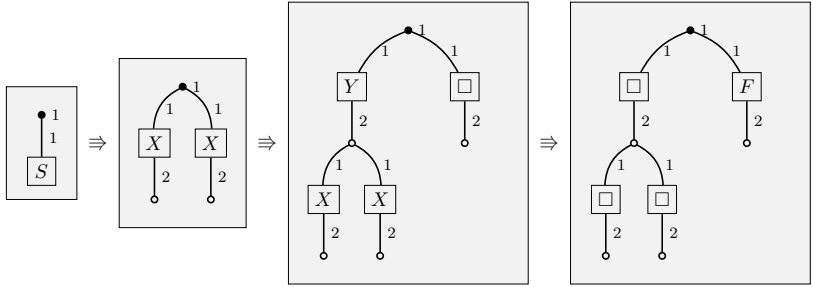

The rule \(r_6\) is responsible for ensuring that whenever terminals are derived, if any non-terminals remained, then the hypergraph can never reach a terminally labelled state in the future, effectively controlling when we stop adding levels to our tree. This trick with the terminals and the \(F\) label will re-appear in Subsection \ref{subsec:subc}, when we discuss synchronisation. Figures \ref{fig:fbtreeder1} and \ref{fig:fbtreeder2} show positive and negative derivations, by which we mean, derivations that lead to a terminally labelled hypergraph, and derivations that are not a prefix of a derivation that leads to a terminally labelled hypergraph, respectively.

\begin{figure}[!ht]
\begin{subfigure}{.325\textwidth}
\centering
\scalebox{0.75}{
\begin{tikzpicture}[every node/.style={align=center}]
    \draw (-1.0,-1.65) -- (-1.0,1.5) -- (3.375,1.5) -- (3.375,-1.65) -- cycle [fill=black!5];
    \node (a) at (-0.5,0)                                                        {$S$};
    \node (b) at (0.0,0)                                                         {$\to$};
    \node (c) at (1.75,1)      [draw, circle, thick, fill=black, scale=0.3]      {\,};
    \node (d) at (2.0,1)                                                         {\footnotesize{1}};
    \node (e) at (1.25,0)      [draw, rectangle, minimum size=5mm]               {$\square$};
    \node (f) at (2.25,0)      [draw, rectangle, minimum size=5mm]               {$\square$};
    \node (g) at (0.75,-1)     [draw, circle, thick, fill=black, scale=0.3]      {\,};
    \node (h) at (0.5,-1)                                                        {\footnotesize{3}};
    \node (i) at (1.75,-1)     [draw, rectangle, minimum size=5mm]               {$\square$};
    \node (j) at (2.75,-1)     [draw, circle, thick, fill=black, scale=0.3]      {\,};
    \node (k) at (3.0,-1)                                                        {\footnotesize{2}};

    \draw (c) edge[-,thick] node [right] {\footnotesize{1}} (f)
          (f) edge[-,thick] node [right] {\footnotesize{2}} (j)
          (k) edge[-,thick] node [below] {\footnotesize{1}} (i)
          (i) edge[-,thick] node [below] {\footnotesize{2}} (g)
          (g) edge[-,thick] node [left]  {\footnotesize{1}} (e)
          (e) edge[-,thick] node [left]  {\footnotesize{2}} (c);
\end{tikzpicture}}
\caption{Rule $r_0$}
\end{subfigure}
\begin{subfigure}{.325\textwidth}
\centering
\scalebox{0.75}{
\begin{tikzpicture}[every node/.style={align=center}]
    \draw (-1.0,-1.5) -- (-1.0,1.5) -- (3.375,1.5) -- (3.375,-1.5) -- cycle [fill=black!5];
    \node (a) at (-0.5,0)                                                        {$S$};
    \node (b) at (0.0,0)                                                         {$\to$};
    \node (c) at (1.75,1)      [draw, circle, thick, fill=black, scale=0.3]      {\,};
    \node (d) at (2.0,1)                                                         {\footnotesize{1}};
    \node (e) at (1.75,0)      [draw, rectangle, minimum size=5mm]               {$X$};
    \node (f) at (0.75,-1)     [draw, circle, thick, fill=black, scale=0.3]      {\,};
    \node (g) at (0.5,-1)                                                        {\footnotesize{3}};
    \node (h) at (2.75,-1)     [draw, circle, thick, fill=black, scale=0.3]      {\,};
    \node (i) at (3.0,-1)                                                        {\footnotesize{2}};

    \draw (c) edge[-,thick] node [right] {\footnotesize{1}} (e)
          (e) edge[-,thick] node [right,yshift=1mm] {\footnotesize{2}} (h)
          (e) edge[-,thick] node [left,yshift=1mm]  {\footnotesize{3}} (f);
\end{tikzpicture}}
\caption{Rule $r_1$}
\end{subfigure}
\begin{subfigure}{.325\textwidth}
\centering
\scalebox{0.75}{
\begin{tikzpicture}[every node/.style={align=center}]
    \draw (-1.0,-2.5) -- (-1.0,2.5) -- (3.375,2.5) -- (3.375,-2.5) -- cycle [fill=black!5];
    \node (a) at (-0.5,0)                                                        {$X$};
    \node (b) at (0.0,0)                                                         {$\to$};
    \node (c) at (1.75,2)      [draw, circle, thick, fill=black, scale=0.3]      {\,};
    \node (d) at (2.0,2)                                                         {\footnotesize{1}};
    \node (e) at (1.75,1)      [draw, rectangle, minimum size=5mm]               {$X$};
    \node (f) at (1.25,0)      [draw, circle, thick, fill=black!5, scale=0.3]    {\,};
    \node (g) at (2.25,0)      [draw, circle, thick, fill=black!5, scale=0.3]    {\,};
    \node (h) at (1.25,-1)     [draw, rectangle, minimum size=5mm]               {$X$};
    \node (i) at (2.25,-1)     [draw, rectangle, minimum size=5mm]               {$X$};
    \node (j) at (0.75,-2)     [draw, circle, thick, fill=black, scale=0.3]      {\,};
    \node (k) at (0.5,-2)                                                        {\footnotesize{3}};
    \node (l) at (1.75,-2)     [draw, circle, thick, fill=black!5, scale=0.3]    {\,};
    \node (m) at (2.75,-2)     [draw, circle, thick, fill=black, scale=0.3]      {\,};
    \node (n) at (3.0,-2)                                                        {\footnotesize{2}};

    \draw (c) edge[-,thick] node [right] {\footnotesize{1}} (e)
          (e) edge[-,thick] node [right] {\footnotesize{2}} (g)
          (e) edge[-,thick] node [left]  {\footnotesize{3}} (f)
          (f) edge[-,thick] node [left]  {\footnotesize{1}} (h)
          (g) edge[-,thick] node [right] {\footnotesize{1}} (i)
          (h) edge[-,thick] node [left]  {\footnotesize{2}} (l)
          (h) edge[-,thick] node [left]  {\footnotesize{3}} (j)
          (i) edge[-,thick] node [right] {\footnotesize{2}} (m)
          (i) edge[-,thick] node [right] {\footnotesize{3}} (l);
\end{tikzpicture}}
\caption{Rule $r_2$}
\vspace{0.75em}
\end{subfigure}
\begin{subfigure}{.325\textwidth}
\centering
\scalebox{0.75}{
\begin{tikzpicture}[every node/.style={align=center}]
    \draw (-1.0,-1.65) -- (-1.0,1.5) -- (3.375,1.5) -- (3.375,-1.65) -- cycle [fill=black!5];
    \node (a) at (-0.5,0)                                                        {$X$};
    \node (b) at (0.0,0)                                                         {$\to$};
    \node (c) at (1.75,1)      [draw, circle, thick, fill=black, scale=0.3]      {\,};
    \node (d) at (2.0,1)                                                         {\footnotesize{1}};
    \node (e) at (1.25,0)      [draw, rectangle, minimum size=5mm]               {$\square$};
    \node (f) at (2.25,0)      [draw, rectangle, minimum size=5mm]               {$\square$};
    \node (g) at (0.75,-1)     [draw, circle, thick, fill=black, scale=0.3]      {\,};
    \node (h) at (0.5,-1)                                                        {\footnotesize{3}};
    \node (i) at (1.75,-1)     [draw, rectangle, minimum size=5mm]               {$\square$};
    \node (j) at (2.75,-1)     [draw, circle, thick, fill=black, scale=0.3]      {\,};
    \node (k) at (3.0,-1)                                                        {\footnotesize{2}};

    \draw (c) edge[-,thick] node [right] {\footnotesize{1}} (f)
          (f) edge[-,thick] node [right] {\footnotesize{2}} (j)
          (k) edge[-,thick] node [below] {\footnotesize{1}} (i)
          (i) edge[-,thick] node [below] {\footnotesize{2}} (g)
          (g) edge[-,thick] node [left]  {\footnotesize{1}} (e)
          (e) edge[-,thick] node [left]  {\footnotesize{2}} (c);
\end{tikzpicture}}
\caption{Rule $r_3$}
\end{subfigure}
\begin{subfigure}{.325\textwidth}
\centering
\scalebox{0.75}{
\begin{tikzpicture}[every node/.style={align=center}]
    \draw (-1.0,-1.5) -- (-1.0,1.5) -- (3.375,1.5) -- (3.375,-1.5) -- cycle [fill=black!5];
    \node (a) at (-0.5,0)                                                        {$\square$};
    \node (b) at (0.0,0)                                                         {$\to$};
    \node (c) at (1.75,1)      [draw, circle, thick, fill=black, scale=0.3]      {\,};
    \node (d) at (2.0,1)                                                         {\footnotesize{1}};
    \node (e) at (1.75,0)      [draw, rectangle, minimum size=5mm]               {$F$};
    \node (f) at (1.75,-1)     [draw, circle, thick, fill=black, scale=0.3]      {\,};
    \node (g) at (2.0,-1)                                                        {\footnotesize{2}};

    \draw (c) edge[-,thick] node [right] {\footnotesize{1}} (e)
          (e) edge[-,thick] node [right] {\footnotesize{2}} (f);
\end{tikzpicture}}
\caption{Rule $r_4$}
\end{subfigure}
\begin{subfigure}{.325\textwidth}
\centering
\scalebox{0.75}{
\begin{tikzpicture}[every node/.style={align=center}]
    \draw (-1.0,-1.5) -- (-1.0,1.5) -- (3.375,1.5) -- (3.375,-1.5) -- cycle [fill=black!5];
    \node (a) at (-0.5,0)                                                        {$F$};
    \node (b) at (0.0,0)                                                         {$\to$};
    \node (c) at (1.75,1)      [draw, circle, thick, fill=black, scale=0.3]      {\,};
    \node (d) at (2.0,1)                                                         {\footnotesize{1}};
    \node (e) at (1.75,0)      [draw, rectangle, minimum size=5mm]               {$F$};
    \node (f) at (1.75,-1)     [draw, circle, thick, fill=black, scale=0.3]      {\,};
    \node (g) at (2.0,-1)                                                        {\footnotesize{2}};

    \draw (c) edge[-,thick] node [right] {\footnotesize{1}} (e)
          (e) edge[-,thick] node [right] {\footnotesize{2}} (f);
\end{tikzpicture}}
\caption{Rule $r_5$}
\end{subfigure}
\caption{Sierpinski triangle rules}
\label{fig:sierpinskirules}
\end{figure}
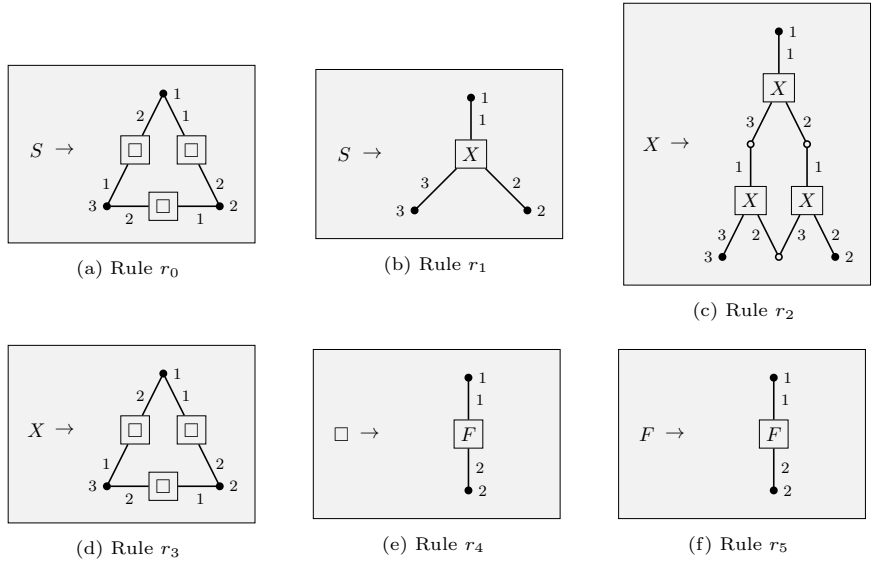

Kreowski showed that his notion of PHR grammars could generate the language of Sierpinski triangles, using two tables \cite{Kreowski92a}. We can improve upon this result, showing that only one table is needed, constructing a repetition-free \((2, 1)\)-PHR grammar \(\mathcal{G}\) generating the language of all directed \(\square\)-labelled Sierpinski triangles. Let \(\mathcal{G} = ((\Sigma, \mathrm{type}), \{\square\}, S, \{T_1\})\) where \(\Sigma = \{S, X, F, \square\}\), \(\mathrm{type}(S) = \mathrm{type}(X) = 3\), \(\mathrm{type}(F) = \mathrm{type}(\square) = 2\), \(T_1 = \{r_0, \dots, r_5\}\), and the fix rules are defined in Figure \ref{fig:sierpinskirules}. Once again, our trick to avoiding a second table is to have the terminal rewrite to \(F\), so that derivations leading to a terminally labelled output must write all the terminals at the same time. Rewriting of terminals was not possible in Kreowski's definition of PHR grammars, and we believe this is precisely what prevented him from using only one table.

\subsection{The Power of Tables}

In this subsection, we show that for each \(k \geq 0\), we in general, have only two possibly distinct classes: the \((k, 1)\)-PHR languages and the \(k\)-PHR languages. That is, the availability of three or more tables does not increase generative power beyond what was possible with only two tables. This is exactly the same situation one finds with E0L and ET0L languages.

\begin{theorem}[PHR Table Power] \label{thm:phrtablepower}
For all \(k \geq 0\), \(\mathcal{P}\mathcal{H}\mathcal{R}_{k,2} = \mathcal{P}\mathcal{H}\mathcal{R}_{k}\) and \(\mathcal{P}\mathcal{H}\mathcal{R}_{k,2}^{\mathrm{rf}} = \mathcal{P}\mathcal{H}\mathcal{R}_{k}^{\mathrm{rf}}\). Moreover, given a (repetition-free) (proper) \((k,l)\)-PHR grammar \(\mathcal{G}\) for any \(k \geq 0\), \(l \geq 2\), one can effectively construct a (repetition-free) (proper) \((k, 2)\)-PHR grammar \(\mathcal{G}'\) such that \(\mathrm{L}(\mathcal{G}) = \mathrm{L}(\mathcal{G}')\).
\end{theorem}

\begin{proof}
\(\mathcal{P}\mathcal{H}\mathcal{R}_{k,2} \subseteq \mathcal{P}\mathcal{H}\mathcal{R}_{k,l}\) and \(\mathcal{P}\mathcal{H}\mathcal{R}_{k,2}^{\mathrm{rf}} \subseteq \mathcal{P}\mathcal{H}\mathcal{R}_{k,l}^{\mathrm{rf}}\) by definition. To see the reverse inclusions, we follow the proof of Theorem V.1.3 of \cite{Rozenberg-Salomaa80a} with modifications lifting the proof from ET0L grammars to PHR grammars. We must show that given some \((k,l)\)-PHR grammar \(\mathcal{G} = ((\Sigma, \mathrm{type}), A, S, \{T_1, \dots. T_l\})\) with \(l \geq 2\), we can construct a \((k,2)\)-PHR grammar \(\mathcal{G} = ((\Sigma', \mathrm{type}'), A, S, \{{T'}_1, {T'}_2\})\), with \(\mathrm{L}(\mathcal{G}) = \mathrm{L}(\mathcal{G}')\).

We setup the signature as follows. Let \(\Sigma' = \{[A, i] \mid A \in \Sigma, i \in \underline{l}\} \cup \Sigma\), \(\mathrm{type}'([A, i]) = \mathrm{type}(A)\) for all \(A \in \Sigma, i \in \underline{l}\), and \(\mathrm{type}'(A) = \mathrm{type}(A)\) for all \(A \in \Sigma\). The first table, \({T'}_1\), contains exactly the rules \((A, [A, 1]^{\bullet})\) for all \(A \in \Sigma\), \(([A, i], [A, i+1]^{\bullet})\) for all \(A \in \Sigma, i \in \underline{l-1}\), and \(([A, l], [A, 1]^{\bullet})\) for all \(A \in \Sigma\). The second table, \({T'}_2\), contains exactly the rules \((A, A^\bullet)\) for all \(A \in \Sigma\) and \(([A, i], R)\) for all \(i \in \underline{l}, (A, R) \in T_i\).

This construction means that the table \({T'}_1\) rewrites a hypergraph \(H \in \mathcal{H}_{\mathcal{C}}\) into the hypergraph \(H_{(1)}\) resulting from \(H\) by replacing every label \(A\) by \([A, 1]\), then \(H_{(1)}\) is rewritten to \(H_{(2)}\) by replacing every \([A, 1]\) by \([A, 2]\), etc, with \([A, l]\) looping back round to \([A, 1]\). If at any moment, table \({T'}_2\) can be applied, it rewrites \(H_{(i)}\) to \(H'\) if and only if \(H \Rrightarrow_{T_i} H'\) where \(H\) is \(H_{(i)}\) with each \([A, i]\) replaced by \(A\). One should also note that the typing function \(\mathrm{type}'\) is set up to be compatible with all these replacements, so all our rules are well-typed.

By repeating the above process, we can see all derivations in \(\mathcal{G}\) are simulated by \(\mathcal{G}'\), so \(\mathrm{L}(\mathcal{G}) \subseteq \mathrm{L}(\mathcal{G}')\), since the terminal labels and start states are the same. The reverse inclusion is obvious, so we have \(\mathrm{L}(\mathcal{G}) = \mathrm{L}(\mathcal{G}')\), as required.

Finally, it is easy to see that all of our transformations preserved if the grammar was repetition-free or proper.
\end{proof}

\subsection{Rational Control of Traces}

It is often convenient to restrict the sequences of allowed traces when defining a language using a PHR grammar, leading to better readability of grammars and possibly shorter proofs. A popular choice in L systems is so-called rational control, and was considered for ET0L in 1975 by Nielsen \cite{Nielsen75a} and later by Asveld \cite{Asveld77a}. We will make precise a notion of rational control for PHR grammars, and show that generational power actually remains the same whenever we have at least two tables, because we can always encode the rational control.

\begin{definition}[Controlled Parallel Derivation]
Given \(\mathcal{C} = (\Sigma, \mathrm{type})\), \(H, H' \in \mathcal{H}_{\mathcal{C}}\), a finite set of tables \(\mathcal{T} = \{T_i \mid i \in I\}\) over \(\Sigma\) indexed by \(I\), and \(\mathcal{M}\) an FSA over \(I\), we say \(H\) (\(\mathcal{M}\)-)parallelly derives \(H'\) if \(H\) parallelly derives \(H'\) with trace \(i_1 i_2 \cdots i_k \in \mathrm{L}(\mathcal{M})\). We write \(H \Rrightarrow_{\mathcal{T}}^{i_1 i_2 \cdots i_k} H'\), \(H \Rrightarrow_{\mathcal{T}}^k H'\), or \(H \Rrightarrow_{\mathcal{T}}^{\mathcal{M}} H'\).
\end{definition}

\begin{definition}[PHR Grammar with Control]
A (repetition-free) parallel hyperedge replacement grammar with control of order \(k\) using at most \(l\) tables is a tuple \(\mathcal{G} = (\mathcal{C}, A, S, \mathcal{T}, \mathcal{M})\) where \((\mathcal{C}, A, S, \mathcal{T})\) is a (repetition-free) \((k,l)\)-PHR grammar (called the underlying grammar) with \(\mathcal{T}\) indexed by \(I\), and \(\mathcal{M}\) is an FSA over \(I\) (called the rational control). The language generated by \(\mathcal{G}\) is \(\mathrm{L}(\mathcal{G}) = \{H \in \mathcal{H}_{\mathcal{C}} \mid S^{\bullet} \Rrightarrow_{\mathcal{T}}^{\mathcal{M}} H \textrm{ with } \mathrm{lab}_H^{-1}(A) = E_H\} \subseteq \mathcal{H}_{\mathcal{C}}\).
\end{definition}

\begin{theorem}[PHR Grammar Control Removal] \label{thm:controlrem}
Given a (repetition-free) \(k\)-PHR grammar with control \(\mathcal{G}\), one can effectively construct a (repetition-free) \(k\)-PHR grammar \(\mathcal{G}'\) such that \(\mathrm{L}(\mathcal{G}) = \mathrm{L}(\mathcal{G}')\).
\end{theorem}

\begin{proof}
Let \(\mathcal{G} = ((\Sigma, \mathrm{type}), A, S, \{T_1, \dots T_l\}, \mathcal{M})\). Without loss of generality, we can assume that \(\mathcal{M} = (Q, \underline{n}, \delta, i, F))\) is deterministic and full, and \(Q \cap \Sigma = \emptyset\).

We construct the (repetition-free) \(k\)-PHR grammar \(\mathcal{G}' = ((\Sigma', \mathrm{type}'), A, S',\) \(\mathcal{T}')\). First, make a disjoint (from \(\Sigma_1\)) copy of \(A\), \(\overbar{A}\), and to each \(X \in \overbar{A}\), associate a unique \(\overbar{X} \in A\). Moreover, given a hypergraph \(H\), denote by \(\overbar{H}\) the same hypergraph but with its labelling function composed with the function that sends \(X \in A\) to \(\overbar{X}\) and leaves everything else in place. Next, choose some additional fresh symbols: \(S'\), \(F_0, \dots, F_k\). Using these, we define \(\Sigma' = \{S'\} \cup Q \cup \Sigma \cup \overbar{A} \cup \{F_0, \dots, F_l\}\) where \(\mathrm{type}(S') = \mathrm{type}(S)\), \(\mathrm{type}(q) = 0\) for all \(q \in Q\), \(\mathrm{type}'(X) = \mathrm{type}(X)\) for all \(X \in \Sigma\), \(\mathrm{type}'(\overbar{X}) = \mathrm{type}(X)\) for all \(X \in A\), and \(\mathrm{type}'(F_j) = j\) for all \(j \in \underline{k}\).

Finally, let \(\mathcal{T}' = \{T'_{0}, T'_1, \dots T'_l\}\) where \(T'_0 = \{(S', \overbar{S^{\bullet}} \sqcup i^{\bullet})\} \cup \{(q, \emptyset) \mid q \in F\} \cup \{(\overbar{X}, X^{\bullet}) \mid X \in A\} \cup \{(X, F_{\mathrm{type}'(X)}) \mid X \in (Q \setminus F) \cup \Sigma \cup \{F_0, \dots, F_k\}\}\) and for each \(j \in \underline{l}\), \(T'_j = \{(\overbar{L}, \overbar{R}) \mid (L, R) \in T_j \land L \in A\} \cup \{(L, \overbar{R}) \mid (L, R) \in T_j \land L \not\in A\} \cup \{(q, \delta(q, j)^{\bullet}) \mid q \in Q\} \cup \{(X, X^{\bullet}) \mid X \in \{S'\} \cup A \cup \{F_0, \dots F_k\}\}\).

We can see that the purpose of table \(0\) is to start and stop the derivation process, with the rest of the tables simulating the original system, while also simulating the automaton. So, if \(S^{\bullet} \Rrightarrow_{\mathcal{T}}^w H\) is a derivation in \(\mathcal{G}\) with \(H\) terminally labelled and \(w \in \mathrm{L}(\mathcal{M})\), then there is a corresponding derivation \(S'^{\bullet} \Rrightarrow_{\mathcal{T}}^{0w0} H\) in \(\mathcal{G}'\). That is \(\mathrm{L}(\mathcal{G}) \subseteq \mathrm{L}(\mathcal{G}')\). To see the reverse inclusion, we analyse all derivations of the form \(S'^{\bullet} \Rrightarrow_{\mathcal{T}}^{w} H\). If \(w\) does not contain \(0\) at least twice, then \(H\) is necessarily not terminally labelled. So, useful derivations must have trace \(x0y0z\) where \(x, y \in \{1, \dots l\}^*\), \(z \in \{0, \dots l\}^*\). Clearly if \(S'^{\bullet} \Rrightarrow_{\mathcal{T}}^{x} H'\), then \(H' \cong S'^{\bullet}\), so we can assume \(x = \epsilon\). Similarly, if \(S'^{\bullet} \Rrightarrow_{\mathcal{T}}^{0y0} H'\) and \(H'\Rrightarrow_{\mathcal{T}}^{z} H\), then either \(H' \cong H\) if \(H\) was terminally labelled, and so we could assume \(z = \epsilon\), or \(H'\) is labelled by at least non-terminal which has no terminally labelled successor hypergraph, and so it doesn't matter what \(H\) is. Finally, we analyse \(y\). If \(y \in \mathrm{L}(\mathcal{M})\), then we proceed as in the analysis of the other direction of inclusion. If \(y \not\in \mathrm{L}(\mathcal{M})\), then the final step sends the type zero symbol tracking the machine state to \(F_0\) which forces all successors of the hypergraph to be not terminally labelled.
\end{proof}

Thus, the hypergraph languages that can be generated by (repetition-free) \(k\)-PHR grammars are exactly those that can be generated by (repetition-free) \(k\)-PHR grammars with control, since certainly no control can be simulated. It is unclear if this also holds for \((k, 1)\)-PHR grammars, but we conjecture that it does not.

\subsection{Synchronisation and Substitution Closure} \label{subsec:subc}

Recall from Subsection \ref{subsec:et0l} that an extremely useful notion in L systems is that of a synchronised system. We will now define this notion of PHR grammars, and show that all PHR grammars can be synchronised. This will allow us to show some substitution closure results for various relevant classes of hypergraph languages and hypergraph substitutions.

\begin{definition}[Synchronised PHR Grammar]
Call a PHR grammar \((\mathcal{C}, A,\) \(S, \mathcal{T})\) synchronised if for all hypergraphs \(H, H'\) over \(A\), there cannot exist a direct derivation \(H \Rrightarrow_T H'\), for any table \(T \in \mathcal{T}\). Moreover, call a symbol \(X \in \Sigma\), \(Y\)-final if every rule in any table in \(\mathcal{T}\) of the form \((X, R)\) is such that \(R \cong Y^{\bullet}\). We call an \(X\)-final symbol \(X\), final.
\end{definition}

The following result shows that grammars can be synchronised (just like ET0L grammars), and that without loss of generality, we can assume the start symbol essentially does not appear in the RHS of any rules (just like context-free grammars).

\begin{lemma}[Synchronisation of PHR Grammars] \label{lem:syncphr}
Given a (repetition-free) (proper) \((k, l)\)-PHR grammar \(\mathcal{G} = ((\Sigma, \mathrm{type}), A, S, \{T_1, \dots T_l\}))\), one can effectively construct a synchronised (repetition-free) (proper) \((k, l)\)-PHR grammar \(\mathcal{G}' = ((\Sigma', \mathrm{type}'), A, S', \mathcal{T}'))\) with \(\Sigma \subseteq \Sigma'\) and fresh symbols \(S', F_0, \dots, F_k \subseteq \Sigma' \setminus \Sigma\) such that:

\begin{enumerate}
    \item \(\restr{\mathrm{type}'}{\Sigma} = \mathrm{type}\) and \(\mathrm{L}(\mathcal{G}) = \mathrm{L}(\mathcal{G}')\);
    \item \(S'\) is such that \(\forall T \in \mathcal{T}', \forall (L, R) \in T, S' \not\in \mathrm{lab}_R(E_R)\);
    \item  \(\forall i \in \underline{k}, \mathrm{type}(F_i) = i\) and each \(F_i\) is \(F_i\)-final;
    \item each \(X \in A\) is \(F_{\mathrm{type}(X)}\)-final.
\end{enumerate}
\end{lemma}

\begin{proof}
We will construct \(\mathcal{G}'\) in multiple phases, each preserving the properties established by the phases prior. Property (1) is already satisfied by \(\mathcal{G}\). It is easy to see that properties (3) and (4) imply that the grammar is synchronised, since any terminally labelled hypergraph is immediately sent to one labelled by only the \(F_i\)s, and cannot escape.

We first construct the grammar \(\mathcal{G}_0 = ((\Sigma_0, \mathrm{type}_0), A, S', \mathcal{T}_0)\). We have \(\Sigma_0 = \Sigma \cup \{S'\}\) where \(S'\) is some fresh label, \(\mathrm{type}_0(X) = \mathrm{type}(X)\) for all \(X \in \Sigma\), and \(\mathrm{type}_0(S') = \mathrm{type}(S)\). Finally, set \(\mathcal{T}_0 = \{{T_0}_1, \dots {T_0}_l\}\) where each \({T_0}_i = T_i \cup \{(S', S^{\bullet})\}\). Clearly \(\mathcal{G}_0\) now additionally satisfies property (2).

Next, we construct the grammar \(\mathcal{G}_1 = ((\Sigma_1, \mathrm{type}_1), A, S', \mathcal{T}_1)\). We have \(\Sigma_1 = \Sigma_0 \cup \{F_0, \dots F_l\}\) where the \(F_i\) are fresh labels, \(\mathrm{type}_1(X) = \mathrm{type}_0(X)\) for all \(X \in \Sigma_0\), and \(\mathrm{type}_1(F_i) = i\) for all \(0 \leq i \leq l\). Finally, set \(\mathcal{T}_1 = \{{T_1}_1, \dots {T_1}_l\}\) where each \({T_1}_i = {T_0}_i \cup \{(F_j, {F_j}^{\bullet}) \mid 0 \leq j \leq l\}\). Clearly \(\mathcal{G}_1\) now additionally satisfies property (3).

Next, we construct the grammar \(\mathcal{G}' = ((\Sigma', \mathrm{type}'), A, S', \mathcal{T}')\). First, make a disjoint (from \(\Sigma_1\)) copy of \(A\), \(\overbar{A}\) and to each \(X \in \overbar{A}\), associate a unique \(\overbar{X} \in \overbar{A}\). Moreover, given a hypergraph \(H\), denote by \(\overbar{H}\) the same hypergraph but with its labelling function composed with the function that sends \(X \in A\) to \(\overbar{X}\) and leaves everything else in place. Now, set \(\Sigma' = \Sigma_1 \cup \overbar{A}\), \(\mathrm{type}'(X) = \mathrm{type}_1(X)\) for all \(X \in \Sigma_1\), and \(\mathrm{type}'(\overbar{X}) = \mathrm{type}_{1}(X)\) for all \(X \in A\). Finally, set \(\mathcal{T}' = \{T'_1, \dots T'_l\}\) where each \(T'_i = \{(X, {F_{\mathrm{type'}(X)}}^{\bullet}), (\overbar{X}, X^{\bullet}) \mid X \in A\}\) \(\cup\) \(\{(\overbar{L}, \overbar{R}) \mid (L, R) \in {T_1}_i \land L \in A\}\) \(\cup\) \(\{(L, \overbar{R}) \mid (L, R) \in {T_1}_i \land L \in \Sigma_1 \setminus A\}\). Clearly \(\mathcal{G}'\) now additionally satisfies property (4).
\end{proof}

Habel observed that every PHR language can be embedded in an HR language \cite{Habel92b}. Using synchronisation, we can quickly establish this result:

\begin{corollary}[PHR Embedding in HR]
Every (repetition-free) (proper) \(k\)-PHR language can be embedded in a (repetition-free) (proper) \(k\)-HR language, for any \(k \geq 0\). Moreover, given a (repetition-free) (proper) \(k\)-PHR grammar \(\mathcal{G}\), we can effectively construct a (repetition-free) (proper) \(k\)-HR grammar \(\mathcal{G}'\) such that \(\mathrm{L}(\mathcal{G}) \subseteq \mathrm{L}(\mathcal{G}')\).
\end{corollary}

\begin{proof}
Lemma \ref{lem:syncphr} tells us how to construct a \(k\)-PHR grammar \(\mathcal{G}_0 = ((\Sigma, \mathrm{type}),\) \(A, S, \{T_1, \dots, T_l\})\) satisfying the conditions of the lemma with \(\mathrm{L}(\mathcal{G}_0) = \mathrm{L}(\mathcal{G})\). We now construct the \(k\)-HR grammar \(\mathcal{G}' = ((\Sigma, \mathrm{type}), N, S, \mathcal{R})\) where \(N = \Sigma \setminus A\) and \(\mathcal{R} = \{(L, R) \mid (L, R) \in \bigcup \mathcal{T} \land L \not\in A\}\). Clearly \(\mathcal{G}'\) is well-defined and \(\mathrm{L}(\mathcal{G}) \subseteq \mathrm{L}(\mathcal{G}')\) via decomposition of parallel derivations. 
\end{proof}

We can also use synchronisation to show that PHR languages are substitution closed, a property also enjoyed by HR languages (Theorem \ref{thm:hrsub}). Specifically, we show that a \(k\)-HR language under a \((k, 1)\)-PHR substitution is a \((k, 1)\)-PHR language, a \((k, 1)\)-PHR language under a finite substitution is a \((k, 1)\)-PHR language, and that a \(k\)-PHR language under a \(k\)-PHR substitution is a \(k\)-PHR language. Our proof is constructive, starting with grammars, and constructing a grammar generating the image under the substitution. Finally, we also note that being proper, a graph language, or a string graph language is preserved under substitutions that send terminals to languages also satisfying these properties. We will later use our result in Theorem \ref{thm:x}.

\begin{theorem}[PHR Languages Substitution Closed] \label{thm:phrsub}
Given \(\mathcal{C} = (\Sigma, \mathrm{type})\) and \(A \subseteq \Sigma\):

\begin{enumerate}
    \item for any \(k \geq 0\), if \(L\) is a \(k\)-HR language over \(A\) and \(s\) is a \(\mathcal{P}\mathcal{H}\mathcal{R}_{k,1}\)-substitution of hypergraphs on \(A\), then \(s(L)\) is a \((k, 1)\)-PHR language;
    \item for any \(k \geq 0\), if \(L\) is a \((k, 1)\)-PHR language over \(A\) and \(s\) is a finite substitution of hypergraphs on \(A\), then \(s(L)\) is a \((k, 1)\)-PHR language;
    \item for any \(k \geq 0\), if \(L\) is a \(k\)-PHR language over \(A\), and \(s\) is a \(\mathcal{P}\mathcal{H}\mathcal{R}_{k}\)-substitution of hypergraphs on \(A\), then \(s(L)\) is a \(k\)-PHR language.
\end{enumerate}

Moreover, if \(\mathcal{P}\) is the property of being repetition-free, proper, a graph language, or a string graph language, if \(L\) is \(\mathcal{P}\) and \(s(X)\) is \(\mathcal{P}\) for each \(X \in A\), then \(s(L)\) is \(\mathcal{P}\).
\end{theorem}

\begin{proof}
Let \(A = \{X_1, \dots X_m\}\) and \(B = \bigcup_{H \in s(A)} \mathrm{lab}_H(E_H)\) throughout this proof, and without loss of generality, we can assume \(A \cap B = \emptyset\), by simply renaming items in \(A\) and adjusting \(s\) accordingly.

To see (1), there must exist \((k, 1)\)-PHR grammars \(\mathcal{G}_i = ((\Sigma_i, \mathrm{type}_i), B, S_i,\) \(\{{T_i}\})\) such that \(\mathrm{L}(\mathcal{G}_i) = s(X_i)\) for all \(i \in \underline{m}\), each \(\mathcal{G}_i\) satisfies the conditions Lemma \ref{lem:syncphr}, the final symbols \(F_0, \dots, F_k\) are shared between the grammars, and the remaining non-terminals are disjoint (for all \(i \neq j \in \underline{m}\), \((\Sigma_i \setminus (\{F_0, \dots, F_k\} \cup A_i)) \cap \Sigma_j \setminus (\{F_0, \dots, F_k\} \cup A_j)) = \emptyset\)). There must also exist a \(k\)-HR grammar \(\mathcal{G} = (\mathcal{C}, N, S, \mathcal{R})\) such that \(\mathrm{L}(\mathcal{G}) = L\), \(\Sigma \setminus N = A\), and \(N \cap (\bigcup_{i \in \underline{k}} \Sigma_i) = \emptyset\).

We now use the proof of Theorem \ref{thm:phrgen} to construct a \((k, 1)\)-PHR grammar \(\mathcal{G}_0 = (\mathcal{C}, A, S, \{T\})\) such that \(\mathrm{L}(\mathcal{G}_0) = \mathrm{L}(\mathcal{G}) = L\). Recall this construction set \(T = \mathcal{R} \cup \{(X, X^{\bullet}) \mid X \in \Sigma\}\). We now construct a new \((k,1)\)-PHR grammar \(\mathcal{G}' = ((\Sigma', \mathrm{type}'), B, S, \{T'\})\). Set \(\Sigma' = \Sigma \cup \Sigma_1 \cup \cdots \cup \Sigma_m\) and the type function accordingly. Set \(T' = T \cup \{(X_i, S_i) \mid 1 \leq i \leq m\} \cup \{(F_i, {F_i}^{\bullet}) \mid 0 \leq i \leq k\} \cup \{(X, {F_{\mathrm{type}'(X)}}^{\bullet}) \mid X \in B\} \cup \bigcup_{i \in \underline{m}}\{(L, R) \in T_i \mid L \in \Sigma \setminus (B \cup \{F_0, \dots, F_k\})\}\) where the \(T_1, \dots, T_m\) are the single tables from the \(\mathcal{G}_1, \dots, \mathcal{G}_m\), respectively. It is now easy to see that derivations in our new grammar simulate derivations in \(\mathcal{G}\), and then arbitrarily re-write terminals (in \(A\)) to the start symbol of the grammar generating the language that terminal is to be substituted for. Note that because the rule \((X, X^{\bullet})\) is in the table for all \(X \in A\), each derivation step need not immediately re-write a terminal to the start symbol for the associated grammar. The effect of this is that when these grammars are then simulated, they need not run for the same number of steps as each other. It is now clear that \(\mathrm{L}(\mathcal{G'}) = s(\mathrm{L}(\mathcal{G})) = s(L)\), as required.

To see (2), let \(\mathcal{G} = (\mathcal{C}, A, S, \{T\})\) be a \((k, 1)\)-PHR grammar generating \(L\). Without loss of generality, we can assume it satisfies the conditions of Lemma \ref{lem:syncphr} with final symbols \(F_0, \dots, F_k\).

We now construct a new \((k, 1)\)-PHR grammar \(\mathcal{G}' = ((\Sigma', \mathrm{type}'), B, S, \{T'\})\) where \(\Sigma' = \Sigma \cup B\) and the typing function is defined in the obvious way. Next, we define the finite substitution \(s'\) on \(\Sigma\) by \(s'(X) = s(X)\) for all \(X \in A\) and \(s'(X) = \{X^{\bullet}\}\) for all \(X \in \Sigma \setminus A\). Finally, let \(T' = (\bigcup \{\{(L, H) \mid H \in s'(R)\} \mid (L, R) \in T \land L \in \Sigma \setminus A\}) \cup \{(X, F_{\mathrm{type}'(X)}) \mid X \in B\}\). It is now clear that this new grammar simulates the original, only with all terminals in RHS replaced with all combinations of their replacements under \(s\), and the synchronisation properties are preserved.

To see (3), there must exist \(k\)-PHR grammars \(\mathcal{G}_i = ((\Sigma_i, \mathrm{type}_i), B, S_i, \mathcal{T}_i)\) such that \(\mathrm{L}(\mathcal{G}_i) = s(X_i)\) for all \(i \in \underline{m}\), each \(\mathcal{G}_i\) satisfies the conditions Lemma \ref{lem:syncphr}, the final symbols \(F_0, \dots, F_k\) are shared between the grammars, and the remaining non-terminals are disjoint (for all \(i \neq j \in \underline{m}\), \((\Sigma_i \setminus (\{F_0, \dots, F_k\} \cup B)) \cap \Sigma_j \setminus (\{F_0, \dots, F_k\} \cup A_j)) = \emptyset\)). There must also exist a \(k\)-PHR grammar \(\mathcal{G} = (\mathcal{C}, A, S, \mathcal{T})\) such that \(\mathrm{L}(\mathcal{G}) = L\), and \(\Sigma \cap (\bigcup_{i \in \underline{k}} \Sigma_i) = \emptyset\). For each \(i \in \underline{m}\), let \(\overbar{\Sigma_i}\) be a copy of \(\Sigma_i\) consisting of fresh symbols, and identify each \(X \in \Sigma_i\) with its copy \(\overbar{X} \in \overbar{\Sigma_i}\). Moreover, given a hypergraph \(H\), by \(\overbar{H}\) we mean \(H\) but with its labelled function composed with the function which takes any \(X \in \Sigma_i\) to \(\overbar{X}\) and leaves everything else fixed.

We now construct the \(k\)-PHR grammar \(\mathcal{G}' = (\mathcal{C}', B, \mathcal{T}', S)\) such that \(\mathrm{L}(\mathcal{G}')\) \(= S(L)\). Let \(\mathcal{C}' = (\Sigma', \mathrm{type}')\) be the union of all the above signatures also including the disjoint copies, where copies are of the same type as their original symbols. Finally, let \(\mathcal{R} = \{(X, X^{\bullet}) \mid X \in \Sigma'\}\), \(\mathcal{F} = \{(X, F_{\mathrm{type}'(X)}^{\bullet}) \mid X \in \Sigma'\}\), and \(\mathcal{T}' = \bigcup_{0 \leq i \leq m} \mathcal{T}'_i\), where \(\mathcal{T}'_0 = \{\mathcal{R} \oplus T \mid T \in \mathcal{T}\} \cup \{\mathcal{F} \oplus \{(X_i, \overbar{S_i}^{\bullet}) \mid i \in \underline{m}\}\}\), and for each \(i \in \underline{m}\) let \(\mathcal{T}_i = \{\mathcal{R} \oplus (\{(\overbar{L}, \overbar{R}) \mid (L, R) \in T\} \cup \{(\overbar{S_i}, \overbar{S_i}^{\bullet})\}) \mid T \in \mathcal{T}_i\} \cup \{\mathcal{R} \oplus (\{(\overbar{X}, X^{\bullet}) \mid X \in B\} \cup \{(\overbar{X}, F_{\mathrm{type}'(X)}^{\bullet}) \mid X \in \Sigma_i \setminus (B \cup \{\overbar{S_i}\})\})\}\).

One can see that derivations that make progress start with \(S^{\bullet}\) then apply tables from the first part of \(\mathcal{T}'_0\), simulating \(\mathcal{G}\). At some point, the final table of \(\mathcal{T}'_0\) may be applied, which immediately rewrites all the terminals to encoded start symbols for their respective grammars for substitution and sends all non-terminals to failure non-terminals. If a hypergraph contains any failure non-terminals at this point, non terminally labelled hypergraph can be derived in future. Derivations can now simulate the \(\mathcal{G}_i\) totally independently, with choice of delaying start, giving total freedom over the simulated derivation sequences for each instance of the encoded start symbol. Finally, the encoded systems can end their simulation at any point, just like before, by sending their encoded terminals to real terminals in \(B\) and their encoded non-terminals to failure non-terminals. It is now clear that \(\mathrm{L}(\mathcal{G}') = s(L)\), as required.

Finally, preservation of the property of being repetition-free, proper, a graph language, or a string graph language is easy to see.
\end{proof}

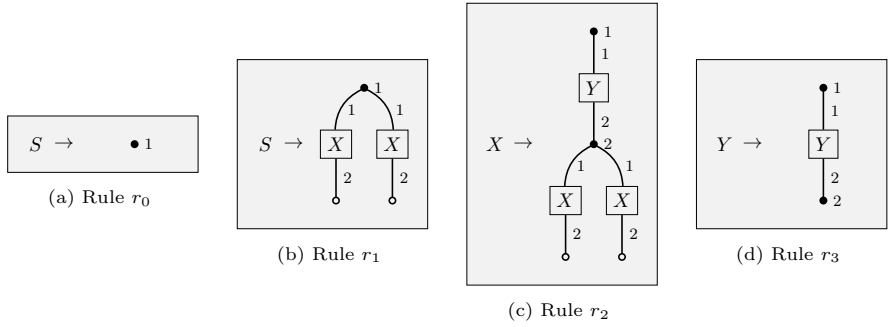
\begin{figure}[!ht]
\begin{subfigure}{.245\textwidth}
\centering
\scalebox{0.75}{
\begin{tikzpicture}[every node/.style={align=center}]
    \draw (-0.5,-0.5) -- (-0.5,0.5) -- (2.875,0.5) -- (2.875,-0.5) -- cycle [fill=black!5];
    \node (a) at (0.0,0)                                                         {$S$};
    \node (b) at (0.5,0)                                                         {$\to$};
    \node (c) at (1.75,0)       [draw, circle, thick, fill=black, scale=0.3]      {\,};
    \node (d) at (2.0,0)                                                        {\footnotesize{1}};
\end{tikzpicture}}
\caption{Rule $r_0$}
\end{subfigure}
\begin{subfigure}{.245\textwidth}
\centering
\scalebox{0.75}{
\begin{tikzpicture}[every node/.style={align=center}]
    \draw (-0.5,-1.5) -- (-0.5,1.5) -- (2.875,1.5) -- (2.875,-1.5) -- cycle [fill=black!5];
    \node (a) at (0.0,0)                                                         {$S$};
    \node (b) at (0.5,0)                                                         {$\to$};
    \node (c) at (1.75,1)      [draw, circle, thick, fill=black, scale=0.3]      {\,};
    \node (d) at (2.0,1)                                                         {\footnotesize{1}};
    \node (e) at (1.25,0)      [draw, rectangle, minimum size=5mm]               {$X$};
    \node (f) at (1.25,-1)     [draw, circle, thick, fill=black!5, scale=0.3]  {\,};
    \node (g) at (2.25,0)      [draw, rectangle, minimum size=5mm]               {$X$};
    \node (h) at (2.25,-1)     [draw, circle, thick, fill=black!5, scale=0.3]  {\,};

    \draw (c) edge[-,thick, bend right=30, pos=0.6] node [right] {\footnotesize{1}} (e)
          (c) edge[-,thick, bend left=30, pos=0.6] node [right] {\footnotesize{1}} (g)
          (e) edge[-,thick] node [right] {\footnotesize{2}} (f)
          (g) edge[-,thick] node [right] {\footnotesize{2}} (h);
\end{tikzpicture}}
\caption{Rule $r_1$}
\end{subfigure}
\begin{subfigure}{.245\textwidth}
\centering
\scalebox{0.75}{
\begin{tikzpicture}[every node/.style={align=center}]
    \draw (-0.5,-2.5) -- (-0.5,2.5) -- (2.875,2.5) -- (2.875,-2.5) -- cycle [fill=black!5];
    \node (a) at (0.0,0)                                                         {$X$};
    \node (b) at (0.5,0)                                                         {$\to$};
    \node (c) at (1.75,2)      [draw, circle, thick, fill=black, scale=0.3]      {\,};
    \node (d) at (2.0,2)                                                         {\footnotesize{1}};
    \node (e) at (1.75,1)      [draw, rectangle, minimum size=5mm]               {$Y$};
    \node (f) at (1.75,0)      [draw, circle, thick, fill=black, scale=0.3]      {\,};
    \node (g) at (2.0,0)                                                         {\footnotesize{2}};
    \node (h) at (1.25,-1)     [draw, rectangle, minimum size=5mm]               {$X$};
    \node (i) at (1.25,-2)     [draw, circle, thick, fill=black!5, scale=0.3]  {\,};
    \node (j) at (2.25,-1)     [draw, rectangle, minimum size=5mm]               {$X$};
    \node (k) at (2.25,-2)     [draw, circle, thick, fill=black!5, scale=0.3]  {\,};

    \draw (c) edge[-,thick] node [right] {\footnotesize{1}} (e)
          (e) edge[-,thick] node [right] {\footnotesize{2}} (f)
          (f) edge[-,thick, bend right=30, pos=0.6] node [right] {\footnotesize{1}} (h)
          (f) edge[-,thick, bend left=30, pos=0.6] node [right] {\footnotesize{1}} (j)
          (h) edge[-,thick] node [right] {\footnotesize{2}} (i)
          (j) edge[-,thick] node [right] {\footnotesize{2}} (k);
\end{tikzpicture}}
\caption{Rule $r_2$}
\end{subfigure}
\begin{subfigure}{.245\textwidth}
\centering
\scalebox{0.75}{
\begin{tikzpicture}[every node/.style={align=center}]
    \draw (-0.5,-1.5) -- (-0.5,1.5) -- (2.875,1.5) -- (2.875,-1.5) -- cycle [fill=black!5];
    \node (a) at (0.0,0)                                                         {$Y$};
    \node (b) at (0.5,0)                                                         {$\to$};
    \node (c) at (1.75,1)      [draw, circle, thick, fill=black, scale=0.3]      {\,};
    \node (d) at (2.0,1)                                                         {\footnotesize{1}};
    \node (e) at (1.75,0)      [draw, rectangle, minimum size=5mm]               {$Y$};
    \node (f) at (1.75,-1)     [draw, circle, thick, fill=black, scale=0.3]      {\,};
    \node (g) at (2.0,-1)                                                        {\footnotesize{2}};

    \draw (c) edge[-,thick] node [right] {\footnotesize{1}} (e)
          (e) edge[-,thick] node [right] {\footnotesize{2}} (f);
\end{tikzpicture}}
\caption{Rule $r_3$}
\end{subfigure}
\caption{Full binary tree rules revisited}
\label{fig:fbtreerulesrev}
\end{figure}

In light of the above result, we can view the language \(\mathrm{FBT}\) from Subsection \ref{subsec:examples} as the image of the repetition-free \((2, 1)\)-PHR grammar \(((\{S, X, Y\}, \langle S \mapsto 1, X \mapsto 2, Y \mapsto 2\rangle), \{X, Y\}, S, \{\{r_0, r_1, r_2, r_3\}\})\) under the repetition-free finite substitution of hypergraphs \(h(X) = h(Y) = \square^{\bullet}\), where the \(r_i\) are defined in Figure \ref{fig:fbtreerulesrev}. Moreover, we can generate the language of full binary trees over an arbitrary label set \(\{X_1, \dots X_m\}\) using the substitution \(h(X) = h(Y) = \{X_1^{\bullet}, \dots X_m^{\bullet}\}\). A similar trick is also possible with the grammar that generated the language of all directed Sierpinski triangles.

Finally, we can generalise Theorem \ref{thm:etolafl} for PHR languages:

\begin{theorem}[PHR Languages Hyper-Algebraically Closed] \label{thm:phrhyper}
Given \(\mathcal{C} = (\Sigma, \mathrm{type})\), \(A \subseteq \Sigma\), and \(k \geq 0\), if \(L\) is a \(k\)-PHR language over \(A\), and \(\mathcal{S}\) is a finite set of \(\mathcal{P}\mathcal{H}\mathcal{R}_{k}\)-substitutions of hypergraphs on \(A\), then \(L' = \bigcup_{n \in \mathbb{N}} \{s_n( \cdots s_1(L) \cdots ) \mid s_1, \dots, s_n \in \mathcal{S}\}\) is a \(k\)-PHR language. Moreover, if \(\mathcal{P}\) is the property of being repetition-free, proper, a graph language, or a string graph language, if \(L\) is \(\mathcal{P}\) and \(s(X)\) is \(\mathcal{P}\) for each \(s \in \mathcal{S}\) and \(X \in A\), then \(L'\) is \(\mathcal{P}\).
\end{theorem}

\begin{proof}
The proof is very similar to that of part (3) of Theorem \ref{thm:phrsub}. The only difference is the addition of another table which can be used to send all terminals to restart the process and all non-terminals to a failure symbol.
\end{proof}

\begin{corollary}[PHR Languages Iterated Substitution Closed]
For all \(k \geq 0\), \(\mathcal{P}\mathcal{H}\mathcal{R}_k\) (\(\mathcal{P}\mathcal{H}\mathcal{R}_k^{\mathrm{rf}}\)) is iterated substitution closed.
\end{corollary}

\subsection{The Emptiness Problem} \label{subsec:emptiness}

An important and useful decision problem is the emptiness problem. That is, given a PHR grammar, decide if it generates a non-empty language or not. In this subsection, we show this problem is decidable, in general. This result also has utility when eliminating unreachable symbols from PHR grammars, which we tackle at the end of this subsection, but also for providing us with an explicit algorithm for solving the membership problem for PHRS languages (Subsection \ref{subsec:phrsmp}).

Before we present and prove Theorem \ref{thm:decemptiness}, we first must define label sets, and derivations on label sets. Soundless and completeness of simulation of PHR derivations by label set derivations, together with the fact that analysis of label sets is easy, enables us to show how to decide emptiness for PHR grammars.

\begin{definition}[Label Set]
Given some signature \(\mathcal{C} = (\Sigma, \mathrm{type})\) let \(\mathcal{L}_{\mathcal{C}} = \mathcal{P}(\Sigma)\) and \(\mathrm{labels}: \mathcal{H}_{\mathcal{C}} \to \mathcal{L}_{\mathcal{C}}\) be the function defined by \(H \mapsto \mathrm{lab}_H(E_H)\). Call \(X \in \mathcal{L}_{\mathcal{C}}\) a label set and define the hypergraph:

\vspace{-0.333333em}
\[
\mathrm{H}(X) = \bigsqcup_{A \in X} \mathrm{untype}(A^{\bullet})
\]
\vspace{-0.666666em}

\noindent
where \(\mathrm{untype}(H)\) is \(H\) with \(\mathrm{ext}_H\) replaced by \(\emptyset\).
\end{definition}

\begin{definition}[Direct Derivation]
If \(X, X' \in \mathcal{L}_{\mathcal{C}}\) and \(T\) is a table over \(\mathcal{C}\), then write \(X \Rrightarrow_T X'\) whenever \(\exists H' \in \mathcal{H}_{\mathcal{C}}, H(X) \Rrightarrow_T H' \land \mathrm{labels}(H') = X'\).
\end{definition}

The following three lemmata are easy to see:

\begin{lemma}[Restriction of Label Set Direct Derivations] \label{lem:ext}
Given \(\mathcal{C}\) and a table \(T\) over \(\mathcal{C}\), then \(\forall X, Y, Y' \in \mathcal{L}_{\mathcal{C}}\):

\vspace{-0.8em}
\[
(Y \Rrightarrow_T Y' \land X \subseteq Y) \Rightarrow (\exists X' \in \mathcal{L}_{\mathcal{C}}, X \Rrightarrow_T X' \land X' \subseteq Y')\textrm{.}
\]
\end{lemma}

\begin{lemma}[Completeness of Label Set Direct Derivations] \label{lem:compl}
Given \(\mathcal{C}\), and a table \(T\) over \(\mathcal{C}\), then \(\forall H, H' \in \mathcal{H}_{\mathcal{C}}\):

\vspace{-1.8em}
\[
(H \Rrightarrow_T H') \Rightarrow (\exists X' \in \mathcal{L}_{\mathcal{C}}, \mathrm{labels}(H) \Rrightarrow_T X' \land X' \subseteq \mathrm{labels}(H'))\textrm{.}
\]
\end{lemma}

\begin{lemma}[Soundness of Label Set Direct Derivations] \label{lem:sound}
Given \(\mathcal{C}\) and a table \(T\) over \(\mathcal{C}\), then \(\forall X, X' \in \mathcal{L}_{\mathcal{C}}, \forall H \in \mathcal{H}_{\mathcal{C}}\):

\vspace{-1.8em}
\[
(X \Rrightarrow_T X' \land \mathrm{labels}(H) = X) \Rightarrow (\exists H' \in \mathcal{H}_{\mathcal{C}}, H \Rrightarrow_T H' \land X' = \mathrm{labels}(H'))\textrm{.}
\]
\end{lemma}

\begin{theorem}[Decidable PHR Emptiness Problem] \label{thm:decemptiness}
The following problem is decidable, with an explicit algorithm:
  \begin{prob}
    \probinstance{A PHR grammar \(\mathcal{G} = (\mathcal{C}, A, S, \mathcal{T})\).}
    \probquestion{Is \(\mathrm{L}(\mathcal{G}) = \emptyset\)?}
  \end{prob}
\end{theorem}

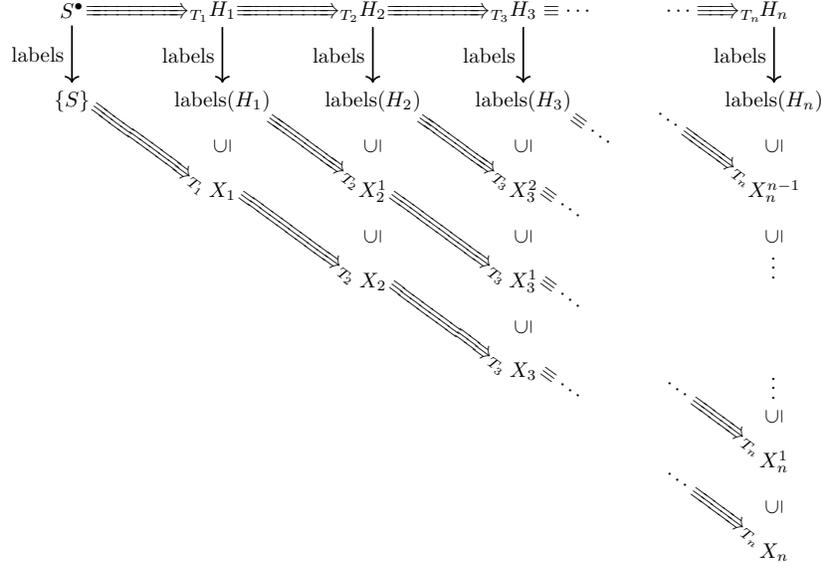
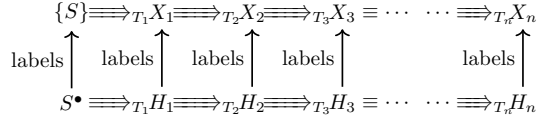
\begin{figure}[!ht]
\begin{subfigure}{.99\textwidth}
\centering
\scalebox{0.8}{
\begin{tikzpicture}
  \node (a0) at (0.0,0) {$S^{\bullet}$};
  \node (b0) at (2.5,0) {$H_1$};
  \node (c0) at (5.0,0) {$H_2$};
  \node (d0) at (7.5,0) {$H_3$};
  \node (e0) at (11.666666,0) {$H_n$};

  \node (a1) at (0.0,-1.5) {$\{S\}$};
  \node (b1) at (2.5,-1.5) {$\mathrm{labels}(H_1)$};
  \node (c1) at (5.0,-1.5) {$\mathrm{labels}(H_2)$};
  \node (d1) at (7.5,-1.5) {$\mathrm{labels}(H_3)$};
  \node (e1) at (11.666666,-1.5) {$\mathrm{labels}(H_n)$};

  \node [rotate=90] (x) at (2.5,-2.25) {$\subseteq$};
  \node [rotate=90] (x) at (5.0,-2.25) {$\subseteq$};
  \node [rotate=90] (x) at (7.5,-2.25) {$\subseteq$};
  \node [rotate=90] (x) at (11.666666,-2.25) {$\subseteq$};

  \node (b2) at (2.5,-3) {$X_1$};
  \node (c2) at (5.0,-3) {$X_2^1$};
  \node (d2) at (7.5,-3) {$X_3^2$};
  \node (e2) at (11.666666,-3) {$X_n^{n-1}$};

  \node [rotate=90] (x) at (5.0,-3.75) {$\subseteq$};
  \node [rotate=90] (x) at (7.5,-3.75) {$\subseteq$};
  \node [rotate=90] (x) at (11.666666,-3.75) {$\subseteq$};
  \node [rotate=90] (x) at (11.666666,-4.25) {$\cdots$};

  \node (c3) at (5,-4.5) {$X_2$};
  \node (d3) at (7.5,-4.5) {$X_3^1$};

  \node [rotate=90] (x) at (7.5,-5.25) {$\subseteq$};

  \node (d4) at (7.5,-6) {$X_3$};

  \node [rotate=90] (x) at (11.666666,-6.25) {$\cdots$};
  \node [rotate=90] (x) at (11.666666,-6.75) {$\subseteq$};

  \node (e5) at (11.666666,-7.5) {$X_n^1$};

  \node [rotate=90] (x) at (11.666666,-8.25) {$\subseteq$};

  \node (e6) at (11.666666,-9) {$X_n$};

  \path[->,draw,thick]
    (a0) edge [left] node {$\mathrm{labels}$} (a1)
    (b0) edge [left] node {$\mathrm{labels}$} (b1)
    (c0) edge [left] node {$\mathrm{labels}$} (c1)
    (d0) edge [left] node {$\mathrm{labels}$} (d1)
    (e0) edge [left] node {$\mathrm{labels}$} (e1)
  ;

  \node (x) at (1.25,0.03) {$\xRrightarrow{\hspace*{15mm}}_{T_1}$};
  \node (x) at (3.75,0.03) {$\xRrightarrow{\hspace*{15mm}}_{T_2}$};
  \node (x) at (6.25,0.03) {$\xRrightarrow{\hspace*{15mm}}_{T_3}$};
  \node (x) at (8.25,0) {$\equiv\cdots$};
  \node (x) at (10.666666,0.03) {$\cdots\xRrightarrow{\hspace*{5mm}}_{T_n}$};

  \node [rotate=-36] (x) at (1.31,-2.24) {$\xRrightarrow{\hspace*{18mm}}_{T_1}$};
  \node [rotate=-36] (x) at (3.78,-3.72) {$\xRrightarrow{\hspace*{18.5mm}}_{T_2}$};
  \node [rotate=-36] (x) at (6.28,-5.22) {$\xRrightarrow{\hspace*{18.5mm}}_{T_3}$};
  \node [rotate=-36] (x) at (8.15,-6.22) {$\equiv\cdots$};
  \node [rotate=-36] (x) at (10.696666,-8.22) {$\cdots\xRrightarrow{\hspace*{8.5mm}}_{T_n}$};

  \node [rotate=-36] (x) at (4.05,-2.3) {$\xRrightarrow{\hspace*{13.5mm}}_{T_2}$};
  \node [rotate=-36] (x) at (6.28,-3.72) {$\xRrightarrow{\hspace*{18.5mm}}_{T_3}$};
  \node [rotate=-36] (x) at (8.15,-4.72) {$\equiv\cdots$};
  \node [rotate=-36] (x) at (10.696666,-6.72) {$\cdots\xRrightarrow{\hspace*{8.5mm}}_{T_n}$};

  \node [rotate=-36] (x) at (6.55,-2.3) {$\xRrightarrow{\hspace*{13.5mm}}_{T_3}$};
  \node [rotate=-36] (x) at (8.15,-3.22) {$\equiv\cdots$};

  \node [rotate=-36] (x) at (8.65,-2.0) {$\equiv\cdots$};

  \node [rotate=-36] (x) at (10.54,-2.24) {$\cdots\xRrightarrow{\hspace*{8.5mm}}_{T_n}$};
\end{tikzpicture}
}
\caption{First part}
\end{subfigure}
\begin{subfigure}{.99\textwidth}
\centering
\vspace{0.75em}
\scalebox{0.8}{
\begin{tikzpicture}
  \node (a0) at (0.0,0) {$\{S\}$};
  \node (b0) at (1.5,0) {$X_1$};
  \node (c0) at (3.0,0) {$X_2$};
  \node (d0) at (4.5,0) {$X_3$};
  \node (e0) at (7.5,0) {$X_n$};

  \node (a1) at (0.0,-1.5) {$S^{\bullet}$};
  \node (b1) at (1.5,-1.5) {$H_1$};
  \node (c1) at (3.0,-1.5) {$H_2$};
  \node (d1) at (4.5,-1.5) {$H_3$};
  \node (e1) at (7.5,-1.5) {$H_n$};

  \path[->,draw,thick]
    (a1) edge [left] node {$\mathrm{labels}$} (a0)
    (b1) edge [left] node {$\mathrm{labels}$} (b0)
    (c1) edge [left] node {$\mathrm{labels}$} (c0)
    (d1) edge [left] node {$\mathrm{labels}$} (d0)
    (e1) edge [left] node {$\mathrm{labels}$} (e0)
  ;

  \node (x) at (0.8,0.03) {$\xRrightarrow{\hspace*{5mm}}_{T_1}$};
  \node (x) at (2.25,0.03) {$\xRrightarrow{\hspace*{6mm}}_{T_2}$};
  \node (x) at (3.75,0.03) {$\xRrightarrow{\hspace*{6mm}}_{T_3}$};
  \node (x) at (5.25,0) {$\equiv\cdots$};
  \node (x) at (6.625,0.03) {$\cdots\xRrightarrow{\hspace*{4mm}}_{T_n}$};

  \node (x) at (0.8,-1.47) {$\xRrightarrow{\hspace*{5mm}}_{T_1}$};
  \node (x) at (2.25,-1.47) {$\xRrightarrow{\hspace*{6mm}}_{T_2}$};
  \node (x) at (3.75,-1.47) {$\xRrightarrow{\hspace*{6mm}}_{T_3}$};
  \node (x) at (5.25,-1.5) {$\equiv\cdots$};
  \node (x) at (6.625,-1.47) {$\cdots\xRrightarrow{\hspace*{4mm}}_{T_n}$};
\end{tikzpicture}
}
\caption{Second part}
\end{subfigure}
\caption{Diagrams for proof of Theorem \ref{thm:decemptiness}}
\label{fig:proofdiag}
\end{figure}

\begin{proof}
Write \(X \Rrightarrow_{\mathcal{T}}^+ X'\) if there exists a sequence of direct derivations \(X \Rrightarrow_{T_1} X_1 \Rrightarrow_{T_2} \cdots \Rrightarrow_{T_n} X'\) with \(n \geq 1\), \(T_i \in \mathcal{T}\), \(X, X', X_i \in \mathcal{L}_{\mathcal{C}}\), and define \(\mathrm{succ}_{\mathcal{T}}(X) = \{X' \in \mathcal{L}_{\mathcal{C}} \mid X \Rrightarrow_\mathcal{T}^+ X'\}\), computed by iteratively computing label set direct successors (breadth first search). Eventually this process must terminate due to the fact that \(\mathcal{L}_{\mathcal{C}}\) is finite. We claim that \(\mathrm{succ}_{\mathcal{T}}(\{S\})\) contains a label set containing only terminals if and only if \(\mathrm{L}(\mathcal{G}) \neq \emptyset\).

Suppose first that \(\mathrm{L}(\mathcal{G})\) is non-empty. Then there must be a hypergraph derivation sequence \(S^{\bullet} \Rrightarrow_{T_1} H_1 \Rrightarrow_{T_2} \cdots \Rrightarrow_{T_n} H_n\) with \(n \geq 1\) and \(\mathrm{lab}_H(E_H) \subseteq A\), which, by Lemmata \ref{lem:ext} and \ref{lem:compl}, induces the corresponding label set derivation sequence \(\{S\} \Rrightarrow_{T_1} X_1 \Rrightarrow_{T_2} \cdots \Rrightarrow_{T_n} X_n\) such that \(X_n \subseteq \mathrm{labels}(H_n) \subseteq A\) and \(X_n \in \mathrm{succ}_{\mathcal{T}}(X)\), as per Figure \ref{fig:proofdiag}(a).

Now suppose that there is a label set containing only terminals in \(\mathrm{succ}_{\mathcal{T}}(S^{\bullet})\). Then there must be a label set derivation sequence \(\{S\} \Rrightarrow_{T_1} X_1 \Rrightarrow_{T_2} \cdots \Rrightarrow_{T_n} X_n\) with \(n \geq 1\) and \(X_n \subseteq A\). By Lemma \ref{lem:sound}, there must be a corresponding derivation sequence \(S^{\bullet} \Rrightarrow_{T_1} H_1 \Rrightarrow_{T_2} \cdots \Rrightarrow_{T_n} H_n\), as per Figure \ref{fig:proofdiag}(b). Since \(\mathrm{labels}(H_n) = X_n \subseteq A\), we know \(H_n \in \mathrm{L}(\mathcal{G})\) so \(\mathrm{L}(\mathcal{G})\) is non-empty.
\end{proof}

Finally, the above view of derivations informs us of how to eliminate unreachable labels from a PHR grammar. Elimination of unreachable symbols will be needed later in the proof of Theorem \ref{thm:phrset0l}.

\begin{definition}[Unreachable Symbol]
Given a PHR grammar over \((\Sigma, \mathrm{type})\), \(X \in \Sigma\) is called unreachable if there is no derivation starting at \(S^{\bullet}\)  containing a hypergraph with a hyperedge labelled by \(X\).
\end{definition}

\begin{lemma}[Elimination of Unreachable Labels] \label{lem:unreachable}
For \(k \geq 2, l \geq 1\), given a (repetition-free) (proper) \((k, l)\)-PHR grammar \(\mathcal{G}\), one can effectively construct a (repetition-free) (proper) \((k, l)\)-PHR grammar \(\mathcal{G}'\) with no unreachable symbols and \(\mathrm{L}(\mathcal{G}) = \mathrm{L}(\mathcal{G}')\).
\end{lemma}

\begin{proof}
Similar to the proof of Theorem \ref{thm:decemptiness}, if a symbol appears in a hypergraph derivation sequence starting at \(S^{\bullet}\), then it appears in a label set derivation sequence starting at \(\{S\}\), and vice versa. So determining if a symbol is unreachable is just a finite process.

Removal of an unreachable symbol from a grammar is an easy process. Suppose that we have identified \(X \in \Sigma\) as an unreachable symbol in the PHR grammar \(\mathcal{G} = ((\Sigma, \mathrm{type}), A, S, \mathcal{T})\). Then the grammar \(\mathcal{G}' = ((\Sigma \setminus \{X\}, \restr{\mathrm{type}}{\Sigma \setminus \{X\}}), A \setminus \{X\}, S, \{\{(L, R) \mid (L, R) \in T \land L \neq X \land X \not\in \mathrm{lab}_R(E_R)\} \mid T \in \mathcal{T}\})\) no longer contains the symbol \(X\) and is such that \(\mathrm{L}(\mathcal{G}) = \mathrm{L}(\mathcal{G}')\). We can repeat this process to remove all unreachable symbols. It is easy to see that after we have completed this process, the resultant grammar will have no unreachable symbols.
\end{proof}

\section{PHRS Languages} \label{sec:phrs}

We now turn our attention to string languages. We believe the family of parallel hyperedge replacement string languages is a genuinely new family, containing all MCF and ET0L languages. It is not simply equal to the (parallel) MCF languages because these are known to be incomparable with ET0L \cite{Nishida-Seki00a}. Recall that the hyperedge replacement string languages are exactly the MCF languages. We will confirm that parallel hyperedge replacement string (PHRS) languages contain all MCF languages and also all of the ET0L languages. We then go on to show various closure properties including that the family of PHRS languages is a hyper-algebraically closed super AFL with decidable membership.

\subsection{Definitions and Foundations}

\begin{definition}[PHR String Language] \label{dfn:phrs}
A string language \(L \subseteq A^*\) is called a (repetition-free) parallel hyperedge replacement string language of order \(k\) with at most \(l\) tables if there is a (repetition-free) \((k, l)\)-PHR grammar \(\mathcal{G}\) such that \(\mathcal{G}\) generates a string graph language and \(\mathrm{STR}(\mathrm{L}(\mathcal{G})) = L \setminus \{\epsilon\}\). The family of (repetition-free) \(k\)-PHRS languages is the union of all (repetition-free) \((k, l)\)-PHRS languages for \(l \geq 1\), and the family of (repetition-free) PHRS languages is the union of all (repetition-free) \(k\)-PHRS languages for \(k \geq 2\). Denote these \(\mathcal{P}\mathcal{H}\mathcal{R}\mathcal{S}_{k,l}\), \(\mathcal{P}\mathcal{H}\mathcal{R}\mathcal{S}_{k}\), and \(\mathcal{P}\mathcal{H}\mathcal{R}\mathcal{S}\) (\(\mathcal{P}\mathcal{H}\mathcal{R}\mathcal{S}_{k,l}^{\mathrm{rf}}\), \(\mathcal{P}\mathcal{H}\mathcal{R}\mathcal{S}_{k}^{\mathrm{rf}}\), and \(\mathcal{P}\mathcal{H}\mathcal{R}\mathcal{S}^{\mathrm{rf}}\)).
\end{definition}

Notice how we exclude the cases where \(k < 2, l < 1\). This is to ensure the classes form a family of string languages, as per the definition in Subsection \ref{subsec:strings}, since \(\mathcal{P}\mathcal{H}\mathcal{R}\mathcal{S}_0^{\mathrm{rf}} = \mathcal{P}\mathcal{H}\mathcal{R}\mathcal{S}_0 = \mathcal{P}\mathcal{H}\mathcal{R}\mathcal{S}_1^{\mathrm{rf}} = \mathcal{P}\mathcal{H}\mathcal{R}\mathcal{S}_1 = \{\emptyset, \{\epsilon\}\}\). Notice also that \(\mathcal{P}\mathcal{H}\mathcal{R}\mathcal{S}_{k,l}^{\mathrm{rf}} \subseteq \mathcal{P}\mathcal{H}\mathcal{R}\mathcal{S}_{k,l}\) for \(k \geq 2, l \geq 1\), and Theorem \ref{thm:phrtablepower} tells us:

\begin{theorem}[PHRS Table Power] \label{thm:phrstablepower}
For all \(k \geq 2\), \(\mathcal{P}\mathcal{H}\mathcal{R}\mathcal{S}_{k,2}^{\mathrm{rf}} = \mathcal{P}\mathcal{H}\mathcal{R}\mathcal{S}_{k}^{\mathrm{rf}}\) and \(\mathcal{P}\mathcal{H}\mathcal{R}\mathcal{S}_{k,2} = \mathcal{P}\mathcal{H}\mathcal{R}\mathcal{S}_{k}\).
\end{theorem}

We now show that the PHRS languages generalise both the ET0L and MCF languages. In particular, the \((2, 1)\)-PHRS languages are exactly the E0L languages, and the \(2\)-PHR languages, ET0L.

\begin{lemma} \label{lem:type2}
Given a \((k, l)\)-PHR grammar \(\mathcal{G}\) generating a string graph language, one can effectively construct a proper \((k, l)\)-PHR grammar \(\mathcal{G}'\) such that there are no unreachable symbols, all terminals are type \(2\), all non-terminals are type at least \(2\), and \(\mathrm{L}(\mathcal{G}) = \mathrm{L}(\mathcal{G}')\).
\end{lemma}

\begin{proof}
Since all string graph languages are proper, Lemmata \ref{lem:properphr} and \ref{lem:unreachable} tell us how to construct a proper \((k, l)\)-PHR grammar \(\mathcal{G}'\) such that there are no unreachable symbols and \(\mathrm{L}(\mathcal{G}) = \mathrm{L}(\mathcal{G}')\). All terminals can be assumed to be type \(2\) since the grammar generates a string graph language. Finally, all type \(0\) non-terminals can be converted to type \(1\), since the start symbol must be of type \(2\), so there will always be a node they can attached to. Next all type \(1\) non-terminals, including those we just converted from type \(0\), can be converted to type \(2\) symbols in the obvious way, and all rules that use them updated so that the extra node is merged upon deletion of the symbol.
\end{proof}

\begin{theorem}[PHRS Generalises ET0L] \label{thm:phrset0l}
For all \(k \geq 2\), \(\mathcal{E}\mathcal{O}\mathcal{L} \subseteq \mathcal{P}\mathcal{H}\mathcal{R}\mathcal{S}_{k,1}^{\mathrm{rf}}\) and \(\mathcal{E}\mathcal{T}\mathcal{O}\mathcal{L} \subseteq \mathcal{P}\mathcal{H}\mathcal{R}\mathcal{S}_k^{\mathrm{rf}}\). When \(k \geq 4\), \(\mathcal{E}\mathcal{O}\mathcal{L} \subsetneq \mathcal{P}\mathcal{H}\mathcal{R}\mathcal{S}_{k,1}^{\mathrm{rf}}\) and \(\mathcal{E}\mathcal{T}\mathcal{O}\mathcal{L} \subsetneq \mathcal{P}\mathcal{H}\mathcal{R}\mathcal{S}_k^{\mathrm{rf}}\). Moreover, \(\mathcal{E}\mathcal{O}\mathcal{L} = \mathcal{P}\mathcal{H}\mathcal{R}\mathcal{S}_{2,1}^{\mathrm{rf}} = \mathcal{P}\mathcal{H}\mathcal{R}\mathcal{S}_{2,1}\) and \(\mathcal{E}\mathcal{T}\mathcal{O}\mathcal{L} = \mathcal{P}\mathcal{H}\mathcal{R}\mathcal{S}_2^{\mathrm{rf}} = \mathcal{P}\mathcal{H}\mathcal{R}\mathcal{S}_2\).
\end{theorem}

\begin{proof}
First we show \(\mathcal{E}\mathcal{O}\mathcal{L} = \mathcal{P}\mathcal{H}\mathcal{R}\mathcal{S}_{2,1}^{\mathrm{rf}}\) and \(\mathcal{E}\mathcal{T}\mathcal{O}\mathcal{L} \subseteq \mathcal{P}\mathcal{H}\mathcal{R}\mathcal{S}_k^{\mathrm{rf}}\) for all \(k \geq 2\). Suppose \(L\) is an E(T)0L language, then by Theorem \ref{thm:emptyrhs} there exists a propagating ET0L grammar \(\mathcal{G} = (\Sigma, A, S, \{T_i \mid i \in I\})\) such that \(L \setminus \{\epsilon\} = \mathrm{L}(\mathcal{G})\). It follows that every rule can be encoded as a hyperedge replacement rule over \(\mathcal{C}' = (\Sigma, \Sigma \times \{2\})\) giving us a repetition-free \(2\)-PHR grammar \(\mathcal{G}' = (\mathcal{C}', A, S, \{\{(L, R^{\bullet}) \mid (L, R) \in T_i\} \mid i \in I\})\) with \(\mathrm{L}(\mathcal{G}) = \mathrm{STR}(\mathrm{L}(\mathcal{G}'))\).

Next, we show that \(\mathcal{P}\mathcal{H}\mathcal{R}\mathcal{S}_{2,1} \subseteq \mathcal{E}\mathcal{O}\mathcal{L}\) and \(\mathcal{P}\mathcal{H}\mathcal{R}\mathcal{S}_2 \subseteq \mathcal{E}\mathcal{T}\mathcal{O}\mathcal{L}\). Suppose \(L\) is a \((2,l)\)-PHRS language, then there is a \((2,l)\)-PHR grammar \(\mathcal{G} = (\mathcal{C}, A, S, \{T_i \mid i \in I\})\) generating a string graph language such that \(L \setminus \{\epsilon\} = \mathrm{STR}(\mathrm{L}(\mathcal{G}))\). Lemma \ref{lem:type2} allows us to assume a lot about the form of RHSs of rules. It is easy to see that all RHSs must actually be string graphs, or could be transformed to string graphs, since any non-conformant pieces can just be inlined into the string graph because it will ultimately be deleted and the nodes merged in any terminally labelled derived hypergraph. Thus, the system can be converted into an ET0L grammar using at most \(l\) tables.

Thus, we have \(\mathcal{P}\mathcal{H}\mathcal{R}\mathcal{S}_{2,1} \subseteq \mathcal{E}\mathcal{O}\mathcal{L} \subseteq \mathcal{P}\mathcal{H}\mathcal{R}\mathcal{S}_{2,1}^{\mathrm{rf}}\) and \(\mathcal{P}\mathcal{H}\mathcal{R}\mathcal{S}_{2,1}^{\mathrm{rf}} \subseteq \mathcal{P}\mathcal{H}\mathcal{R}\mathcal{S}_{2,1}\), so all these inclusions must be equalities. Similarly, we have \(\mathcal{P}\mathcal{H}\mathcal{R}\mathcal{S}_{2,l} \subseteq \mathcal{E}\mathcal{T}\mathcal{O}\mathcal{L} \subseteq \mathcal{P}\mathcal{H}\mathcal{R}\mathcal{S}_{2,l}^{\mathrm{rf}}\) and \(\mathcal{P}\mathcal{H}\mathcal{R}\mathcal{S}_{2,l}^{\mathrm{rf}} \subseteq \mathcal{P}\mathcal{H}\mathcal{R}\mathcal{S}_{2,l}\), so all these inclusions must be equalities.

Finally, strictness follows from Theorems \ref{thm:mcfequiv}, \ref{thm:incompar}, and \ref{thm:phrgen}. That is, we can construct a repetition-free \((4, 1)\)-PHR grammar \(\mathcal{G}'\) generating a string graph language with \(\mathrm{STR}(\mathrm{L}(\mathcal{G}')) = K \setminus \{\epsilon\}\) (from Theorem \ref{thm:incompar}) which is not E0L.
\end{proof}

\begin{corollary}
There are repetition-free \((2,1)\)-PHRS languages that are not semilinear.
\end{corollary}

\begin{proof}
Theorem \ref{thm:incompar} tells us \(L = \{a^{2^n} \mid n \in \mathbb{N}\}\) is an E0L language which is not semilinear. Theorem \ref{thm:phrset0l} tells us \(L\) is a repetition-free \((2,1)\)-PHRS language.
\end{proof}

\begin{theorem}[PHRS Generalises MCF] \label{thm:hrsmcf}
For all \(k \geq 2\), \(\mathcal{H}\mathcal{R}\mathcal{S}_k^{\mathrm{rf}} \subsetneq \mathcal{P}\mathcal{H}\mathcal{R}\mathcal{S}_k^{\mathrm{rf}}\).
\end{theorem}

\begin{proof}
This follows from Theorem \ref{thm:mcfequiv}, and Theorem \ref{thm:phrgen} and its proof. We get strictness from Theorem \ref{thm:incompar} together with Theorem \ref{thm:phrset0l}.
\end{proof}

\subsection{Formal Language Closure Properties} \label{subsec:closure}

In this subsection, we show that the family of PHRS languages is a hyper-algebraically closed super AFL, and that the family of repetition-free PHRS languages is closed under non-erasing (iterated) substitution, with closure under rational operations and non-erasing homomorphisms following from this as a corollary.

\begin{theorem}[PHRS Closed Under Substitutions] \label{thm:x}
Let \(L \subseteq A^*\). Then:

\begin{enumerate}
    \item for any \(k \geq 2\), if \(L\) is a \(k\)-HRS language and \(h\) is a \(\mathcal{P}\mathcal{H}\mathcal{R}\mathcal{S}_{k,1}\)-substitution (non-erasing \(\mathcal{P}\mathcal{H}\mathcal{R}\mathcal{S}_{k,1}^{\mathrm{rf}}\)-substitution) on \(A\), then \(h(L)\) is a \((k, 1)\)-PHRS language (rep.-free \((k, 1)\)-PHRS language);
    \item for any \(k \geq 2\), if \(L\) is a \((k, 1)\)-PHRS language (rep.-free \((k, 1)\)-PHRS language) and \(h\) is a finite substitution (non-erasing finite substitution) on \(A\), then \(h(L)\) is a \((k, 1)\)-PHRS language (rep.-free \((k, 1)\)-PHRS language);
    \item for any \(k \geq 2\), if \(L\) is a \(k\)-PHRS language (rep.-free \(k\)-PHRS language) and \(h\) is a \(\mathcal{P}\mathcal{H}\mathcal{R}\mathcal{S}_k\)-substitution (non-erasing \(\mathcal{P}\mathcal{H}\mathcal{R}\mathcal{S}_k^{\mathrm{rf}}\)-substitution) on \(A\), then \(h(L)\) is a \(k\)-PHRS language (rep.-free \(k\)-PHRS language).
\end{enumerate}
\end{theorem}

\begin{proof}
First, recall that all \(k\)-HRS languages can be generated by repetition-free \(k\)-HRS grammars by Theorem \ref{thm:mcfequiv}. Now, for all \(k \geq 2, l \geq 1\), all (repetition-free) \(\mathcal{P}\mathcal{H}\mathcal{R}\mathcal{S}_{k,l}\)-substitutions \(h\) of strings on \(A\) can be converted to (repetition-free) \(\mathcal{P}\mathcal{H}\mathcal{R}_{k,l}\)-substitutions \(s\) of hypergraphs on \(A\) viewing A as part of the signature \((A, \mathrm{type})\) where \(\mathrm{type}(X) = 2\) for all \(X \in A\), defining \(s(X) = \{w^{\bullet} \mid w \in h(X)\}\) for all \(X \in A\). The rest then follows from Theorem \ref{thm:phrsub}, translating back from string graphs to strings.
\end{proof}

\begin{corollary}[PHRS Closed Under Homomorphisms] \label{cor:phrshom}
Let \(L \subseteq A^*\) be a \((k, l)\)-PHRS language (repetition-free \((k, l)\)-PHRS language) for any \(k \geq 2, l \geq 1\) and \(\varphi: A^* \to B^*\) be a homomorphism (non-erasing homomorphism). Then \(\varphi(L)\) is a \((k, l)\)-PHRS language (repetition-free \((k, l)\)-PHRS language).
\end{corollary}

\begin{proof}
If \(l = 1\), then we can use Theorem \ref{thm:x}(2) since a (non-erasing) homomorphism can be trivially viewed as a (non-erasing) finite substitution. If \(l \geq 2\), then we can similarly use Theorem \ref{thm:x}(3).
\end{proof}

\begin{theorem}[PHRS Hyper-Algebraically Closed] \label{thm:phrshyper}
Let \(L \subseteq A^*\). Then, for any \(k \geq 2\), if \(L\) is a \(k\)-PHRS language (repetition-free \(k\)-PHRS language) and \(H\) is a finite set of \(\mathcal{P}\mathcal{H}\mathcal{R}\mathcal{S}_k\)-substitutions (non-erasing \(\mathcal{P}\mathcal{H}\mathcal{R}\mathcal{S}_k^{\mathrm{rf}}\)-substitutions) on \(A\), then \(\mathrm{ITER}_{H}(L)\) is a \(k\)-PHRS language (repetition-free \(k\)-PHRS language).
\end{theorem}

\begin{proof}
This follows directly from Theorem \ref{thm:phrhyper}, realising substitutions as in the proof of Theorem \ref{thm:x}.
\end{proof}

Next, we show closure under rational operations, which can be seen via the following general result:

\begin{lemma} \label{lem:ratclosure}
Let \(\mathcal{F}\) be a family of string languages such that if \(L \subseteq A^*\) is regular and \(h\) is a non-erasing \(\mathcal{F}\)-substitution on \(A\), then \(h(L) \in \mathcal{F}\). Then for \(\mathcal{F}\) languages \(L_1, L_2 \subseteq A^*\):
\begin{enumerate}
\item \(L_1 \cup L_2\) is an \(\mathcal{F}\) language; \tabto{6cm} (closure under union)
\item \(L_1 L_2\) is an \(\mathcal{F}\) language;      \tabto{6cm} (closure under concatenation)
\item \(L_1^+\) is an \(\mathcal{F}\) language.        \tabto{6cm} (closure under Kleene plus)
\end{enumerate}
\end{lemma}

\begin{proof}
To see (1), notice that \(L_1 \cup L_2\) is simply \(h(K)\) where \(K = \text{ \underline{if} } \epsilon \in L_1 \cup L_2 \text{ \underline{then} } \{X, Y, \epsilon\} \text{ \underline{else} } \{X, Y\}\), and \(h\) is a non-erasing \(\mathcal{F}\)-substitution defined by \(h(X) = L_1 \setminus \{\epsilon\}\) and \(h(Y) = L_2 \setminus \{\epsilon\}\). Thus we have \(L_1 \cup L_2 = h(K)\), and since, in either case, \(K\) is a regular language, \(h(K) \in \mathcal{F}\) due to our assumptions about \(\mathcal{F}\), as required.

To see (2), we again construct a regular language \(K\). If \(\epsilon \in X\) and \(\epsilon \in Y\), then set \(K = \{XY, X, Y, \epsilon\}\). If only \(\epsilon \in X\), set \(K = \{XY, Y\}\). If only \(\epsilon \in Y\), set \(K = \{XY, X\}\). Otherwise, set \(K = \{XY\}\). Finally, \(h\) is exactly as before, giving us \( L_1 L_2 = h(K)\in \mathcal{F}\) as required.

Finally, to see (3), set \(K = \text{ \underline{if} } \epsilon \in L_1 \text{ \underline{then} } \{X^n \mid n \geq 0\} \text{ \underline{else} } \{X^n \mid n \geq 1\}\), and set \(h(X) = L_1\), giving us \(L_1^+ = h(K) \in \mathcal{F}\) as required.
\end{proof}

\begin{theorem}[PHRS Closed Under Rational Operations] \label{thm:phrrat}
Let \(L_1, L_2 \subseteq A_1^*\) be (repetition-free) \((k, l)\)-PHRS languages for any \(k \geq 2, l \geq 1\). Then:
\begin{enumerate}
\item \(L_1 \cup L_2\) is a (r.-free) \((k, l)\)-PHRS language; \tabto{6.75cm} (closure under union)
\item \(L_1 L_2\) is a (r.-free) \((k, l)\)-PHRS language;      \tabto{6.75cm} (closure under concatenation)
\item \(L_1^+\) is a (r.-free) \((k, l)\)-PHRS language.        \tabto{6.75cm} (closure under Kleene plus)
\end{enumerate}
\end{theorem}

\begin{proof}
If \(l = 1\), then Theorem \ref{thm:x}(1) tells us that \(\mathcal{P}\mathcal{H}\mathcal{R}\mathcal{S}_{k,1}\) satisfies the condition of Lemma \ref{lem:ratclosure} since \(\mathcal{R}\mathcal{E}\mathcal{G} \subseteq \mathcal{H}\mathcal{R}\mathcal{S}_k\). If \(l \geq 2\) then Theorem \ref{thm:x}(3) tells us that \(\mathcal{P}\mathcal{H}\mathcal{R}\mathcal{S}_{k}\) satisfies the condition of Lemma \ref{lem:ratclosure} since \(\mathcal{R}\mathcal{E}\mathcal{G} \subseteq \mathcal{P}\mathcal{H}\mathcal{R}\mathcal{S}_k\).
\end{proof}

\begin{corollary}[PHRS More Than ET0L and MCF]
We have the following strict inclusions: \(\mathcal{M}\mathcal{C}\mathcal{F}_2 \cup \mathcal{E}\mathcal{O}\mathcal{L} \subsetneq \mathcal{P}\mathcal{H}\mathcal{R}\mathcal{S}_{4,1}^{\mathrm{rf}}\) and \(\mathcal{M}\mathcal{C}\mathcal{F}_2 \cup \mathcal{E}\mathcal{T}\mathcal{O}\mathcal{L} \subsetneq \mathcal{P}\mathcal{H}\mathcal{R}\mathcal{S}_{4}^{\mathrm{rf}}\).
\end{corollary}

\begin{proof}
Inclusion follows from Theorems \ref{thm:phrset0l} and \ref{thm:hrsmcf}, but is not the interesting part of this Theorem. The interesting part is that this there are repetition-free \((2, 1)\)-PHRS (repetition-free \(2\)-PHRS) languages that are not just E0L (ET0L) or \(2\)-MCF. Take \(K \subseteq A^*\) and \(L \subseteq B^*\) to be as in Theorem \ref{thm:incompar}, over disjoint alphabets. Then \(L' = K \cup L\) is repetition-free \((2, 1)\)-PHRS by Theorem \ref{thm:phrrat}, but is neither E(T)0L nor MCF since if it were, then \(L' \cap A^*\) and \(L' \cap B^*\) would be both \(2\)-MCF and E0L since these language families are closed under rational intersection (Theorems \ref{thm:mcfafl} and \ref{thm:etolafl}), but Theorem \ref{thm:incompar} tells us that \(L' \cap A^*\) is not ET0L and \(L' \cap B^*\) is not MCF.
\end{proof}

We now show closure under rational intersection, inspired by the proof of Theorem V.1.7(iv) of \cite{Rozenberg-Salomaa80a}:

\begin{theorem}[PHRS Closed Under Rational Intersection] \label{thm:rat}
Let \(L \subseteq A^*\) be a (repetition-free) \((k, l)\)-PHRS language and \(K \subseteq B^*\) be a regular language, for any \(k \geq 2, l \geq 1\). Then \(L \cap K\) is a (repetition-free) \((k, l)\)-PHRS language.
\end{theorem}

\begin{proof}
By Theorem \ref{thm:phrstablepower}, it will suffice to show that when \(L\) is a (repetition-free) \((k, 1)\)-PHRS language, \(L \cap K\) is a (repetition-free) \((k, 1)\)-PHRS language, and when \(L\) is a (repetition-free) \((k, 2)\)-PHRS language, \(L \cap K\) is a (repetition-free) \(k\)-PHRS language. Our proof will set up some commonalities for both cases, and then we will analyse the cases \(l = 1\) and \(l = 2\) separately.

There must exist a (repetition-free) \((k, l)\)-PHR grammar \(\mathcal{G} = ((\Sigma, \mathrm{type}), A,\) \(S, \{T_1, \dots, T_l\}, S)\) such that \(\mathrm{L}(\mathcal{G})\) is a string graph language and \(\mathrm{STR}(\mathrm{L}(\mathcal{G})) = L \setminus \{\epsilon\}\). We assume without loss of generality that \((\Sigma \setminus A) \cap B = \emptyset\) and that \(\mathcal{G}\) is of the form specified by Lemma \ref{lem:syncphr}. There must also be a deterministic full FSA \(\mathcal{M} = (Q, B, \delta, p, F)\) such that \(\mathrm{L}(\mathcal{M}) = K\).

We introduce the intermediate notion of a hypergraph with node labels because we want to label the nodes by states of \(\mathcal{M}\). A node labelled hypergraph over \((\mathcal{C}, Q)\) is a pair \((H, l)\) where \(H\) is a hypergraph over \(\mathcal{C} = (\Sigma, \mathrm{type})\) and \(l\) is a function \(V_H \to Q\). Notice that any such node labelled hypergraph can be encoded as a hypergraph over \((\Delta, \restr{\mathrm{type}'}{\Delta})\): for each \(e \in E_H\), the new hyperedge labelling function is defined by sending \(e\) to  \((\mathrm{lab}_H(e), q_1, \dots, q_i)\) where \(i = \mathrm{type}_H(e)\) and \(q_j = l(\mathrm{att}_H(e)(j))\) for \(1 \leq j \leq i\). Call this injective encoding function \(\mathrm{enc}\). Next, given a type \(t\) hypergraph \(H\) over \(\mathcal{C}\) and a sequence \(\sigma: \underline{t} \to Q\), define \(\mathrm{CHOICES}_Q(H, \sigma) = \{\mathrm{enc}((H, l)) \mid l: V_H \to Q, l \circ \mathrm{ext}_H = \sigma\}\). Given a rule \((L, R)\) over \(\mathcal{C}\), define \(\mathrm{AUGMENT}_Q(L, R) = \{((L, \sigma(1), \dots, \sigma(t)), H) \mid t = \mathrm{type}(L), \sigma: \underline{t} \to Q, H \in \mathrm{CHOICES}_Q(R, \sigma)\}\).

We now handle the case \(l = 1\). We will construct a (repetition-free) \((k, 1)\)-PHR grammar \(\mathcal{G}' = ((\Sigma', \mathrm{type}'), A \cap B, \{T'\}, S)\) such that \(\mathrm{L}(\mathcal{G}')\) is a string graph language and \(\mathrm{STR}(\mathrm{L}(\mathcal{G}')) = (L \cap K) \setminus \{\epsilon\}\), thus proving that \(L \cap K\) is a (repetition-free) \((k, 1)\)-PHRS language. Let \(\Delta = (\bigcup_{0 \leq i \leq k} \{\mathrm{type}^{-1}(\{i\}) \times Q^i\}) \cup \{F_0, \dots, F_k\}\), \(\Sigma' = \Delta \cup \{S\} \cup (A \cap B) \cup \{F_0, \dots, F_k\}\), \(\mathrm{type}'((X, q_1, \dots q_i)) = \mathrm{type}(X)\) for all \((X, q_1, \dots q_i) \in \Delta\), \(\mathrm{type}'(S) = 2\), \(\mathrm{type}'(X) = 2\) for all \(X \in A \cap B\), and \(\mathrm{type}'(F_i) = i\) for all \(0 \leq i \leq k\). Define \(T' = \mathcal{F} \oplus (T'_1 \cup T'_2 \cup T'_3)\) where:
\begin{enumerate}
\item \(T'_1 = \{((X, q_1, q_2), Y^{\bullet}) \mid X \in A \cap B, \delta(q_1, X) = q_2\}\);
\item \(T'_2 = \bigcup_{(L, R) \in T} \mathrm{AUGMENT}_Q(L, R)\);
\item \(T'_3 = \{(S, (S, p, q)^{\bullet}) \mid q \in F\}\);
\item \(\mathcal{F} = \{(X, F_{\mathrm{type}'(X)}^{\bullet}) \mid X \in \Sigma'\}\).
\end{enumerate}
It is not hard to check this grammar will produce a terminal string graph \((x_1 x_2 \cdots x_m)^{\bullet}\) if and only if the previous hypergraph in the derivation was a string graph of the form \(((x_1, q_1, q_2) (x_2, q_2, q_3) \cdots\) \((x_m, q_m, q_{m+1}))^{\bullet}\) and \(\delta(q_i, x_i)\) \(= q_{i+1}\) for \(1 \leq i \leq m\), \(q_1 = p\), and \(q_{m+1} \in F\), that is, traced out an accepting path in the FSA \(\mathcal{M}\), having simulated \(\mathcal{G}\).

Finally, we handle the case \(l = 2\). We will construct a (repetition-free) \(k\)-PHR grammar \(\mathcal{G}' = ((\Sigma', \mathrm{type}'), A \cap B, \{T_0', T_1', T_2'\}, S)\) such that \(\mathrm{L}(\mathcal{G}')\) is a string graph language and \(\mathrm{STR}(\mathrm{L}(\mathcal{G}')) = (L \cap K) \setminus \{\epsilon\}\), thus proving that \(L \cap K\) is a (repetition-free) \(k\)-PHRS language. Let \(\Delta\) and \((\Sigma, \mathrm{type}')\) be as before. Define (for \(i = 1, 2\)):
\begin{enumerate}
\item \(T_0' = \mathcal{F} \oplus \{((X, q_1, q_2), Y^{\bullet}) \mid X \in A \cap B, \delta(q_1, X) = q_2\}\);
\item \(T_i' = \mathcal{R} \oplus (\bigcup_{(L, R) \in T_i} \mathrm{AUGMENT}_Q(L, R) \cup \{(S, (S, p, q)^{\bullet}) \mid q \in F\})\);
\end{enumerate}
where \(\mathcal{F} = \{(X, F_{\mathrm{type}'(X)}^{\bullet}) \mid X \in \Sigma'\}\) and \(\mathcal{R} = \{(X, X^{\bullet}) \mid X \in \Sigma'\}\). Clearly progress in derivations is made when the tables \(T'_1\) and \(T'_2\) are used, and application of the table \(T_0\) will produce a terminal string graph \((x_1 x_2 \cdots x_m)^{\bullet}\) if and only if the previous hypergraph was a string graph of the form \(((x_1, q_1, q_2)\) \((x_2, q_2, q_3) \cdots (x_m, q_m, q_{m+1}))^{\bullet}\) and \(\delta(q_i, x_i) = q_{i+1}\) for \(1 \leq i \leq m\), \(q_1 = p\), and \(q_{m+1} \in F\), as before.
\end{proof}

Finally, we show closure under inverse homomorphisms, via the following general result:

\begin{lemma} \label{lem:closure}
Let \(\mathcal{F}\) be a family of string languages which is closed under rational substitution and rational intersection. Let \(L \subseteq A^*\) be an \(\mathcal{F}\) language and \(\varphi: B^* \to A^*\) a homomorphism. Then  \(\varphi^{-1}(L)\) is an \(\mathcal{F}\) language too.
\end{lemma}

\begin{proof}
Let \(\overbar{B}\) be a copy of \(B\) such that \((A \cup B) \cap \overbar{B} = \emptyset\), and let \(\overbar{\,\cdot\,}: B \to \overbar{B}\) identify each \(b \in B\) with its copy \(\overbar{b} \in \overbar{B}\). For each \(a \in A\), define the regular language \(L_a = \{w_1 a w_2 \mid w_1, w_2 \in \overbar{B}^*\} \subseteq (A \cup \overbar{B})^*\) and the rational substitution \(h\) on \(A\) by \(a \mapsto L_a\). Also define \(K = \bigcup_{n \in \mathbb{N}} \{\varphi(x_1) \overbar{x_1} \varphi(x_2) \overbar{x_2} \cdot\cdot\cdot \varphi(x_n) \overbar{x_n} \mid x_1, x_2, \dots x_n \in B\}\) and the homomorphism \(\psi: (A \cup \overbar{B})^* \to B^*\) by \(\psi(a) = \epsilon\) for each \(a \in A\) and \(\psi(\overbar{b}) = b\) for each \(b \in B\).

Notice \(h(L) \cap K = \bigcup_{n \in \mathbb{N}} \{\varphi(x_1) \overbar{x_1} \varphi(x_2) \overbar{x_2} \cdot\cdot\cdot \varphi(x_n) \overbar{x_n} \mid x_1, \dots, x_n \in B\) and \( \varphi(x_1) \varphi(x_2) \cdots \varphi(x_n) \in L\}\), so we have \(\varphi^{-1}(L) = \psi(h(L) \cap K)\). Now, \(h(L)\) is an \(\mathcal{F}\) language since \(\mathcal{F}\) is closed under rational substitution, \(h(L) \cap K\) is an \(\mathcal{F}\) language since \(\mathcal{F}\) is closed under rational intersection, and \(\psi(h(L) \cap K)\) is an \(\mathcal{F}\) language since \(\mathcal{F}\) is closed under homomorphisms (a special case of rational substitution). Thus, \(\varphi^{-1}(L)\) is an \(\mathcal{F}\) language, as required.
\end{proof}

\begin{theorem}[PHRS Closed Under Inverse Homomorphisms] \label{thm:invhom}
For all \(k \geq 2\), \(\mathcal{P}\mathcal{H}\mathcal{R}\mathcal{S}_{k}\) is closed under inverse homomorphisms.
\end{theorem}

\begin{proof}
The result follows from Theorems \ref{thm:x} and \ref{thm:rat} and Lemma \ref{lem:closure}.
\end{proof}

\subsection{Group Word Problem Closure Properties}

Since the family of \(k\)-PHRS languages is a full AFL for any \(k \geq 2\), it satisfies the following important properties:

\begin{theorem}[WP Independent Of Presentation \cite{Herbst-Thomas93a}]
Let \(\mathcal{F}\) be a family of string languages which is closed under inverse homomorphisms, and let \(\langle X \mid R \rangle\) be a presentation of a group \(G\) such that \(\mathrm{WP}_X(G)\) is an \(\mathcal{F}\) language. Then all presentations \(\langle X' \mid R' \rangle\) of \(G\) are such that \(\mathrm{WP}_{X'}(G)\) is an \(\mathcal{F}\) language.
\end{theorem}

\begin{theorem}[WP Subgroup and Supergroup Closure \cite{Gilman-Kropholler-Schleimer18a}] \label{thm:wpafl}
Let \(\mathcal{F}\) be a full AFL and \(G\) be a group with word problem in \(\mathcal{F}\). Then every finitely generated subgroup and every finite index supergroup of \(G\) has word problem in \(\mathcal{F}\).
\end{theorem}

In 2019, Kropholler and Spriano showed that a graph of groups with vertex groups with MCF word problem and edge groups finite, yields a group with an MCF word problem \cite{Kropholler-Spriano19a}. A special case of this construction is a free product of groups. We now show that a free product of groups with (repetition-free) PHRS word problems is a group with a (repetition-free) PHRS word problem. Our strategy is entirely different to Kropholler and Spriano's approach, which relied on Denkinger's automata characterisation of MCF languages (Theorem \ref{thm:mcfequiv}).

The following easy lemma, where presentations of groups are written as monoid presentations, gives us a recursive description of the word problem of free products, enabling us to prove Theorem \ref{thm:freeprod}.

\begin{lemma} \label{lem:fp}
Let \(G_1\), \(G_2\) be finitely generated groups over disjoint alphabets \(A_1 = \{a_1, \dots a_n\}\), \(A_2 = \{b_1, \dots b_m\}\), respectively. If \(X = A_1 \cup A_2\) and \(L_i = \mathrm{WP}_{A_i}(G_i)\) for \(i = 1, 2\), then \(\mathrm{WP}_{X}(G_1 * G_2)\) is the smallest set \(L\) such that \(\epsilon \in L\) and \(\forall i \in \{1, 2\}, \forall w \in L_i, \forall u, v \in X^*, u v \in L \Rightarrow u w v \in L\).
\end{lemma}

\begin{theorem}[WP Free Product Closure] \label{thm:freeprod}
Let \(\mathcal{F}\) be a family of string languages containing all finite languages, closed under union and concatenation, and closed under nested iterated substitution. Then if \(G_1\), \(G_2\) are groups with presentations admitting a \(\mathcal{F}\) word problem, \(G_1 * G_2\) has a presentation admitting a \(\mathcal{F}\) word problem.
\end{theorem}

\begin{proof}
Let \(A_1\), \(A_2\), \(X\), \(L_1\), \(L_2\), \(L\) be as in Lemma \ref{lem:fp}, then it is immediate that iterated application of the nested non-erasing \(\mathcal{F}\)-substitution \(h\) of strings on \(A_1 \cup A_2\), defined by \(h(a_i) = \{a_i\} \cup a_i L_1 \cup L_1 a_i\) and \(h(b_j) = \{b_j\} \cup b_j L_2 \cup L_2 b_j\) for all \(i \in \underline{n}, j \in \underline{m}\), to \(L\), gives us exactly \(\mathrm{WP}_{X}(G_1 * G_2)\). The result them follows from the assumed closure properties.
\end{proof}

\subsection{The Membership Problem} \label{subsec:phrsmp}

We show that the universal membership problem for PHRS grammars is decidable. That is, there is an algorithm that, when given a PHR grammar \(\mathcal{G}\) generating a string graph language and a string \(w\), can decide if \(w \in \mathrm{STR}(\mathrm{L}(\mathcal{G}))\).

\begin{theorem}[Decidable PHRS Membership] \label{thm:decmembership}
The following problem is decidable, with an explicit algorithm:
  \begin{prob}
    \probinstance{A PHR grammar \(\mathcal{G} = (\mathcal{C}, A, S, \mathcal{T})\) generating a string graph language and a string \(w \in A^*\).}
    \probquestion{Is \(w \in \mathrm{STR}(\mathrm{L}(\mathcal{G}))\)?}
  \end{prob}
\end{theorem}

\begin{proof}
Theorem \ref{thm:rat} tells us how to turn the PHR grammar \(\mathcal{G}\) and the regular language \(\{w\}\) into a PHR grammar \(\mathcal{G}'\) generating a string graph language such that \(\mathrm{STR}(\mathrm{L}(\mathcal{G}')) = \mathrm{STR}(\mathrm{L}(\mathcal{G})) \cap \{w\}\). Then:

\vspace{-2em}
\begin{multline*}
w \not\in \mathrm{STR}(\mathrm{L}(\mathcal{G})) \\
\shoveleft \,\,\,\, \Leftrightarrow \mathrm{STR}(\mathrm{L}(\mathcal{G})) \cap \{w\} = \emptyset \\
\shoveleft \,\,\,\, \Leftrightarrow \mathrm{STR}(\mathrm{L}(\mathcal{G}')) = \emptyset \\
\shoveleft \,\,\,\, \Leftrightarrow \mathrm{L}(\mathcal{G}') = \emptyset\textrm{.} \\
\end{multline*}
\vspace{-3em}

Finally, Theorem \ref{thm:decemptiness} tells us how to decide emptiness of \(\mathrm{L}(\mathcal{G}')\).
\end{proof}

\begin{corollary}[PHRS is Recursive]
\(\mathcal{P}\mathcal{H}\mathcal{R}\mathcal{S} \subsetneq \mathcal{R}\mathcal{E}\mathcal{C}\).
\end{corollary}

\begin{proof}
Inclusion follows from decidable membership (Theorem \ref{thm:decmembership}). Strictness from decidable emptiness (Theorem \ref{thm:decemptiness}).
\end{proof}

\subsection{Weak-Coded PHRS Languages} \label{subsec:wphrs}

Theorem \ref{thm:hrrf} tells us that the string generational power of HR grammars is not restricted by requiring rules to be repetition-free. It is not clear if a similar result holds in the parallel replacement setting. If it turns out that there is no such result, there is still a middle-ground where one can obtain all of the closure properties we have shown in this section, but without allowing merging of nodes by derivations:

\begin{definition}[WPHR String Language]
Call the below equivalent families the repetition-free weak-coded \((k,l)\)-PHR string (\((k,l)\)-WPHRS) languages:

\begin{enumerate}
\item The family of string languages generated by repetition-free \((k,l)\)-PHR grammars under the image of some weak coding.
\item The family of string languages generated by repetition-free \((k,l)\)-PHR grammars with a special type \(2\) label \texttt{empty}, interpreted as the empty string by \(\mathrm{STR}\).
\end{enumerate}

\noindent
For \(k \geq 2\) and \(l \geq 1\), we define \(\mathcal{W}\mathcal{P}\mathcal{H}\mathcal{R}\mathcal{S}_{k, l}^{\mathrm{rf}}\), \(\mathcal{W}\mathcal{P}\mathcal{H}\mathcal{R}\mathcal{S}_{k}^{\mathrm{rf}}\), and \(\mathcal{W}\mathcal{P}\mathcal{H}\mathcal{R}\mathcal{S}^{\mathrm{rf}}\) in the obvious way.
\end{definition}

\begin{theorem}[WPHRS Hierarchy Position]
For all \(k \geq 2\), we have:

\begin{enumerate}
\item \(\mathcal{W}\mathcal{P}\mathcal{H}\mathcal{R}\mathcal{S}_{k,2}^{\mathrm{rf}} = \mathcal{W}\mathcal{P}\mathcal{H}\mathcal{R}\mathcal{S}_{k}^{\mathrm{rf}}\);

\item \(\mathcal{E}\mathcal{O}\mathcal{L} \subseteq \mathcal{P}\mathcal{H}\mathcal{R}\mathcal{S}_{k,1}^{\mathrm{rf}} \subseteq \mathcal{W}\mathcal{P}\mathcal{H}\mathcal{R}\mathcal{S}^{\mathrm{rf}}_{k,1} \subseteq \mathcal{P}\mathcal{H}\mathcal{R}\mathcal{S}_{k,1}\)

\item \(\mathcal{E}\mathcal{T}\mathcal{O}\mathcal{L} \subseteq \mathcal{P}\mathcal{H}\mathcal{R}\mathcal{S}_{k}^{\mathrm{rf}} \subseteq \mathcal{W}\mathcal{P}\mathcal{H}\mathcal{R}\mathcal{S}^{\mathrm{rf}}_{k} \subseteq \mathcal{P}\mathcal{H}\mathcal{R}\mathcal{S}_{k}\).
\end{enumerate}

\noindent
Moreover:

\begin{enumerate}
\item \(\mathcal{E}\mathcal{O}\mathcal{L} = \mathcal{P}\mathcal{H}\mathcal{R}\mathcal{S}_{2,1}^{\mathrm{rf}} = \mathcal{W}\mathcal{P}\mathcal{H}\mathcal{R}\mathcal{S}^{\mathrm{rf}}_{2,1} = \mathcal{P}\mathcal{H}\mathcal{R}\mathcal{S}_{2,1}\);
\item \(\mathcal{E}\mathcal{T}\mathcal{O}\mathcal{L} = \mathcal{P}\mathcal{H}\mathcal{R}\mathcal{S}_{2}^{\mathrm{rf}} = \mathcal{W}\mathcal{P}\mathcal{H}\mathcal{R}\mathcal{S}^{\mathrm{rf}}_{2} = \mathcal{P}\mathcal{H}\mathcal{R}\mathcal{S}_{2}\).
\end{enumerate}
\end{theorem}

\begin{proof}
The first part follows from Theorem \ref{thm:phrtablepower}. For the second part, \(\mathcal{E}\mathcal{O}\mathcal{L} \subseteq \mathcal{P}\mathcal{H}\mathcal{R}\mathcal{S}_{k,1}^{\mathrm{rf}}\) follows from Theorem \ref{thm:phrset0l}, \(\mathcal{P}\mathcal{H}\mathcal{R}\mathcal{S}_{k,1}^{\mathrm{rf}} \subseteq \mathcal{W}\mathcal{P}\mathcal{H}\mathcal{R}\mathcal{S}^{\mathrm{rf}}_{k,1}\) follows by definition using the identity coding, and \(\mathcal{W}\mathcal{P}\mathcal{H}\mathcal{R}\mathcal{S}^{\mathrm{rf}}_{k,1} \subseteq \mathcal{P}\mathcal{H}\mathcal{R}\mathcal{S}_{k,1}\) follows using the fact that \(\mathcal{P}\mathcal{H}\mathcal{R}\mathcal{S}_{k,1}\) is closed under weak codings (Theorem \ref{cor:phrshom}). The third part follows in a similar way to the second. The final two statements follow from Theorem \ref{thm:phrset0l} and the earlier inclusions we established in this proof.
\end{proof}

Using the results and proofs from Subsection \ref{subsec:closure}, it is not too difficult to see that \(\mathcal{W}\mathcal{P}\mathcal{H}\mathcal{R}\mathcal{S}_k^{\mathrm{rf}}\) is a hyper-algebraically closed full AFL, for all \(k \geq 2\). We finish this subsection by giving the fine details of the closure properties.

\begin{theorem}[WPHRS Closed Under Substitutions]
Let \(L \subseteq A^*\). Then:

\begin{enumerate}
    \item for any \(k \geq 2\), if \(L\) is a \(k\)-HRS language and \(h\) is a \(\mathcal{W}\mathcal{P}\mathcal{H}\mathcal{R}\mathcal{S}_{k,1}^{\mathrm{rf}}\)-substitution on \(A\), then \(h(L)\) is a repetition-free \((k, 1)\)-WPHRS language;
    \item for any \(k \geq 2\), if \(L\) is a repetition-free \((k, 1)\)-WPHRS language and \(h\) is a finite substitution on \(A\), then \(h(L)\) is a repetition-free \((k, 1)\)-WPHRS language;
    \item for any \(k \geq 2\), if \(L\) is a repetition-free \(k\)-PHRS language and \(h\) is a \(\mathcal{W}\mathcal{P}\mathcal{H}\mathcal{R}\mathcal{S}_k^{\mathrm{rf}}\)-substitution on \(A\), then \(h(L)\) is a repetition-free \(k\)-WPHRS language.
\end{enumerate}
\end{theorem}

\begin{corollary}[WPHRS Closed Under Homomorphisms]
Let \(L \subseteq A^*\) be a repetition-free \((k, l)\)-WPHRS language for any \(k \geq 2, l \geq 1\) and \(\varphi: A^* \to B^*\) be a homomorphism. Then \(\varphi(L)\) is a repetition-free \((k, l)\)-WPHRS language.
\end{corollary}

\begin{theorem}[WPHRS Hyper-Algebraically Closed]
Let \(L \subseteq A^*\). Then, for any \(k \geq 2\), if \(L\) is a repetition-free \(k\)-WPHRS language and \(H\) is a finite set of \(\mathcal{W}\mathcal{P}\mathcal{H}\mathcal{R}\mathcal{S}_k^{\mathrm{rf}}\)-substitutions on \(A\), then \(\mathrm{ITER}_{H}(L)\) is a repetition-free \(k\)-WPHRS language.
\end{theorem}

\begin{theorem}[WPHRS Closed Under Rational Operations]
For all \(k \geq 2, l \geq 1\), \(\mathcal{W}\mathcal{P}\mathcal{H}\mathcal{R}\mathcal{S}_{k}^{\mathrm{rf}}\) is closed under rational operations.
\end{theorem}

\begin{theorem}[WPHRS Closed Under Rational Intersection]
For all \(k \geq 2, l \geq 1\), \(\mathcal{W}\mathcal{P}\mathcal{H}\mathcal{R}\mathcal{S}_{k}^{\mathrm{rf}}\) is closed under rational intersection.
\end{theorem}

\begin{theorem}[WPHRS Closed Under Inverse Homomorphisms]
For all \(k \geq 2, l \geq 1\), \(\mathcal{W}\mathcal{P}\mathcal{H}\mathcal{R}\mathcal{S}_{k}^{\mathrm{rf}}\) is closed under inverse homomorphisms.
\end{theorem}

\section{Conclusion and Future Work}

We have shown some foundational properties of parallel hyperedge replacement grammars, with a focus on string generational power, showing that the family of PHRS languages is a hyper-algebraically closed super AFL, containing all MCF and ET0L languages. In Subsection \ref{subsec:wphrs} we discuss the possible gap between the string generative power of repetition-free and not necessarily repetition-free PHR grammars. We conjecture that there is no gap:

\begin{conjecture}[PHR String Generational Power] \label{conj:phrgenpower}
For all \(k \geq 2, l \geq 1\), we have the following: \(\mathcal{P}\mathcal{H}\mathcal{R}\mathcal{S}_{k,l}^{\mathrm{rf}} = \mathcal{W}\mathcal{P}\mathcal{H}\mathcal{R}\mathcal{S}^{\mathrm{rf}}_{k,l} = \mathcal{P}\mathcal{H}\mathcal{R}\mathcal{S}_{k,l}\).
\end{conjecture}

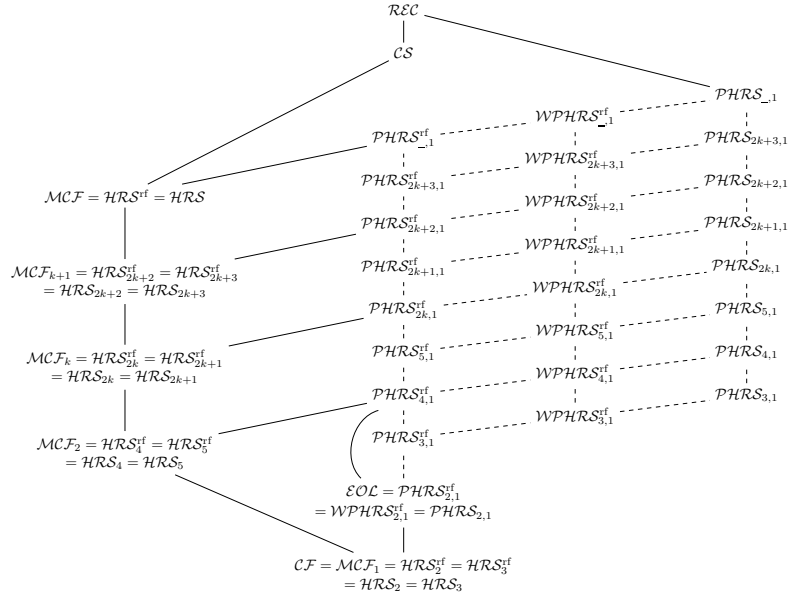
\begin{figure}[!ht]
\centering
\scalebox{0.57}{
\begin{tikzpicture}
  \node[align=center] (a) at (-9,1) {\,};
  \node[align=center] (a) at (9,1) {\,};

  \node[align=center] (d) at (-6.5,3.666667) {$\mathcal{M}\mathcal{C}\mathcal{F}_2 = \mathcal{H}\mathcal{R}\mathcal{S}_4^{\mathrm{rf}} = \mathcal{H}\mathcal{R}\mathcal{S}_5^{\mathrm{rf}}$\\$= \mathcal{H}\mathcal{R}\mathcal{S}_4 = \mathcal{H}\mathcal{R}\mathcal{S}_5$};
  \node[align=center] (h) at (-6.5,5.666667) {$\mathcal{M}\mathcal{C}\mathcal{F}_k = \mathcal{H}\mathcal{R}\mathcal{S}_{2k}^{\mathrm{rf}} = \mathcal{H}\mathcal{R}\mathcal{S}_{2k+1}^{\mathrm{rf}}$\\$= \mathcal{H}\mathcal{R}\mathcal{S}_{2k} = \mathcal{H}\mathcal{R}\mathcal{S}_{2k+1}$};
  \node[align=center] (l) at (-6.5,7.666667) {$\mathcal{M}\mathcal{C}\mathcal{F}_{k+1} = \mathcal{H}\mathcal{R}\mathcal{S}_{2k+2}^{\mathrm{rf}} = \mathcal{H}\mathcal{R}\mathcal{S}_{2k+3}^{\mathrm{rf}}$\\$= \mathcal{H}\mathcal{R}\mathcal{S}_{2k+2} = \mathcal{H}\mathcal{R}\mathcal{S}_{2k+3}$};
  \node[align=center] (r) at (-6.5,9.666667) {$\mathcal{M}\mathcal{C}\mathcal{F} = \mathcal{H}\mathcal{R}\mathcal{S}^{\mathrm{rf}} = \mathcal{H}\mathcal{R}\mathcal{S}$};

  \node[align=center] (c) at (0.0,0.833333) {$\mathcal{C}\mathcal{F} = \mathcal{M}\mathcal{C}\mathcal{F}_1 = \mathcal{H}\mathcal{R}\mathcal{S}_2^{\mathrm{rf}} = \mathcal{H}\mathcal{R}\mathcal{S}_3^{\mathrm{rf}}$\\$= \mathcal{H}\mathcal{R}\mathcal{S}_2 = \mathcal{H}\mathcal{R}\mathcal{S}_3$};
  \node[align=center] (e) at (0.0,2.5) {$\mathcal{E}\mathcal{O}\mathcal{L} = \mathcal{P}\mathcal{H}\mathcal{R}\mathcal{S}_{2,1}^{\mathrm{rf}}$\\$= \mathcal{W}\mathcal{P}\mathcal{H}\mathcal{R}\mathcal{S}_{2,1}^{\mathrm{rf}} = \mathcal{P}\mathcal{H}\mathcal{R}\mathcal{S}_{2,1}$};
  \node[align=center] (f) at (0.0,4.0) {$\mathcal{P}\mathcal{H}\mathcal{R}\mathcal{S}_{3,1}^{\mathrm{rf}}$};
  \node[align=center] (i) at (0.0,5.0) {$\mathcal{P}\mathcal{H}\mathcal{R}\mathcal{S}_{4,1}^{\mathrm{rf}}$};
  \node[align=center] (j) at (0.0,6.0) {$\mathcal{P}\mathcal{H}\mathcal{R}\mathcal{S}_{5,1}^{\mathrm{rf}}$};
  \node[align=center] (m) at (0.0,7.0) {$\mathcal{P}\mathcal{H}\mathcal{R}\mathcal{S}_{2k,1}^{\mathrm{rf}}$};
  \node[align=center] (o) at (0.0,8.0) {$\mathcal{P}\mathcal{H}\mathcal{R}\mathcal{S}_{2k+1,1}^{\mathrm{rf}}$};
  \node[align=center] (y) at (0.0,9.0) {$\mathcal{P}\mathcal{H}\mathcal{R}\mathcal{S}_{2k+2,1}^{\mathrm{rf}}$};
  \node[align=center] (yy) at (0.0,10.0) {$\mathcal{P}\mathcal{H}\mathcal{R}\mathcal{S}_{2k+3,1}^{\mathrm{rf}}$};
  \node[align=center] (s) at (0.0,11.0) {$\mathcal{P}\mathcal{H}\mathcal{R}\mathcal{S}_{\underline{\,\,\,},1}^{\mathrm{rf}}$};
  \node[align=center] (v) at (0.0,13.0) {$\mathcal{C}\mathcal{S}$};
  \node[align=center] (w) at (0.0,14.0) {$\mathcal{R}\mathcal{E}\mathcal{C}$};

  \node[align=center] (g) at (4.0,4.5) {$\mathcal{W}\mathcal{P}\mathcal{H}\mathcal{R}\mathcal{S}_{3,1}^{\mathrm{rf}}$};
  \node[align=center] (k) at (4.0,5.5) {$\mathcal{W}\mathcal{P}\mathcal{H}\mathcal{R}\mathcal{S}_{4,1}^{\mathrm{rf}}$};
  \node[align=center] (n) at (4.0,6.5) {$\mathcal{W}\mathcal{P}\mathcal{H}\mathcal{R}\mathcal{S}_{5,1}^{\mathrm{rf}}$};
  \node[align=center] (p) at (4.0,7.5) {$\mathcal{W}\mathcal{P}\mathcal{H}\mathcal{R}\mathcal{S}_{2k,1}^{\mathrm{rf}}$};
  \node[align=center] (q) at (4.0,8.5) {$\mathcal{W}\mathcal{P}\mathcal{H}\mathcal{R}\mathcal{S}_{2k+1,1}^{\mathrm{rf}}$};
  \node[align=center] (x) at (4.0,9.5) {$\mathcal{W}\mathcal{P}\mathcal{H}\mathcal{R}\mathcal{S}_{2k+2,1}^{\mathrm{rf}}$};
  \node[align=center] (xx) at (4.0,10.5) {$\mathcal{W}\mathcal{P}\mathcal{H}\mathcal{R}\mathcal{S}_{2k+3,1}^{\mathrm{rf}}$};
  \node[align=center] (u) at (4.0,11.5) {$\mathcal{W}\mathcal{P}\mathcal{H}\mathcal{R}\mathcal{S}_{\underline{\,\,\,},1}^{\mathrm{rf}}$};

  \node[align=center] (yg) at (8.0,5.0) {$\mathcal{P}\mathcal{H}\mathcal{R}\mathcal{S}_{3,1}$};
  \node[align=center] (yk) at (8.0,6.0) {$\mathcal{P}\mathcal{H}\mathcal{R}\mathcal{S}_{4,1}$};
  \node[align=center] (yn) at (8.0,7.0) {$\mathcal{P}\mathcal{H}\mathcal{R}\mathcal{S}_{5,1}$};
  \node[align=center] (yp) at (8.0,8.0) {$\mathcal{P}\mathcal{H}\mathcal{R}\mathcal{S}_{2k,1}$};
  \node[align=center] (yq) at (8.0,9.0) {$\mathcal{P}\mathcal{H}\mathcal{R}\mathcal{S}_{2k+1,1}$};
  \node[align=center] (yx) at (8.0,10.0) {$\mathcal{P}\mathcal{H}\mathcal{R}\mathcal{S}_{2k+2,1}$};
  \node[align=center] (yxx) at (8.0,11.0) {$\mathcal{P}\mathcal{H}\mathcal{R}\mathcal{S}_{2k+3,1}$};
  \node[align=center] (yu) at (8.0,12.0) {$\mathcal{P}\mathcal{H}\mathcal{R}\mathcal{S}_{\underline{\,\,\,},1}$};

  \draw (c) -- (d);
  \draw (c) -- (e);
  \draw (d) -- (h);
  \draw (h) -- (l);
  \draw (l) -- (r);
  
  \draw (d) -- (i);
  \draw (h) -- (m);
  \draw (l) -- (y);
  \draw (r) -- (s);

  \draw (r) -- (v);
  \draw (v) -- (w);

  \draw (e) to [bend left=60] (i);

  \draw[dashed] (e) -- (f);
  \draw[dashed] (f) -- (i);
  \draw[dashed] (i) -- (j);
  \draw[dashed] (j) -- (m);
  \draw[dashed] (m) -- (o);
  \draw[dashed] (o) -- (y);
  \draw[dashed] (y) -- (yy);
  \draw[dashed] (yy) -- (s);

  \draw[dashed] (f) -- (g);
  \draw[dashed] (i) -- (k);
  \draw[dashed] (j) -- (n);
  \draw[dashed] (m) -- (p);
  \draw[dashed] (o) -- (q);
  \draw[dashed] (s) -- (u);
  \draw[dashed] (y) -- (x);
  \draw[dashed] (yy) -- (xx);

  \draw[dashed] (g) -- (yg);
  \draw[dashed] (k) -- (yk);
  \draw[dashed] (n) -- (yn);
  \draw[dashed] (p) -- (yp);
  \draw[dashed] (q) -- (yq);
  \draw[dashed] (u) -- (yu);
  \draw[dashed] (x) -- (yx);
  \draw[dashed] (xx) -- (yxx);

  \draw[dashed] (g) -- (k);
  \draw[dashed] (k) -- (n);
  \draw[dashed] (n) -- (p);
  \draw[dashed] (p) -- (q);
  \draw[dashed] (q) -- (x);
  \draw[dashed] (x) -- (xx);
  \draw[dashed] (xx) -- (u);

  \draw[dashed] (yg) -- (yk);
  \draw[dashed] (yk) -- (yn);
  \draw[dashed] (yn) -- (yp);
  \draw[dashed] (yp) -- (yq);
  \draw[dashed] (yq) -- (yx);
  \draw[dashed] (yx) -- (yxx);
  \draw[dashed] (yxx) -- (yu);
  \draw (yu) -- (w);
\end{tikzpicture}
}
\caption{Proved single table string language hierarchy ($k \geq 3$)}
\label{fig:str-hierarchy-1}
\vspace{0.5em}
\end{figure}

It remains future work to show that the family of PHRS languages is a strict subclass of the context-sensitive languages. We conjecture this to be true, and we also conjecture that only even increments in order increase string generative power. Figure \ref{fig:closuretable} summarises the formal language closure properties we know, and Figures \ref{fig:str-hierarchy-1}, \ref{fig:str-hierarchy-2}, \ref{fig:conj-str-hierarchy} and \ref{fig:hypergraph-hierarchy} summarise the key language hierarchies, where \(\mathcal{P}\mathcal{H}\mathcal{R}_{\underline{\,\,\,},l} = \bigcup_{k \geq 2} \mathcal{P}\mathcal{H}\mathcal{R}_{k,l}\) and \(\mathcal{P}\mathcal{H}\mathcal{R}\mathcal{S}_{\underline{\,\,\,},l} = \bigcup_{k \geq 2} \mathcal{P}\mathcal{H}\mathcal{R}\mathcal{S}_{k,l}\) for any \(l \geq 1\).

\begin{conjecture}[CS Generalises PHRS]
\(\mathcal{P}\mathcal{H}\mathcal{R}\mathcal{S} \subsetneq \mathcal{C}\mathcal{S}\).
\end{conjecture}

\begin{conjecture}[PHRS Grouping]
For all \(k \geq 1\), \(\mathcal{P}\mathcal{H}\mathcal{R}\mathcal{S}_{2k} = \mathcal{P}\mathcal{H}\mathcal{R}\mathcal{S}_{2k+1}\).
\end{conjecture}

Because \(\mathcal{P}\mathcal{H}\mathcal{R}\mathcal{S}\) is closed under inverse homomorphisms, we know that the property of having a PHRS word problem is independent of the presentation. We have additionally shown that PHRS groups are closed under free product. We also conjecture the following, which has a wide-reaching corollary:

\begin{conjecture}[PHRS WP Double Torus] \label{conj:wpdt}
The fundamental group of the double torus admits a PHRS word problem which is neither an MCF nor ET0L language.
\end{conjecture}

\begin{corollary}
If Conjecture \ref{conj:wpdt} is true, then the word problem of any surface group is PHRS.
\end{corollary}

\begin{proof}
By a \emph{surface} here, we mean a closed, connected, orientable, 2-manifold, and by a \emph{surface group}, we mean the fundamental group of a surface. Any surface always has a finite genus. The genus \(0\) surface (the sphere) gives us the trivial group, and \(1\) (the torus), \(\mathbb{Z}^2\) (see for example \cite{Massey77a}). We know both of these groups are regular, \(2\)-MCF \cite{Ho18a}, respectively, so certainly PHRS (Theorem \ref{thm:hrsmcf}). For higher genuses, it follows from the Fundamental Theorem of Covering Spaces (Theorem 1.38 of \cite{Hatcher02a}) that the fundamental group appears as a finitely generated subgroup of the fundamental group of a genus \(2\) surface such as a double torus. Because \(\mathcal{P}\mathcal{H}\mathcal{R}\mathcal{S}\) is a full AFL, if the double torus has fundamental group with PHRS word problem, then all its finitely generated subgroups do too (Theorem \ref{thm:wpafl}).
\end{proof}

Highly related to the word problem is the consideration of sets of solutions of more general equations over groups or other structures. It is a recent result that solutions set (of fixed normal forms) of finite systems of equations in hyperbolic groups are EDT0L languages \cite{Ciobanu-Elder19a}. We are yet to consider deterministic parallel hyperedge replacement, but it may be possible to establish that other classes of groups have solution sets that are deterministic parallel hyperedge replacement string languages.

Recall from Theorem \ref{thm:emptyrhs} that the ET0L languages are the smallest hyper-algebraically closed super AFL. It is possible that the PHRS languages could be the smallest hyper-algebraically closed super AFL containing the MCF languages. Other more general future work would include investigating both the tree and graph generational power of PHR grammars, and investigating more decidability and complexity results for basic problems relating to PHR grammars. We do not know if the finiteness problem is decidable, but we conjecture that this can be decided, in general. If Conjecture \ref{conj:phrgenpower} fails, it will also be worthwhile to fill out some of the question marks in Figure \ref{fig:closuretable}.

\begin{conjecture}[Decidable PHR Finiteness]
The following problem is decidable, with an explicit algorithm:
  \begin{prob}
    \probinstance{A PHR grammar \(\mathcal{G} = (\mathcal{C}, A, S, \mathcal{T})\).}
    \probquestion{Does \(\mathrm{L}(\mathcal{G})\) contain only finitely many non-isomorphic hypergraphs?}
  \end{prob}
\end{conjecture}

\begin{figure}[!ht]
\centering
\scalebox{0.5}{
\begin{tabular}{|l|c|c|c|c|c|c|c|}
\hline
Operation/Family & \multicolumn{1}{>{\centering\arraybackslash}p{1.9cm}|}{$\mathcal{H}\mathcal{R}\mathcal{S}_k^{\mathrm{rf}}$ $= \mathcal{H}\mathcal{R}\mathcal{S}_k$} & \multicolumn{1}{>{\centering\arraybackslash}p{1.9cm}|}{$\mathcal{P}\mathcal{H}\mathcal{R}\mathcal{S}_{k,1}^{\mathrm{rf}}$} & \multicolumn{1}{>{\centering\arraybackslash}p{1.9cm}|}{$\mathcal{W}\mathcal{P}\mathcal{H}\mathcal{R}\mathcal{S}_{k,1}^{\mathrm{rf}}$} & \multicolumn{1}{>{\centering\arraybackslash}p{1.9cm}|}{$\mathcal{P}\mathcal{H}\mathcal{R}\mathcal{S}_{k,1}$} & \multicolumn{1}{>{\centering\arraybackslash}p{1.9cm}|}{$\mathcal{P}\mathcal{H}\mathcal{R}\mathcal{S}_k^{\mathrm{rf}}$} & \multicolumn{1}{>{\centering\arraybackslash}p{1.9cm}|}{$\mathcal{W}\mathcal{P}\mathcal{H}\mathcal{R}\mathcal{S}_k^{\mathrm{rf}}$} & \multicolumn{1}{>{\centering\arraybackslash}p{1.9cm}|}{$\mathcal{P}\mathcal{H}\mathcal{R}\mathcal{S}_k$} \\ \hline
Rational Operations & \ding{51} & \ding{51} & \ding{51} & \ding{51} & \ding{51} & \ding{51} & \ding{51} \\ \hline
Rational Intersection & \ding{51} & \ding{51} & \ding{51} & \ding{51} & \ding{51} & \ding{51} & \ding{51} \\ \hline
Inverse Homomorphisms & \ding{51} & ? & ? & ? & ? & \ding{51} & \ding{51} \\ \hline
Non-Erasing Homomorphisms & \ding{51} & \ding{51} & \ding{51} & \ding{51} & \ding{51} & \ding{51} & \ding{51} \\ \hline
Arbitrary Homomorphisms & \ding{51} & ? & \ding{51} & \ding{51} & ? & \ding{51} & \ding{51} \\ \hline
Non-Erasing Finite Substitutions & \ding{51} & \ding{51} & \ding{51} & \ding{51} & \ding{51} & \ding{51} & \ding{51} \\ \hline
Arbitrary Finite Substitutions & \ding{51} & ? & \ding{51} & \ding{51} & ? & \ding{51} & \ding{51} \\ \hline
Non-Erasing Substitutions & \ding{51} & ? & ? & ? & \ding{51} & \ding{51} & \ding{51} \\ \hline
Arbitrary Substitutions & \ding{51} & ? & ? & ? & ? & \ding{51} & \ding{51} \\ \hline
Iterated Nested Non-Erasing Substitutions & \ding{51} & ? & ? & ? & \ding{51} & \ding{51} & \ding{51} \\ \hline
Iterated Nested Arbitrary Substitutions & \ding{51} & ? & ? & ? & ? & \ding{51} & \ding{51} \\ \hline
Iterated Non-Erasing Substitutions & \ding{55} & ? & ? & ? & \ding{51} & \ding{51} & \ding{51} \\ \hline
Iterated Arbitrary Substitutions & \ding{55} & ? & ? & ? & ? & \ding{51} & \ding{51} \\ \hline
Hyper-Algebraic Closure & \ding{55} & \ding{55} & \ding{55} & \ding{55} & ? & \ding{51} & \ding{51} \\ \hline
\end{tabular}
}
\caption{Summary of formal language closure properties ($k \geq 2$)}
\label{fig:closuretable}
\end{figure}

\begin{figure}[!ht]
\centering
\scalebox{0.57}{
\begin{tikzpicture}
  \node[align=center] (d) at (-7.25,3.666667) {$\mathcal{M}\mathcal{C}\mathcal{F}_2 = \mathcal{H}\mathcal{R}\mathcal{S}_4^{\mathrm{rf}} = \mathcal{H}\mathcal{R}\mathcal{S}_5^{\mathrm{rf}}$\\$= \mathcal{H}\mathcal{R}\mathcal{S}_4 = \mathcal{H}\mathcal{R}\mathcal{S}_5$};
  \node[align=center] (h) at (-7.25,5.666667) {$\mathcal{M}\mathcal{C}\mathcal{F}_k = \mathcal{H}\mathcal{R}\mathcal{S}_{2k}^{\mathrm{rf}} = \mathcal{H}\mathcal{R}\mathcal{S}_{2k+1}^{\mathrm{rf}}$\\$= \mathcal{H}\mathcal{R}\mathcal{S}_{2k} = \mathcal{H}\mathcal{R}\mathcal{S}_{2k+1}$};
  \node[align=center] (l) at (-7.25,7.666667) {$\mathcal{M}\mathcal{C}\mathcal{F}_{k+1} = \mathcal{H}\mathcal{R}\mathcal{S}_{2k+2}^{\mathrm{rf}} = \mathcal{H}\mathcal{R}\mathcal{S}_{2k+3}^{\mathrm{rf}}$\\$= \mathcal{H}\mathcal{R}\mathcal{S}_{2k+2} = \mathcal{H}\mathcal{R}\mathcal{S}_{2k+3}$};
  \node[align=center] (r) at (-7.25,9.666667) {$\mathcal{M}\mathcal{C}\mathcal{F} = \mathcal{H}\mathcal{R}\mathcal{S}^{\mathrm{rf}} = \mathcal{H}\mathcal{R}\mathcal{S}$};

  \node[align=center] (c) at (0.0,0.833333) {$\mathcal{C}\mathcal{F} = \mathcal{M}\mathcal{C}\mathcal{F}_1 = \mathcal{H}\mathcal{R}\mathcal{S}_2^{\mathrm{rf}} = \mathcal{H}\mathcal{R}\mathcal{S}_3^{\mathrm{rf}}$\\$= \mathcal{H}\mathcal{R}\mathcal{S}_2 = \mathcal{H}\mathcal{R}\mathcal{S}_3$};
  \node[align=center] (e) at (0.0,2.5) {$\mathcal{E}\mathcal{T}\mathcal{O}\mathcal{L} = \mathcal{P}\mathcal{H}\mathcal{R}\mathcal{S}_{2,2}^{\mathrm{rf}} = \mathcal{W}\mathcal{P}\mathcal{H}\mathcal{R}\mathcal{S}_{2,2}^{\mathrm{rf}} = \mathcal{P}\mathcal{H}\mathcal{R}\mathcal{S}_{2,2}$\\$= \mathcal{P}\mathcal{H}\mathcal{R}\mathcal{S}_2^{\mathrm{rf}} = \mathcal{W}\mathcal{P}\mathcal{H}\mathcal{R}\mathcal{S}_2^{\mathrm{rf}} = \mathcal{P}\mathcal{H}\mathcal{R}\mathcal{S}_2$};

  \node[align=center] (f) at (-2.25,4.0) {$\mathcal{P}\mathcal{H}\mathcal{R}\mathcal{S}_{3,2}^{\mathrm{rf}} = \mathcal{P}\mathcal{H}\mathcal{R}\mathcal{S}_{3}^{\mathrm{rf}}$};
  \node[align=center] (i) at (-2.25,5.0) {$\mathcal{P}\mathcal{H}\mathcal{R}\mathcal{S}_{4,2}^{\mathrm{rf}} = \mathcal{P}\mathcal{H}\mathcal{R}\mathcal{S}_{4}^{\mathrm{rf}}$};
  \node[align=center] (j) at (-2.25,6.0) {$\mathcal{P}\mathcal{H}\mathcal{R}\mathcal{S}_{5,2}^{\mathrm{rf}} = \mathcal{P}\mathcal{H}\mathcal{R}\mathcal{S}_{5}^{\mathrm{rf}}$};
  \node[align=center] (m) at (-2.25,7.0) {$\mathcal{P}\mathcal{H}\mathcal{R}\mathcal{S}_{2k,2}^{\mathrm{rf}} = \mathcal{P}\mathcal{H}\mathcal{R}\mathcal{S}_{2k}^{\mathrm{rf}}$};
  \node[align=center] (o) at (-2.25,8.0) {$\mathcal{P}\mathcal{H}\mathcal{R}\mathcal{S}_{2k+1,2}^{\mathrm{rf}} = \mathcal{P}\mathcal{H}\mathcal{R}\mathcal{S}_{2k+1}^{\mathrm{rf}}$};
  \node[align=center] (y) at (-2.25,9.0) {$\mathcal{P}\mathcal{H}\mathcal{R}\mathcal{S}_{2k+2,2}^{\mathrm{rf}} = \mathcal{P}\mathcal{H}\mathcal{R}\mathcal{S}_{2k+2}^{\mathrm{rf}}$};
  \node[align=center] (yy) at (-2.25,10.0) {$\mathcal{P}\mathcal{H}\mathcal{R}\mathcal{S}_{2k+3,2}^{\mathrm{rf}} = \mathcal{P}\mathcal{H}\mathcal{R}\mathcal{S}_{2k+3}^{\mathrm{rf}}$};
  \node[align=center] (s) at (-2.25,11.0) {$\mathcal{P}\mathcal{H}\mathcal{R}\mathcal{S}_{\underline{\,\,\,},2}^{\mathrm{rf}} = \mathcal{P}\mathcal{H}\mathcal{R}\mathcal{S}^{\mathrm{rf}}$};

  \node[align=center] (v) at (0.0,13.5) {$\mathcal{C}\mathcal{S}$};
  \node[align=center] (w) at (0.0,14.5) {$\mathcal{R}\mathcal{E}\mathcal{C}$};

  \node[align=center] (ee) at (0.55,2.95) {$\,$};

  \node[align=center] (g) at (3.25,4.5) {$\mathcal{W}\mathcal{P}\mathcal{H}\mathcal{R}\mathcal{S}_{3,2}^{\mathrm{rf}} = \mathcal{W}\mathcal{P}\mathcal{H}\mathcal{R}\mathcal{S}_{3}^{\mathrm{rf}}$};
  \node[align=center] (k) at (3.25,5.5) {$\mathcal{W}\mathcal{P}\mathcal{H}\mathcal{R}\mathcal{S}_{4,2}^{\mathrm{rf}} = \mathcal{W}\mathcal{P}\mathcal{H}\mathcal{R}\mathcal{S}_{4}^{\mathrm{rf}}$};
  \node[align=center] (n) at (3.25,6.5) {$\mathcal{W}\mathcal{P}\mathcal{H}\mathcal{R}\mathcal{S}_{5,2}^{\mathrm{rf}} = \mathcal{W}\mathcal{P}\mathcal{H}\mathcal{R}\mathcal{S}_{5}^{\mathrm{rf}}$};
  \node[align=center] (p) at (3.25,7.5) {$\mathcal{W}\mathcal{P}\mathcal{H}\mathcal{R}\mathcal{S}_{2k,2}^{\mathrm{rf}} = \mathcal{W}\mathcal{P}\mathcal{H}\mathcal{R}\mathcal{S}_{2k}^{\mathrm{rf}}$};
  \node[align=center] (q) at (3.25,8.5) {$\mathcal{W}\mathcal{P}\mathcal{H}\mathcal{R}\mathcal{S}_{2k+1,2}^{\mathrm{rf}} = \mathcal{W}\mathcal{P}\mathcal{H}\mathcal{R}\mathcal{S}_{2k+1}^{\mathrm{rf}}$};
  \node[align=center] (x) at (3.25,9.5) {$\mathcal{W}\mathcal{P}\mathcal{H}\mathcal{R}\mathcal{S}_{2k+2,2}^{\mathrm{rf}} = \mathcal{W}\mathcal{P}\mathcal{H}\mathcal{R}\mathcal{S}_{2k+2}^{\mathrm{rf}}$};
  \node[align=center] (xx) at (3.25,10.5) {$\mathcal{W}\mathcal{P}\mathcal{H}\mathcal{R}\mathcal{S}_{2k+3,2}^{\mathrm{rf}} = \mathcal{W}\mathcal{P}\mathcal{H}\mathcal{R}\mathcal{S}_{2k+3}^{\mathrm{rf}}$};
  \node[align=center] (u) at (3.25,11.5) {$\mathcal{W}\mathcal{P}\mathcal{H}\mathcal{R}\mathcal{S}_{\underline{\,\,\,},2}^{\mathrm{rf}} = \mathcal{W}\mathcal{P}\mathcal{H}\mathcal{R}\mathcal{S}^{\mathrm{rf}}$};

  \node[align=center] (yg) at (8.5,5.0) {$\mathcal{P}\mathcal{H}\mathcal{R}\mathcal{S}_{3,2} = \mathcal{P}\mathcal{H}\mathcal{R}\mathcal{S}_{3}$};
  \node[align=center] (yk) at (8.5,6.0) {$\mathcal{P}\mathcal{H}\mathcal{R}\mathcal{S}_{4,2} = \mathcal{P}\mathcal{H}\mathcal{R}\mathcal{S}_{4}$};
  \node[align=center] (yn) at (8.5,7.0) {$\mathcal{P}\mathcal{H}\mathcal{R}\mathcal{S}_{5,2} = \mathcal{P}\mathcal{H}\mathcal{R}\mathcal{S}_{5}$};
  \node[align=center] (yp) at (8.5,8.0) {$\mathcal{P}\mathcal{H}\mathcal{R}\mathcal{S}_{2k,2} = \mathcal{P}\mathcal{H}\mathcal{R}\mathcal{S}_{2k}$};
  \node[align=center] (yq) at (8.5,9.0) {$\mathcal{P}\mathcal{H}\mathcal{R}\mathcal{S}_{2k+1,2} = \mathcal{P}\mathcal{H}\mathcal{R}\mathcal{S}_{2k+1}$};
  \node[align=center] (yx) at (8.5,10.0) {$\mathcal{P}\mathcal{H}\mathcal{R}\mathcal{S}_{2k+2,2} = \mathcal{P}\mathcal{H}\mathcal{R}\mathcal{S}_{2k+2}$};
  \node[align=center] (yxx) at (8.5,11.0) {$\mathcal{P}\mathcal{H}\mathcal{R}\mathcal{S}_{2k+3,2} = \mathcal{P}\mathcal{H}\mathcal{R}\mathcal{S}_{2k+3}$};
  \node[align=center] (yu) at (8.5,12.0) {$\mathcal{P}\mathcal{H}\mathcal{R}\mathcal{S}_{\underline{\,\,\,},2} = \mathcal{P}\mathcal{H}\mathcal{R}\mathcal{S}$};

  \draw (c) -- (d);
  \draw (c) -- (e);
  \draw (d) -- (h);
  \draw (h) -- (l);
  \draw (l) -- (r);
  
  \draw (d) -- (i);
  \draw (h) -- (m);
  \draw (l) -- (y);
  \draw (r) -- (s);

  \draw (r) -- (v);
  \draw (v) -- (w);

  \draw (ee) to [bend right=30] (i);

  \draw[dashed] (e) -- (f);
  \draw[dashed] (f) -- (i);
  \draw[dashed] (i) -- (j);
  \draw[dashed] (j) -- (m);
  \draw[dashed] (m) -- (o);
  \draw[dashed] (o) -- (y);
  \draw[dashed] (y) -- (yy);
  \draw[dashed] (yy) -- (s);

  \draw[dashed] (f) -- (g);
  \draw[dashed] (i) -- (k);
  \draw[dashed] (j) -- (n);
  \draw[dashed] (m) -- (p);
  \draw[dashed] (o) -- (q);
  \draw[dashed] (s) -- (u);
  \draw[dashed] (y) -- (x);
  \draw[dashed] (yy) -- (xx);

  \draw[dashed] (g) -- (yg);
  \draw[dashed] (k) -- (yk);
  \draw[dashed] (n) -- (yn);
  \draw[dashed] (p) -- (yp);
  \draw[dashed] (q) -- (yq);
  \draw[dashed] (u) -- (yu);
  \draw[dashed] (x) -- (yx);
  \draw[dashed] (xx) -- (yxx);

  \draw[dashed] (g) -- (k);
  \draw[dashed] (k) -- (n);
  \draw[dashed] (n) -- (p);
  \draw[dashed] (p) -- (q);
  \draw[dashed] (q) -- (x);
  \draw[dashed] (x) -- (xx);
  \draw[dashed] (xx) -- (u);

  \draw[dashed] (yg) -- (yk);
  \draw[dashed] (yk) -- (yn);
  \draw[dashed] (yn) -- (yp);
  \draw[dashed] (yp) -- (yq);
  \draw[dashed] (yq) -- (yx);
  \draw[dashed] (yx) -- (yxx);
  \draw[dashed] (yxx) -- (yu);
  \draw (yu) -- (w);
\end{tikzpicture}
}
\caption{Proved multiple tables string language hierarchy ($k \geq 3$)}
\label{fig:str-hierarchy-2}
\vspace{0.5em}
\end{figure}
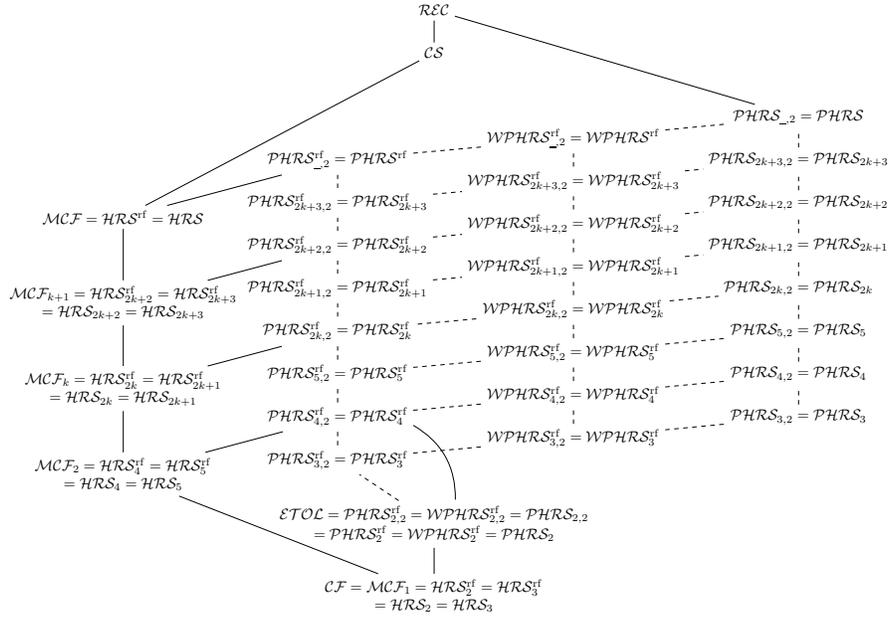

\begin{figure}[!ht]
\centering
\scalebox{0.57}{
\begin{tikzpicture}
  \node[align=center] (a) at (-6.75,5.0) {$\mathcal{M}\mathcal{C}\mathcal{F}_2 = \mathcal{H}\mathcal{R}\mathcal{S}_4^{\mathrm{rf}} = \mathcal{H}\mathcal{R}\mathcal{S}_5^{\mathrm{rf}}$\\$= \mathcal{H}\mathcal{R}\mathcal{S}_4 = \mathcal{H}\mathcal{R}\mathcal{S}_5$};
  \node[align=center] (b) at (-6.75,8.0) {$\mathcal{M}\mathcal{C}\mathcal{F}_k = \mathcal{H}\mathcal{R}\mathcal{S}_{2k}^{\mathrm{rf}} = \mathcal{H}\mathcal{R}\mathcal{S}_{2k+1}^{\mathrm{rf}}$\\$= \mathcal{H}\mathcal{R}\mathcal{S}_{2k} = \mathcal{H}\mathcal{R}\mathcal{S}_{2k+1}$};
  \node[align=center] (c) at (-6.75,11.0) {$\mathcal{M}\mathcal{C}\mathcal{F}_{k+1} = \mathcal{H}\mathcal{R}\mathcal{S}_{2k+2}^{\mathrm{rf}} = \mathcal{H}\mathcal{R}\mathcal{S}_{2k+3}^{\mathrm{rf}}$\\$= \mathcal{H}\mathcal{R}\mathcal{S}_{2k+2} = \mathcal{H}\mathcal{R}\mathcal{S}_{2k+3}$};
  \node[align=center] (d) at (-6.75,14.0) {$\mathcal{M}\mathcal{C}\mathcal{F} = \mathcal{H}\mathcal{R}\mathcal{S}^{\mathrm{rf}} = \mathcal{H}\mathcal{R}\mathcal{S}$};

  \node[align=center] (e) at (0.0,1.0) {$\mathcal{C}\mathcal{F} = \mathcal{M}\mathcal{C}\mathcal{F}_1 = \mathcal{H}\mathcal{R}\mathcal{S}_2^{\mathrm{rf}} = \mathcal{H}\mathcal{R}\mathcal{S}_3^{\mathrm{rf}}$\\$= \mathcal{H}\mathcal{R}\mathcal{S}_2 = \mathcal{H}\mathcal{R}\mathcal{S}_3$};
  \node[align=center] (f) at (0.0,4.0) {$\mathcal{E}\mathcal{O}\mathcal{L} = \mathcal{P}\mathcal{H}\mathcal{R}\mathcal{S}_{2,1}^{\mathrm{rf}} = \mathcal{P}\mathcal{H}\mathcal{R}\mathcal{S}_{3,1}^{\mathrm{rf}} = \mathcal{W}\mathcal{P}\mathcal{H}\mathcal{R}\mathcal{S}_{2,1}^{\mathrm{rf}}$\\$= \mathcal{W}\mathcal{P}\mathcal{H}\mathcal{R}\mathcal{S}_{3,1}^{\mathrm{rf}} = \mathcal{P}\mathcal{H}\mathcal{R}\mathcal{S}_{2,1} = \mathcal{P}\mathcal{H}\mathcal{R}\mathcal{S}_{3,1}$};
  \node[align=center] (g) at (0.0,7.0) {$\mathcal{P}\mathcal{H}\mathcal{R}\mathcal{S}_{4,1}^{\mathrm{rf}} = \mathcal{P}\mathcal{H}\mathcal{R}\mathcal{S}_{5,1}^{\mathrm{rf}} = \mathcal{W}\mathcal{P}\mathcal{H}\mathcal{R}\mathcal{S}_{4,1}^{\mathrm{rf}}$\\$= \mathcal{W}\mathcal{P}\mathcal{H}\mathcal{R}\mathcal{S}_{5,1}^{\mathrm{rf}} = \mathcal{P}\mathcal{H}\mathcal{R}\mathcal{S}_{4,1} = \mathcal{P}\mathcal{H}\mathcal{R}\mathcal{S}_{5,1}$};
  \node[align=center] (h) at (0.0,10.0) {$\mathcal{P}\mathcal{H}\mathcal{R}\mathcal{S}_{2k,1}^{\mathrm{rf}} = \mathcal{P}\mathcal{H}\mathcal{R}\mathcal{S}_{2k,1+1}^{\mathrm{rf}} = \mathcal{W}\mathcal{P}\mathcal{H}\mathcal{R}\mathcal{S}_{2k,1}^{\mathrm{rf}}$\\$= \mathcal{W}\mathcal{P}\mathcal{H}\mathcal{R}\mathcal{S}_{2k+1,1}^{\mathrm{rf}} = \mathcal{P}\mathcal{H}\mathcal{R}\mathcal{S}_{2k,1} = \mathcal{P}\mathcal{H}\mathcal{R}\mathcal{S}_{2k+1,1}$};
  \node[align=center] (i) at (0.0,13.0) {$\mathcal{P}\mathcal{H}\mathcal{R}\mathcal{S}_{2k+2,1}^{\mathrm{rf}} = \mathcal{P}\mathcal{H}\mathcal{R}\mathcal{S}_{2k+3,1}^{\mathrm{rf}} = \mathcal{W}\mathcal{P}\mathcal{H}\mathcal{R}\mathcal{S}_{2k+2,1}^{\mathrm{rf}}$\\$= \mathcal{W}\mathcal{P}\mathcal{H}\mathcal{R}\mathcal{S}_{2k+3,1}^{\mathrm{rf}} = \mathcal{P}\mathcal{H}\mathcal{R}\mathcal{S}_{2k+2,1} = \mathcal{P}\mathcal{H}\mathcal{R}\mathcal{S}_{2k+3,1}$};
  \node[align=center] (j) at (0.0,16.0) {$\mathcal{P}\mathcal{H}\mathcal{R}\mathcal{S}_{\underline{\,\,\,},1}^{\mathrm{rf}} = \mathcal{W}\mathcal{P}\mathcal{H}\mathcal{R}\mathcal{S}^{\mathrm{rf}}_{\underline{\,\,\,},1} = \mathcal{P}\mathcal{H}\mathcal{R}\mathcal{S}_{\underline{\,\,\,},1}$};
  \node[align=center] (k) at (0.0,20.5) {$\mathcal{C}\mathcal{S}$};
  \node[align=center] (l) at (0.0,21.5) {$\mathcal{R}\mathcal{E}\mathcal{C}$};

  \node[align=center] (m) at (7.75,6.0) {$\mathcal{E}\mathcal{T}\mathcal{O}\mathcal{L} = \mathcal{P}\mathcal{H}\mathcal{R}\mathcal{S}_{2,2}^{\mathrm{rf}} = \mathcal{P}\mathcal{H}\mathcal{R}\mathcal{S}_{3,2}^{\mathrm{rf}} = \mathcal{W}\mathcal{P}\mathcal{H}\mathcal{R}\mathcal{S}_{2,2}^{\mathrm{rf}}$\\$= \mathcal{W}\mathcal{P}\mathcal{H}\mathcal{R}\mathcal{S}_{3,2}^{\mathrm{rf}} = \mathcal{P}\mathcal{H}\mathcal{R}\mathcal{S}_{2,2} = \mathcal{P}\mathcal{H}\mathcal{R}\mathcal{S}_{3,2}$\\$= \mathcal{P}\mathcal{H}\mathcal{R}\mathcal{S}_2^{\mathrm{rf}} = \mathcal{P}\mathcal{H}\mathcal{R}\mathcal{S}_3^{\mathrm{rf}} = \mathcal{W}\mathcal{P}\mathcal{H}\mathcal{R}\mathcal{S}_2^{\mathrm{rf}}$\\$= \mathcal{W}\mathcal{P}\mathcal{H}\mathcal{R}\mathcal{S}_3^{\mathrm{rf}} = \mathcal{P}\mathcal{H}\mathcal{R}\mathcal{S}_2 = \mathcal{P}\mathcal{H}\mathcal{R}\mathcal{S}_3$};
  \node[align=center] (n) at (7.75,9.0) {$\mathcal{P}\mathcal{H}\mathcal{R}\mathcal{S}_{4,2}^{\mathrm{rf}} = \mathcal{P}\mathcal{H}\mathcal{R}\mathcal{S}_{5,2}^{\mathrm{rf}} = \mathcal{W}\mathcal{P}\mathcal{H}\mathcal{R}\mathcal{S}_{4,2}^{\mathrm{rf}}$\\$= \mathcal{W}\mathcal{P}\mathcal{H}\mathcal{R}\mathcal{S}_{5,2}^{\mathrm{rf}} = \mathcal{P}\mathcal{H}\mathcal{R}\mathcal{S}_{4,2} = \mathcal{P}\mathcal{H}\mathcal{R}\mathcal{S}_{5,2}$\\$= \mathcal{P}\mathcal{H}\mathcal{R}\mathcal{S}_4^{\mathrm{rf}} = \mathcal{P}\mathcal{H}\mathcal{R}\mathcal{S}_5^{\mathrm{rf}} = \mathcal{W}\mathcal{P}\mathcal{H}\mathcal{R}\mathcal{S}_4^{\mathrm{rf}}$\\$= \mathcal{W}\mathcal{P}\mathcal{H}\mathcal{R}\mathcal{S}_5^{\mathrm{rf}} = \mathcal{P}\mathcal{H}\mathcal{R}\mathcal{S}_4 = \mathcal{P}\mathcal{H}\mathcal{R}\mathcal{S}_5$};
  \node[align=center] (o) at (7.75,12.0) {$\mathcal{P}\mathcal{H}\mathcal{R}\mathcal{S}_{2k,2}^{\mathrm{rf}} = \mathcal{P}\mathcal{H}\mathcal{R}\mathcal{S}_{2k,2+1}^{\mathrm{rf}} = \mathcal{W}\mathcal{P}\mathcal{H}\mathcal{R}\mathcal{S}_{2k,2}^{\mathrm{rf}}$\\$= \mathcal{W}\mathcal{P}\mathcal{H}\mathcal{R}\mathcal{S}_{2k+1,2}^{\mathrm{rf}} = \mathcal{P}\mathcal{H}\mathcal{R}\mathcal{S}_{2k,2} = \mathcal{P}\mathcal{H}\mathcal{R}\mathcal{S}_{2k+1,2}$\\$= \mathcal{P}\mathcal{H}\mathcal{R}\mathcal{S}_{2k}^{\mathrm{rf}} = \mathcal{P}\mathcal{H}\mathcal{R}\mathcal{S}_{2k+1}^{\mathrm{rf}} = \mathcal{W}\mathcal{P}\mathcal{H}\mathcal{R}\mathcal{S}_{2k}^{\mathrm{rf}}$\\$= \mathcal{W}\mathcal{P}\mathcal{H}\mathcal{R}\mathcal{S}_{2k+1}^{\mathrm{rf}} = \mathcal{P}\mathcal{H}\mathcal{R}\mathcal{S}_{2k} = \mathcal{P}\mathcal{H}\mathcal{R}\mathcal{S}_{2k+1}$};
  \node[align=center] (p) at (7.75,15.0) {$\mathcal{P}\mathcal{H}\mathcal{R}\mathcal{S}_{2k+2,2}^{\mathrm{rf}} = \mathcal{P}\mathcal{H}\mathcal{R}\mathcal{S}_{2k+3,2}^{\mathrm{rf}} = \mathcal{W}\mathcal{P}\mathcal{H}\mathcal{R}\mathcal{S}_{2k+2,2}^{\mathrm{rf}}$\\$= \mathcal{W}\mathcal{P}\mathcal{H}\mathcal{R}\mathcal{S}_{2k+3,2}^{\mathrm{rf}} = \mathcal{P}\mathcal{H}\mathcal{R}\mathcal{S}_{2k+2,2} = \mathcal{P}\mathcal{H}\mathcal{R}\mathcal{S}_{2k+3,2}$\\$= \mathcal{P}\mathcal{H}\mathcal{R}\mathcal{S}_{2k+2}^{\mathrm{rf}} = \mathcal{P}\mathcal{H}\mathcal{R}\mathcal{S}_{2k+3}^{\mathrm{rf}} = \mathcal{W}\mathcal{P}\mathcal{H}\mathcal{R}\mathcal{S}_{2k+2}^{\mathrm{rf}}$\\$= \mathcal{W}\mathcal{P}\mathcal{H}\mathcal{R}\mathcal{S}_{2k+3}^{\mathrm{rf}} = \mathcal{P}\mathcal{H}\mathcal{R}\mathcal{S}_{2k+2} = \mathcal{P}\mathcal{H}\mathcal{R}\mathcal{S}_{2k+3}$};
  \node[align=center] (q) at (7.75,18.0) {$\mathcal{P}\mathcal{H}\mathcal{R}\mathcal{S}_{\underline{\,\,\,},2}^{\mathrm{rf}} = \mathcal{W}\mathcal{P}\mathcal{H}\mathcal{R}\mathcal{S}_{\underline{\,\,\,},2}^{\mathrm{rf}} = \mathcal{P}\mathcal{H}\mathcal{R}\mathcal{S}_{\underline{\,\,\,},2}$\\$= \mathcal{P}\mathcal{H}\mathcal{R}\mathcal{S}^{\mathrm{rf}} = \mathcal{W}\mathcal{P}\mathcal{H}\mathcal{R}\mathcal{S}^{\mathrm{rf}} = \mathcal{P}\mathcal{H}\mathcal{R}\mathcal{S}$};

  \draw (a) -- (b);
  \draw (b) -- (c);
  \draw (c) -- (d);

  \draw (e) -- (f);
  \draw (f) -- (g);
  \draw (g) -- (h);
  \draw (h) -- (i);
  \draw (i) -- (j);

  \draw (m) -- (n);
  \draw (n) -- (o);
  \draw (o) -- (p);
  \draw (p) -- (q);

  \draw (e) -- (a);
  \draw (f) -- (m);

  \draw (a) -- (g);
  \draw (g) -- (n);

  \draw (b) -- (h);
  \draw (h) -- (o);

  \draw (c) -- (i);
  \draw (i) -- (p);

  \draw (d) -- (j);
  \draw (j) -- (q);

  \draw (q) -- (k);
  \draw (k) -- (l);
\end{tikzpicture}
}
\caption{Conjectured string language hierarchy ($k \geq 3$)}
\label{fig:conj-str-hierarchy}
\vspace{0.5em}
\end{figure}

\begin{figure}[!ht]\centering
\scalebox{0.57}{
\begin{tikzpicture}
  \node[align=center] (a) at (0.0,0.0) {$\mathcal{H}\mathcal{R}_{0}$};
  \node[align=center] (b) at (4.0,0.8) {$\mathcal{P}\mathcal{H}\mathcal{R}_{0,1}$};
  \node[align=center] (c) at (0.0,1.6) {$\mathcal{H}\mathcal{R}_{k}$};
  \node[align=center] (d) at (8.0,1.6) {$\mathcal{P}\mathcal{H}\mathcal{R}_{0,2} = \mathcal{P}\mathcal{H}\mathcal{R}_{0}$};
  \node[align=center] (e) at (4.0,2.4) {$\mathcal{P}\mathcal{H}\mathcal{R}_{k,1}$};
  \node[align=center] (f) at (0.0,3.2) {$\mathcal{H}\mathcal{R}_{k+1}$};
  \node[align=center] (g) at (8.0,3.2) {$\mathcal{P}\mathcal{H}\mathcal{R}_{k,2} = \mathcal{P}\mathcal{H}\mathcal{R}_{k}$};
  \node[align=center] (h) at (4.0,4.0) {$\mathcal{P}\mathcal{H}\mathcal{R}_{k+1,1}$};
  \node[align=center] (i) at (0.0,4.8) {$\mathcal{H}\mathcal{R}$};
  \node[align=center] (j) at (8.0,4.8) {$\mathcal{P}\mathcal{H}\mathcal{R}_{k+1,2} = \mathcal{P}\mathcal{H}\mathcal{R}_{k+1}$};
  \node[align=center] (k) at (4.0,5.6) {$\mathcal{P}\mathcal{H}\mathcal{R}_{\underline{\,\,\,},1}$};
  \node[align=center] (l) at (8.0,6.4) {$\mathcal{P}\mathcal{H}\mathcal{R}_{\underline{\,\,\,},2} = \mathcal{P}\mathcal{H}\mathcal{R}$};

  \draw (a) -- (b);
  \draw[dashed] (b) -- (d);

  \draw (c) -- (e);
  \draw[dashed] (e) -- (g);

  \draw (f) -- (h);
  \draw[dashed] (h) -- (j);

  \draw (i) -- (k);
  \draw[dashed] (k) -- (l);

  \draw (a) -- (c);
  \draw (c) -- (f);
  \draw (f) -- (i);

  \draw (b) -- (e);
  \draw (e) -- (h);
  \draw (h) -- (k);

  \draw (d) -- (g);
  \draw (g) -- (j);
  \draw (j) -- (l);
\end{tikzpicture}
}
\caption{Proved hypergraph language hierarchy ($k \geq 1$)}
\label{fig:hypergraph-hierarchy}
\vspace{0.5em}
\end{figure}

\section*{Acknowledgements}

I should like to thank Detlef Plump for introducing me to graph transformation and teaching me to write papers for this audience, my supervisors Sarah Rees and Andrew Duncan for their guidance, Annegret Habel and Meng-Che Ho for their helpful email discussions regarding hyperedge replacement and surface groups, respectively, and Murray Elder for introducing me to MCF languages. I am also grateful to the anonymous reviewers for their comments on the earlier TERMGRAPH workshop proceedings version of this paper.

\bibliography{ms}

\end{document}